\documentclass[varenna]{cimento}

\newcommand{\eV}{{\rm eV}}
\newcommand{\keV}{{\rm keV}}

\newcommand{\Kel}{{\rm K}}

\newcommand{\cm}{{\rm cm}}
\newcommand{\Hz}{{\rm Hz}}
\newcommand{\ergs}{{\rm ergs}}
\newcommand{\sr}{{\rm sr}}

\newcommand{\MJysr}{{\rm MJy\,sr^{-1}}}

\newcommand{\Mpc}{{\rm Mpc}}
\newcommand{\GHz}{{\rm GHz}}

\newcommand{\expf}[1]{{{\rm e}^{#1}}}
\newcommand{\Jbb}{\mathcal{J}}

\newcommand{\zh}{{z_{\rm h}}}

\newcommand{\zmudc}{{z_{\rm dc}}}

\newcommand{\nS}{n_{\rm S}}

\newcommand{\nrun}{n_{\rm run}}

\newcommand{\kD}{k_{\rm D}}

\newcommand{\xe}{x_{\rm e}}

\newcommand{\xc}{x_{\rm c}}

\newcommand{\id}{{\,\rm d}}

\newcommand{\beq}{\begin{equation}}   %

\newcommand{\eeq}{\end{equation}}   %

\newcommand{\beqa}{\begin{eqnarray}}   %

\newcommand{\eeqa}{\end{eqnarray}}   %

\newcommand{\beal}[1]{\begin{align} #1 \end{align}}

\newcommand{\bspl}{\begin{split}}

\newcommand{\espl}{\end{split}}

\newcommand{\bsub}{\begin{subequations}}

\newcommand{\esub}{\end{subequations}}

\newcommand{\bmulti}{\begin{multline}}   %

\newcommand{\beqm}{\begin{mathletters}}   %

\newcommand{\eeqm}{\end{mathletters}}   %

\newcommand{\me}{m_{\rm e}}

\newcommand{\Ne}{N_{\rm e}}

\newcommand{\Te}{T_{\rm e}}

\newcommand{\Tg}{T_{\gamma}}

\newcommand{\The}{\theta_{\rm e}}

\newcommand{\Thg}{\theta_{\gamma}}

\newcommand{\sigT}{\sigma_{\rm T}}

\newcommand{\nPl}{n_{\rm bb}}

\newcommand{\pot}[2]{#1 \times 10^{#2}}

\newcommand{\Yp}{Y_{\rm p}}

\newcommand{\zeq}{z_{\rm eq}}

\usepackage{amsmath}
\usepackage{txfonts}
\usepackage{epsfig}
\usepackage{color}
\usepackage{grffile}
\usepackage{graphics}

\newcommand{\citep}[1]{\cite{#1}}
\newcommand{\citet}[1]{\cite{#1}}

\title{Future Steps in Cosmology using Spectral Distortions of the Cosmic Microwave Background}

\author{Jens Chluba \thanks{Jens.Chluba@Manchester.ac.uk}}

\institute{Jodrell Bank Centre for Astrophysics, School of Physics and Astronomy, \\ 
University of Manchester, Oxford Road, Manchester M13 9PL, UK}

\begin{document}

\maketitle

\begin{abstract}
Since the measurements of {COBE/FIRAS} in the mid-90's we know that the energy spectrum of the cosmic microwave background (CMB) is extremely close to that of a perfect blackbody at an average temperature $T_0\simeq2.726\,{\rm K}$. However, a number of early-universe processes are expected to create {\it CMB spectral distortions} -- departures of the average CMB energy spectrum from a blackbody -- at a level that is within reach of present-day technology. This provides strong motivation to study the physics of CMB spectral distortions and ask what these small signals might be able to tell us about the Universe we live in. 
In this lecture, I will give a broad-brush overview of recent theoretical and experimental developments, explaining why future spectroscopic measurements of the CMB will open an unexplored new window to early-universe and particle physics. I will give an introduction about the different types of distortions, how they evolve and thermalize and highlight some of the physical processes that can cause them. 
I hope to be able to convince you that CMB spectral distortions could open an exciting new path forward in CMB cosmology, which is complementary to planned and ongoing searches for primordial B-mode polarization signals. Spectral distortions should thus be considered very seriously as part of the activities in the next decades.
\end{abstract}

\section{Overview and motivation}
\vspace{-1mm}
Cosmology is now a precise scientific discipline, with detailed theoretical models that fit a wealth of very accurate measurements. Of the many cosmological data sets, the cosmic microwave background (CMB) temperature and polarization anisotropies provide the most stringent and robust constraints to theoretical models, allowing us to determine the key parameters of our Universe with unprecedented precision and address fundamental questions about inflation and early-universe physics. Clearly, by looking at the statistics of the CMB anisotropies with different experiments over the past decades we have learned a lot about the Universe we live in, entering the {\it era of precision cosmology} and establishing the $\Lambda$CDM concordance model \cite{Smoot1992, WMAP_params, Planck2013params}.

But the quest continues. Today we are in the position to ask exciting questions about extensions of the standard cosmological model \cite{PRISM2013WPII,Abazajian2015Inf, Abazajian2015,Abazajian2016S4SB}. For instance, what do the CMB anisotropies tell us about Big Bang Nucleosynthesis (BBN) and in particular the primordial helium abundance, $Y_{\rm p}$? How many neutrino species are there in our Universe? This question is often addressed through the effective number of relativistic degree's of freedom, $N_{\rm eff}$. What are the neutrino masses and their hierarchy? Are there some decaying or annihilating particles? What about dark radiation? 
And regarding the initial conditions of our Universe: what is the running of the power spectrum of curvature perturbations? How about the gravitational wave background, parametrized through the tensor-to-scalar ratio, $r$, which determines the energy scale of inflation, at least when assuming the standard inflation scenario. And to top it up, what about dark energy and the accelerated expansion of our Universe?

All these questions are extremely exciting and define todays cutting-edge research in cosmology, driving present-day theoretical  and experimental efforts. The CMB anisotropies in combination with large-scale structure, weak lensing and supernova observations deliver ever more precise answers to these questions \cite{Calabrese2013, Planck2015params}. 
But the CMB holds another, complementary and independent piece of invaluable information: its {\it frequency spectrum}. Departures of the CMB frequency spectrum from a pure blackbody -- commonly referred to as {\it spectral distortion} -- encode information about the thermal history of the early Universe (from when it was a few month old until today). Since the measurements with COBE/FIRAS in the early 90's, the average CMB spectrum is known to be extremely close to a perfect blackbody at a temperature $T_0=(2.726\pm 0.001)\,\Kel$ \cite{Fixsen1996, Fixsen2009} at redshift $z=0$, with possible distortions limited to one part in $10^5$. This impressive measurement was awarded the Nobel Prize in Physics 2006 and already rules out cosmologies with extended periods of significant energy release, disturbing the thermal equilibrium between matter and radiation in the Universe.

\begin{figure} 
   \centering
   \includegraphics[width=0.74\columnwidth]{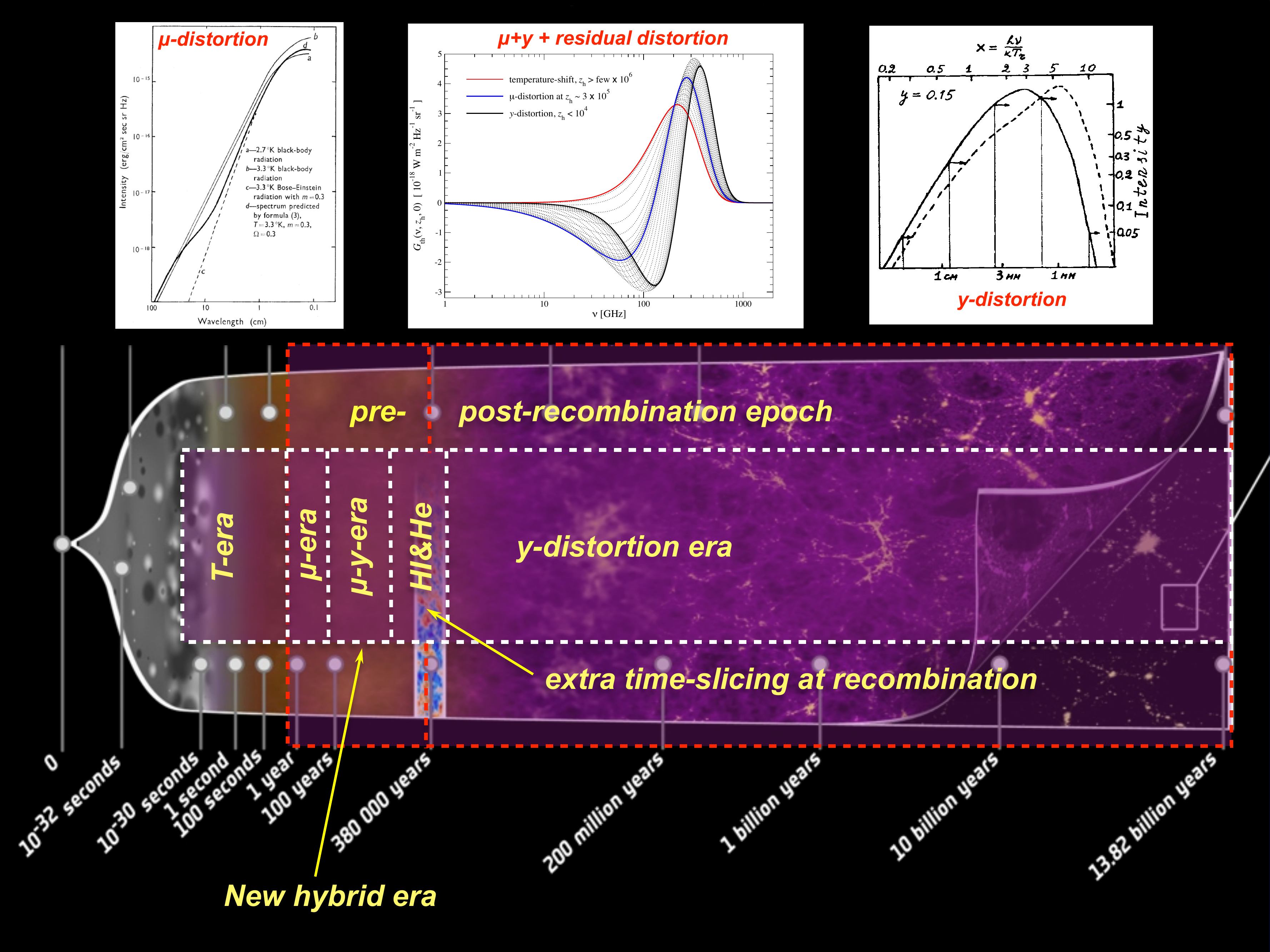}
   \caption{CMB spectral distortions probe the thermal history of the Universe at various stages during the pre- and post-recombination era. Energy release at $z\gtrsim \pot{\rm few}{6}$, when the Universe was only a few month old, merely causes a change of the CMB temperature. A $\mu$-type distortion arises from energy release at $\pot{3}{5}\lesssim z\lesssim \pot{\rm few}{6}$, while $y$-type distortions are created at $z\lesssim 10^4$. The signal caused during the $\mu/y$-transition era ($10^{4}\lesssim z\lesssim \pot{3}{5}$) is described by a superposition of $\mu$- and $y$-distortion with some small {\it residual} distortion that allows probing the time-dependence of the energy-release mechanism. In the recombination era ($10^3\lesssim z\lesssim 10^4$), additional spectral features appear due to atomic transitions of hydrogen and helium. These could allow us to distinguish pre- from post-recombination $y$-distortions.}
   \label{fig:history}
\end{figure}

\begin{figure} 
   \centering
   \vspace{1mm}
   \includegraphics[width=0.74\columnwidth]{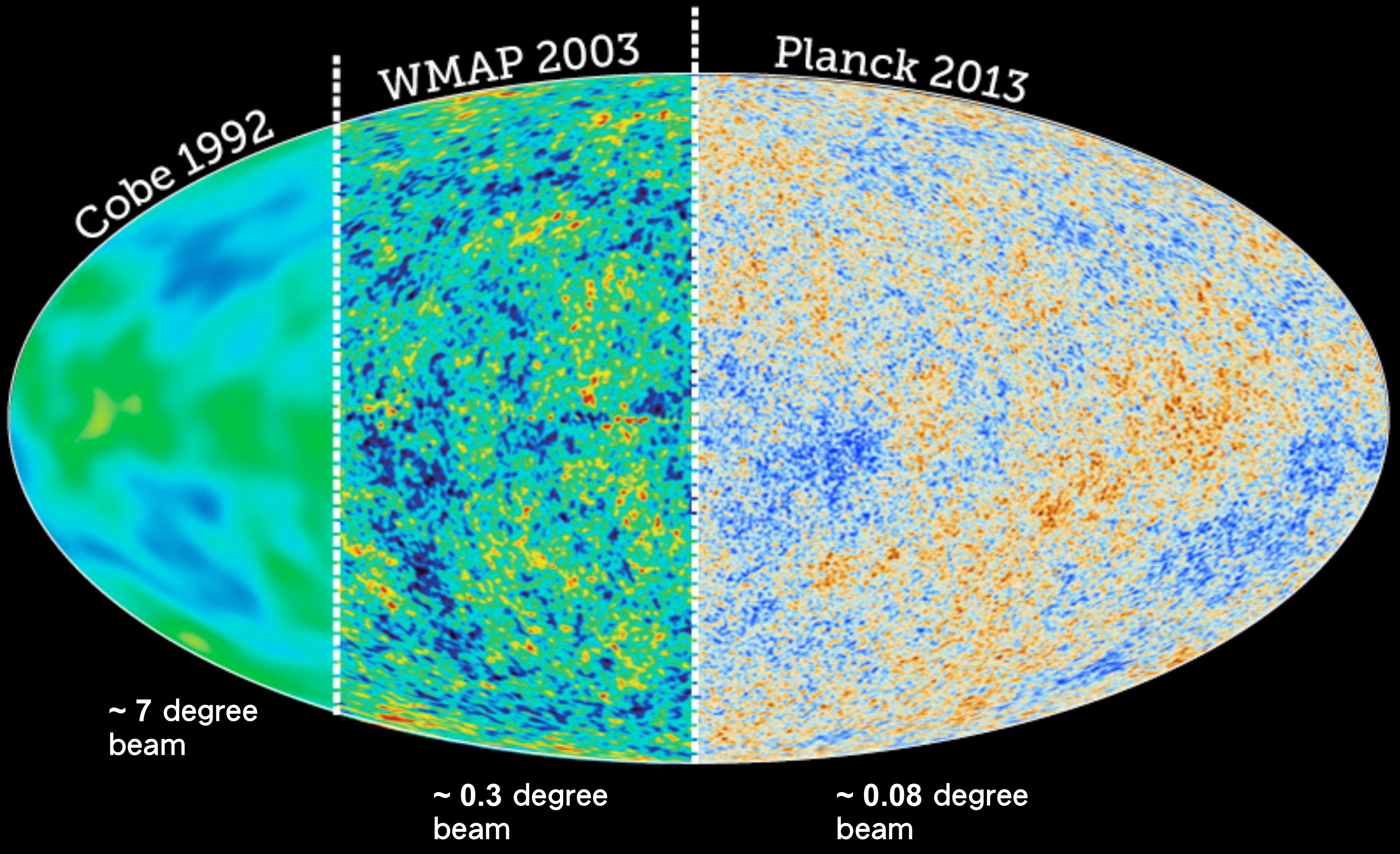}
   \caption{Over the past decades, CMB experiments have seen a dramatic improvement in sensitivity and angular resolution, illustrated here with a comparison of COBE, WMAP and PLANCK. In contrast, CMB spectral distortion measurements are still in the state of some 25 years ago, with COBE/FIRAS defining the unchallenged standard.}
   \label{fig:progress}
\end{figure}

\vspace{-1mm}
\subsection{Why are spectral distortions so interesting today}
So far no spectral distortion of the average CMB spectrum was found. Thus, why is it at all interesting to think about spectral distortions now? 
First of all, there is a long list of processes that could lead to spectral distortions. These include: {\it reionization} and {\it structure formation}; {\it decaying} or {\it annihilating particles}; {\it dissipation of primordial density fluctuations}; {\it cosmic strings}; {\it primordial black holes}; {\it small-scale magnetic fields}; {\it adiabatic cooling of matter}; {\it cosmological recombination}; and several new physics examples \cite{Chluba2011therm, Sunyaev2013, Chluba2013fore, Tashiro2014, deZotti2015, Chluba2016}. This certainly makes  theorists very happy, but most importantly, many of these processes (e.g., reionization and cosmological recombination) are part of our standard cosmological model and therefore should lead to guaranteed signals to search for. This shows that studies of spectral distortions offer both the possibility to {\it constrain well-known physics} but also to open up a {\it discovery space} for non-standard physics, potentially adding new {\it time-dependent information} to the picture (Fig.~\ref{fig:history}).

The second reason for spectral distortion being interesting is due to impressive technological advances since COBE. Although measurements of the CMB temperature and polarization anisotropies have improved significantly in terms of angular resolution and sensitivity since COBE/DMR, our knowledge of the CMB energy spectrum is still in a similar state as more than 25 years ago (Fig.~\ref{fig:progress}). Already in 2002, improvements by a factor of $\simeq 100$ over COBE/FIRAS were deemed feasible \cite{Fixsen2002}, and today even more ambitious experimental concepts like PIXIE \cite{Kogut2011PIXIE, Kogut2016SPIE} and PRISM \cite{PRISM2013WPII}, possibly reaching $\gtrsim 10^3$ in spectral sensitivity, are being seriously considered. These types of experiments provide a unique way to learn about processes that are otherwise hidden from us. At this stage, CMB spectral distortion measurements at high frequencies are furthermore only possible from space, so that, in contrast to $B$-mode polarization science, competition from the ground is largely excluded, making CMB spectral distortions a unique target for future CMB space missions \cite{Chluba2014Science}. 
These efforts could be complemented from the ground at low frequencies ($\nu \lesssim 10\,\GHz$), targeting the cosmological recombination ripples, as suggested for APSERa \cite{Mayuri2015}, or $\mu$ and $y$-distortions using COSMO.

The immense potential of spectral distortions was realized in the NASA 30-year Roadmap study, where improved characterization of the CMB spectrum was declared as one of the future targets \cite{Roadmap2014}. The strong synergy between spectral distortion and B-mode polarization measurements in terms of challenges related to foregrounds and systematic effects further motivate serious consideration of both science cases as part of the future experimental activities.

\vspace{-1mm}
\subsection{Overview and goal of the lecture}
The main goal of the lectures is to convince you that CMB spectral distortion studies provide us with a new and immensely rich probe of early-universe physics, making it an exciting direction of cosmology for the future. These notes are based on extensive lectures on thermalization physics given as part of the CUSO lecture series in 2014, with extended lecture notes available at {\tt www.chluba.de/science}.
I will briefly review the physics of CMB spectral distortions, explaining the different types of distortions and how to compute them for different scenarios. I will then highlight different sources of distortions and what we might learn by measuring distortion signals in the future. Particular attention will be payed to the dissipation of small-scale perturbations and decaying particle scenarios, which illustrate the potential of distortion science. I will also briefly talk about the recombination era and the associated distortion signals and then mention a few of the challenges related to CMB foregrounds. This will also emphasize some of the synergies of distortion and B-mode searches.

\vspace{-2mm}
\section{The physics of CMB spectral distortions}
In this section, I briefly review the main ingredients to describe CMB spectral distortions. The pioneering works on this topic are mainly due to Yakov Zeldovich and Rashid Sunyaev in the 60's and 70's \cite{Zeldovich1969, Sunyaev1970mu, Illarionov1974, Illarionov1975}. These early works were later extended by \cite{Danese1977, Danese1982}, to include the effect of double Compton emission, and \cite{Burigana1991, Hu1993}, with refined numerical and analytical treatments. Latest considerations of spectral distortion and their science can be found in \cite{Chluba2011therm, Sunyaev2013, Chluba2013fore, Tashiro2014, deZotti2015, Chluba2016} and \cite{Jose2006, Sunyaev2009, Chluba2016Rec} for the recombination radiation.

\subsection{Simple Blackbody relations}
\label{sec:blackbody}
Before talking about CMB spectral distortions, let us briefly remind ourselves of a few important blackbody relations. We shall denote the blackbody intensity or {\it Planckian} as, $B_\nu(T)$, where $\nu$ is the frequency and $T$ the blackbody temperature. The {\it Planck} law reads:
\beal{
\label{eq:planck_formula}
B_\nu(T)=\frac{2 h}{c^2}\frac{\nu^3}{\expf{h\nu/kT}-1}=\frac{2 h\,\nu^3}{c^2}\,n^{\rm bb}_\nu(T)=I_0\,\frac{x^3}{\expf{x}-1},
}
having units $[B_\nu(T)]=\ergs  \sec^{-1}\cm^{-2} \Hz^{-1} \sr^{-1}= 10^{17}\,\MJysr$. 
The spectrum of the Sun is approximately represented by this expression (let's be a theorists and forget about all the Fraunhofer lines and existence of the atmosphere with all its absorption bands) with a temperature $T_{\rm ph}\simeq 6000\,\Kel$ (photosphere). Also, we already heard about the CMB blackbody spectrum, which is unbelievably close to a blackbody at a temperature $T_0=2.726\,\Kel$ \cite{Fixsen1996, Fixsen2009}. 

In Eq.~\eqref{eq:planck_formula}, we also indicate the connection of $B_\nu$ to the blackbody occupation number, $n^{\rm bb}_\nu(T)=1/(\expf{h\nu/kT}-1)=1/(\expf{x}-1)$, and transformed to the dimensionless frequency, $x=h\nu/kT$ (redshift-independent), introducing $I_0(T)=(2h/c^2)(kT/h)^3\approx 270\,\MJysr (T/2.726\Kel)^3$. It is useful to remember that $x=1$ corresponds to $\nu \approx 56.8\,\GHz$ for the CMB. Also, the maximum of the blackbody spectrum (Wien's displacement law) is located at $\nu_{\rm max} \approx 160\,\GHz \left[\frac{T}{2.726\,\Kel}\right]$ or $x_{\rm max}\approx 2.821$. We furthermore have the important limiting cases
\beal{
B_\nu(T)\approx 
\begin{cases}
\frac{2 \nu^2}{c^2}  k T   & \text{for}\;h\nu \ll kT\qquad\text{(Rayleigh-Jeans limit)},
\\[2mm]
\frac{2 h \nu^3}{c^2}  \expf{-h\nu/kT}   & \text{for}\;h\nu \gg kT\qquad\text{(Wien law)},
\end{cases}
}
for the blackbody spectrum. In the Wien part of the spectrum, very few photons are found but their energy is large. The opposite is true in the Rayleigh-Jeans part.

\subsection{Photon energy and number density}
For our discussions, the total photon number and energy densities, $\rho_\gamma$ and $N_\gamma$, will be important. These are defined by the integrals, $\rho_\gamma=\int \frac{I_\nu}{c} \id \nu \id \Omega$ and $N_\gamma=\int \frac{I_\nu}{c\,h\nu} \id \nu \id \Omega$, over all photon energies and directions. Here, $I_\nu$ is the photon intensity. For blackbody radiation, this simply gives 
\bsub
\beal{
\rho^{\rm Pl}_\gamma&=\frac{2h}{c^3}\int \frac{\nu^3}{\expf{x}-1} \id \nu \id \Omega
=\frac{8\pi h}{c^3}\left(\frac{kT}{h}\right)^4\int \frac{x^3\id x}{\expf{x}-1} = \frac{8\pi^5 (kT)^4}{15\,c^3 h^3}
\nonumber
\\
&=a_R T^4
\approx \pot{5.10}{-7}\,\me c^2 \cm^{-3} \left(\frac{T}{2.726\Kel}\right)^4\approx 0.26\,\eV\,\cm^{-3} \left(\frac{T}{2.726\Kel}\right)^4
\\[2mm]
N^{\rm Pl}_\gamma&=\frac{2}{c^3}\int \frac{\nu^2}{\expf{x}-1} \id \nu \id \Omega
=\frac{8\pi}{c^3}\left(\frac{kT}{h}\right)^3 \int \frac{x^2 \id x}{\expf{x}-1}
= \frac{16 \pi \zeta_3 (kT)^3}{c^3 h^3}
\nonumber
\\
&= b_R T^3\approx 410\,\cm^{-3} \left(\frac{T}{2.726\Kel}\right)^3,
}
\esub
where $\zeta_i$ denotes the Riemann $\zeta$-function. Here, $a_R=4\sigma/c \approx \pot{7.566}{-15}\,\ergs \,\cm^{-3} \,\Kel^{-4}$ is the {\it radiation constant}, where $\sigma$ is the {\it Stefan-Boltzmann constant}.
We also have the useful relation $\rho^{\rm Pl}_\gamma\approx 2.701 k T N^{\rm Pl}_\gamma$. In particular, we have $\rho^{\rm Pl}_\gamma\propto T^4$ and $N^{\rm Pl}_\gamma\propto T^3$, the crucial blackbody relations.

\subsection{What do we need to do to change the blackbody temperature}
The blackbody spectrum is fully characterized by one number, its temperature $T$. Thus, one simple question is, {\it what do we have to do to shift the temperature to $T'\neq T$}?
Let's suppose we increase the temperature by adding some energy to the photon field (let's say we just move all photons upwards in frequency in some way; no change of the volume or photon number), $\epsilon=\Delta \rho_\gamma/\rho_\gamma^{\rm Pl}(T)\equiv (T'/T)^4-1$, then the expected change in the photon temperature is 
\beal{
\frac{\Delta T}{T}=\left(1+\epsilon\right)^{1/4}-1 \approx \frac{1}{4}\frac{\Delta \rho_\gamma}{\rho_\gamma^{\rm Pl}},
}
for small $\Delta \rho_\gamma/\rho_\gamma^{\rm Pl}$. Clearly, if we stopped here, the new spectrum cannot be a blackbody anymore, since we did not change the photon number density. Thus, pure energy release/extraction inevitably leads to a spectral distortion, no matter how the photons are distributed in energy.

To keep the blackbody relation, $N^{\rm Pl}_\gamma\propto T^3$, unchanged we {\it simultaneously} need to add 
\beal{
\label{eq:blackbody_cond}
\frac{\Delta N_\gamma}{N_\gamma^{\rm Pl}}=(T'/T)^3-1=(1+\epsilon)^{3/4}-1\approx 3 \frac{\Delta T}{T} 
\quad\Longrightarrow \quad
\frac{\Delta N_\gamma}{N_\gamma^{\rm Pl}}\approx \frac{3}{4} \frac{\Delta \rho_\gamma}{\rho_\gamma^{\rm Pl}}
}
of photons to avoid creating a non-blackbody spectrum. This condition is {\it necessary but not sufficient}, since it does not specify {\it how} the missing photons are distributed in energy! 
For example, let us assume we add photons to the blackbody spectrum at one frequency only. Then $\Delta \rho_\gamma=h \nu \Delta N_\gamma$ and $\epsilon \approx (h \nu/2.701kT) \, \Delta  N_\gamma/N_\gamma^{\rm Pl}$. To satisfy the condition Eq.~\eqref{eq:blackbody_cond}, we just need to tune the frequency to $h \nu/kT \approx (4/3) \, 2.70\approx 3.60$. Clearly, a blackbody spectrum with a single narrow line at $h \nu \simeq 3.6 \, kT$ is no longer a blackbody even if Eq.~\eqref{eq:blackbody_cond} is satisfied. We thus also need to add photons to the CMB spectrum in just the right way and the question is how?

To go from one blackbody with temperature $T$ to another at temperature $T'$, we need to have a change of the photon occupation number by
\beal{
\nonumber
\Delta n_\nu=\nPl(T')-\nPl(T)=\frac{1}{\expf{x'}-1}-\frac{1}{\expf{x}-1}
= -x \partial_x \nPl \,\frac{\Delta T}{T}+\mathcal{O}\left(\frac{\Delta T}{T}\right)^2
= \frac{x\expf{x}}{(\expf{x}-1)^2} \,\frac{\Delta T}{T}+\mathcal{O}\left(\frac{\Delta T}{T}\right)^2
}
with $x'=x \,T/T'$. In what follows, we will frequently use the definition 
\beal{
\label{eq:def_temperature_shift}
G(x)=-x\partial_x \nPl=\frac{x\,\expf{x}}{(\expf{x}-1)^2}
\approx
\begin{cases}
\frac{1}{x} &\text{for}\qquad x\ll 1
\\
x\, \expf{-x} &\text{for}\qquad x\gg 1,
\end{cases}
}
which determines the {\it spectrum of a temperature shift}: $T \partial_T B_\nu\propto x^3 G(x)$, for small $\Delta T/T$. Its spectral shape is shown in Fig.~\ref{fig:Y_SZ_and_mu_distortion}. It is easy to prove that a change with this spectral distribution does not lead to any distortion as long as $\Delta T/T$ is sufficiently small.
We will thus refer to $G(x)$ as the spectrum of a temperature shift. In the thermalization problem, it is created through the combined action of Compton scattering and photon creation processes (i.e., double Compton and Bremsstrahlung emission).

\subsection{What is the thermalization problem all about}
When considering the cosmological thermalization problem we are asking: {\it how was the present CMB spectrum really created?} Assuming that everything starts off with a pure blackbody spectrum, the uniform adiabatic expansion of the Universe alone (absolutely no collisions and spatial perturbations here!) leaves this spectrum unchanged -- a blackbody thus remains a blackbody at all times. 
However, as the simple discussion in the proceeding section already showed, processes leading to photon production/destruction or energy release/extraction should inevitably introduce momentary distortions to the CMB spectrum. Then the big question is: {\it was there enough time from the creation of the distortion until today to fully restore the blackbody shape, pushing distortions below any observable level?} 
For this, we need to redistribute photons in energy through Compton scattering by free electrons. However, this is not enough to restore the blackbody spectrum. We also need to adjust the number of photons through double Compton and Bremsstrahlung. 
By understanding the thermalization problem and studying the CMB spectrum in fine detail we can thus learn about different early-universe processes and the thermal history of our Universe.
This can open a new window to the early Universe, allowing us to peek behind the last scattering surface which is so important for the formation of the CMB temperature and polarization anisotropies.

\subsection{General conditions relevant to the thermalization problem}
In the early Universe, photons undergo many interactions with the other particles. We shall mainly concern ourselves with the {\it average} CMB spectrum and neglect distortion anisotropies when describing their evolution.\footnote{Some distortion anisotropies are created by SZ clusters \citep{Carlstrom2002}. Primordial distortion anisotropies can also be created by anisotropic acoustic heating \citep{Pajer2012, Ganc2012}.} Distortion anisotropies can be created through anisotropic energy release processes; however, these are usually very small, such that we only briefly touch on them below. We also assume that the distortions are always minor in amplitude, so that the problem can be linearized. This allows us to resort to a Green's function approach when solving the thermalization problem \cite{Chluba2013Green, Chluba2015GreenII}, which greatly simplifies explicit thermalization calculations for different energy release scenarios as can be carried out using the full thermalization code {\tt CosmoTherm} \cite{Chluba2011therm}.

We furthermore assume the standard $\Lambda$CDM background cosmology \cite{Planck2015params} with standard ionization history computed using {\tt CosmoRec} \cite{Chluba2010b}. Also, the electron and baryon distribution functions are given by Maxwellians at a common temperature, $\Te$, down to very low redshifts ($z\lesssim 10$), when thermalization process is already extremely inefficient. We furthermore need not worry about the evolution of distortions before the electron-positron annihilation era ($z\gtrsim 10^7-10^8$), since in this regime rapid thermalization processes always ensure that the CMB spectrum is very close to that of a blackbody. 
We are thus just dealing with non-relativistic electrons, protons and helium nuclei immersed in a bath of CMB photons. We can also neglect the traces of other light elements for the thermalization problem and usually assume that neutrinos and dark matter are only important for determining the expansion rate of the Universe.

\subsection{Photon Boltzmann equation for average spectrum}
\label{Ch:Boltzmann_eq}
The study of the formation and evolution of CMB fluctuations in both real and frequency space begins with the radiative transport, or {\it Boltzmann equation} for the photon phase space distribution, $n(x^\mu, p^\mu)$. Here, we are only interested in the evolution of the average spectrum. In this case, perturbations can be neglected, such that $n(x^\mu, p^\mu)\rightarrow n(t, p)$ and we may express the photon Boltzmann equation as
\beal{
\label{eq:gen_Boltzmann_equation_II_final}
\frac{\partial n}{\partial t}- H \,p \frac{\partial n}{\partial p}={\rm C}[n],
}
omitting the any spatial dependence. Here, $H(t)$ is the standard Hubble expansion rate and ${\rm C}[n]$ denotes the collision term, which accounts for interactions of photons with the other species in the Universe. The collision term incorporates several important effects. Most importantly, Compton scattering couples photons and electrons, keeping the two in close thermal contact until low redshifts, $z\lesssim 100-200$. Bremsstrahlung and double Compton emission allow adjusting the photon number and are especially fast at low frequencies, as we explain below.

Neglecting collisions (${\rm C}[n]=0$), we directly recover $n(t, p)=n[t_0, p \,a(t)/a(t_0)]$, which means that the shape of the photon distribution is conserved by the universal expansion and only the photon momenta are redshifted. Introducing the variable $x=p/kT_\gamma(t)=h\nu/kT_\gamma(t)$, with $T_\gamma(t)=T_\gamma(t_0)\,a(t_0)/a(t)\propto (1+z)$, the photon Boltzmann equation takes the more compact form $\partial n(t, x)/\partial t={\rm C}[n(t, x)]$ (see \cite{Chluba2011therm} for more details), which highlights the conservation law.

\subsection{Collision term for Compton scattering}
\label{sec:CS_phys}
We already mentioned that Compton scattering is responsible for redistributing photons in energy. This problem has been studied a lot in connection with X-rays from compact objects \citep{Pozdniakov1979, Sunyaev1980Comptonization} and the cosmological context \citep{Zeldovich1969, Sunyaev1970mu}. In reality, electron-photon scattering also helps isotropizing the photon field (Thomson scattering limit), although for this energy exchange is not as crucial \citep{Chluba2012, Chluba2014mSZII}. 

To account for the Comptonization of photons by free thermal electrons, we can use the so-call {\it Kompaneets equation} \citep{Kompa56}:
\beal{
\label{eq:Kompaneets}
\left.\frac{\partial n}{\partial \tau}\right|_{\rm CS}
&\approx \frac{\The}{\xe^2}\frac{\partial}{\partial \xe} \xe^4 \left[\frac{\partial}{\partial \xe} n +n(1+n) \right]
\equiv \frac{\The}{x^2}\frac{\partial}{\partial x} x^4 \left[\frac{\partial}{\partial x} n +\frac{\Tg}{\Te}\,n(1+n) \right],
}
where $\id \tau = \Ne \sigT c \id t$ is the Thomson optical depth, $\The=k\Te/\me c^2$ is the dimensionless electron temperature and $\xe=h\nu/k\Te$ is the chosen frequency variable. This expression can be obtained by computing the Compton collision term in the limit $h\nu\ll k\Te$ and $k\Te \ll \me c^2$, keeping only terms up to first order in $\The$ and $h\nu/\me c^2$ \citep{Rybicki1979}. This is equivalent to considering the first two moments of the photons frequency shift, $\Delta \nu/\nu$ over the scattering kernel \citep{Sazonov2000}.
The Kompaneets equation can be used to describe the repeated scattering of photons by thermal electrons in the isotropic medium. The first term in the brackets describes {\it Doppler broadening} and {\it Doppler boosting} and the last term accounts for the {\it recoil effect} and {\it stimulated recoil}. These latter terms are especially important for reaching full equilibrium in the limit of many scatterings.

Below we discuss some analytic solutions of the Kompaneets equation in limiting cases. Here, a couple of words about limitations of this equation. First of all, we assumed that the change in the energy of the photon by the scattering is small. For hot electrons this is no longer correct and one has to go beyond the lowest orders in $\Delta \nu/\nu$. This is for example important for the {\it Sunyaev-Zeldovich effect} of very hot clusters \citep{Itoh98, Sazonov1998, Challinor1998}, but this procedure only converges {\it asymptotically} \citep{ChlubaSZpack, Chluba2012moments}. The second limitation is that if the photon distribution has sharp features (more narrow than the width of the scattering kernel) then the shape of the scattered photon distribution is not well represented with the diffusion approximation. In this case, a scattering kernel approach can be used to describe the scattering problem \citep{Sazonov2000}, although efficient numerical scheme for many scatterings are cumbersome.

\begin{figure}
\centering
\includegraphics[width=0.7\columnwidth]{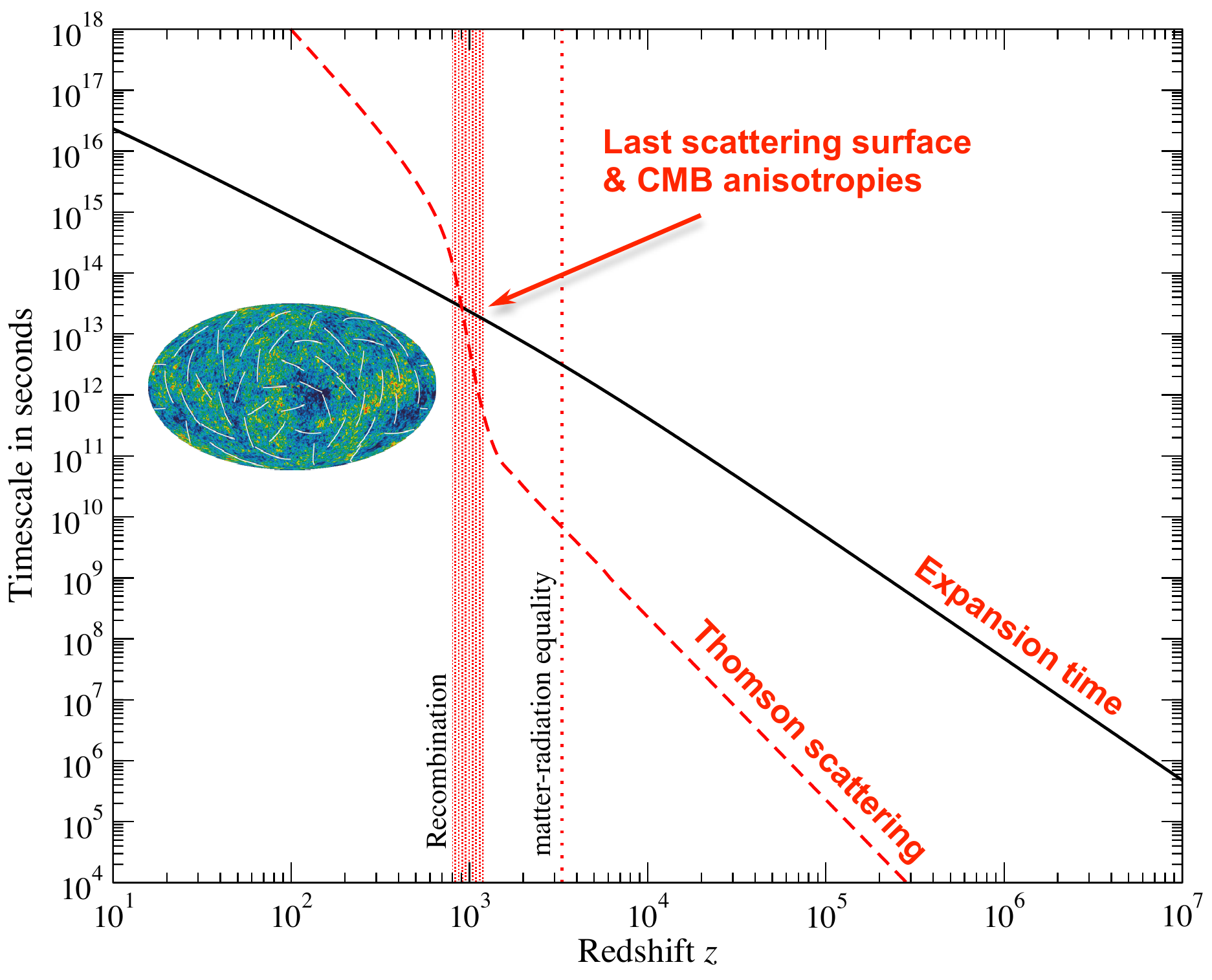}
\caption{Comparison of the Thomson scattering time-scale with the Hubble expansion time-scale.}
\label{fig:Thomson_Hubble_times}
\end{figure}

\subsubsection{Comptonization efficiency}
With the Kompaneets equation, we can already understand some of the important aspects of Comptonization, by simply looking at characteristic time-scales. One important quantity is the Thomson scattering time-scale, $t_{\rm T}=(\sigT \Ne c)^{-1}$. It describes how rapidly photons scatter with electrons. For the standard cosmology with $24\%$ of helium (by mass), we have
\vspace{-1mm}
\beal{
\label{eq:t_Thomson}
t_{\rm T}=(\sigT \Ne c)^{-1} \simeq \pot{2.7}{20} \,X_{\rm e}^{-1} (1+z)^{-3} \, \sec \simeq  \pot{4.0}{4} \left[\frac{X_{\rm e}}{0.16}\right]^{-1}\left[\frac{1+z}{1100}\right]^{-3}\,{\rm years},
}
where $X_{\rm e}=\Ne/N_{\rm H}$ is the free electron fraction relative to the number of hydrogen nuclei. At $z=1100$, this corresponds to $\simeq 40\,000$ years between scatterings! To put this into perspective we have to compare with the typical expansion time-scale given by the inverse Hubble rate:
\vspace{-1mm}
\beal{
\label{eq:t_exp}
t_{\rm exp}=H^{-1} \simeq 
\begin{cases}
\pot{4.8}{19} \,(1+z)^{-2} \, \sec &\qquad \text{(radiation domination)}
\\
\pot{8.4}{17} \,(1+z)^{-3/2} \, \sec  &\qquad \text{(matter domination)},
\end{cases}
}
where the transition between matter and radiation (photons + neutrinos) domination occurs around $\zeq\simeq 3400$. From Fig.~\ref{fig:Thomson_Hubble_times} we see that the Thomson scattering rate (shorter time-scales) is much higher than the Hubble expansion rate until after decoupling around $z\simeq 10^3$. But even then, the time-scale for scattering only exceeds the expansion time by a factor of $\simeq 10^2-10^4$. However, this is when the isotropization process of CMB temperature and polarization anisotropies becomes inefficient and we start seeing the primordial CMB fluctuations.

\begin{figure}
\centering
\includegraphics[width=0.7\columnwidth]{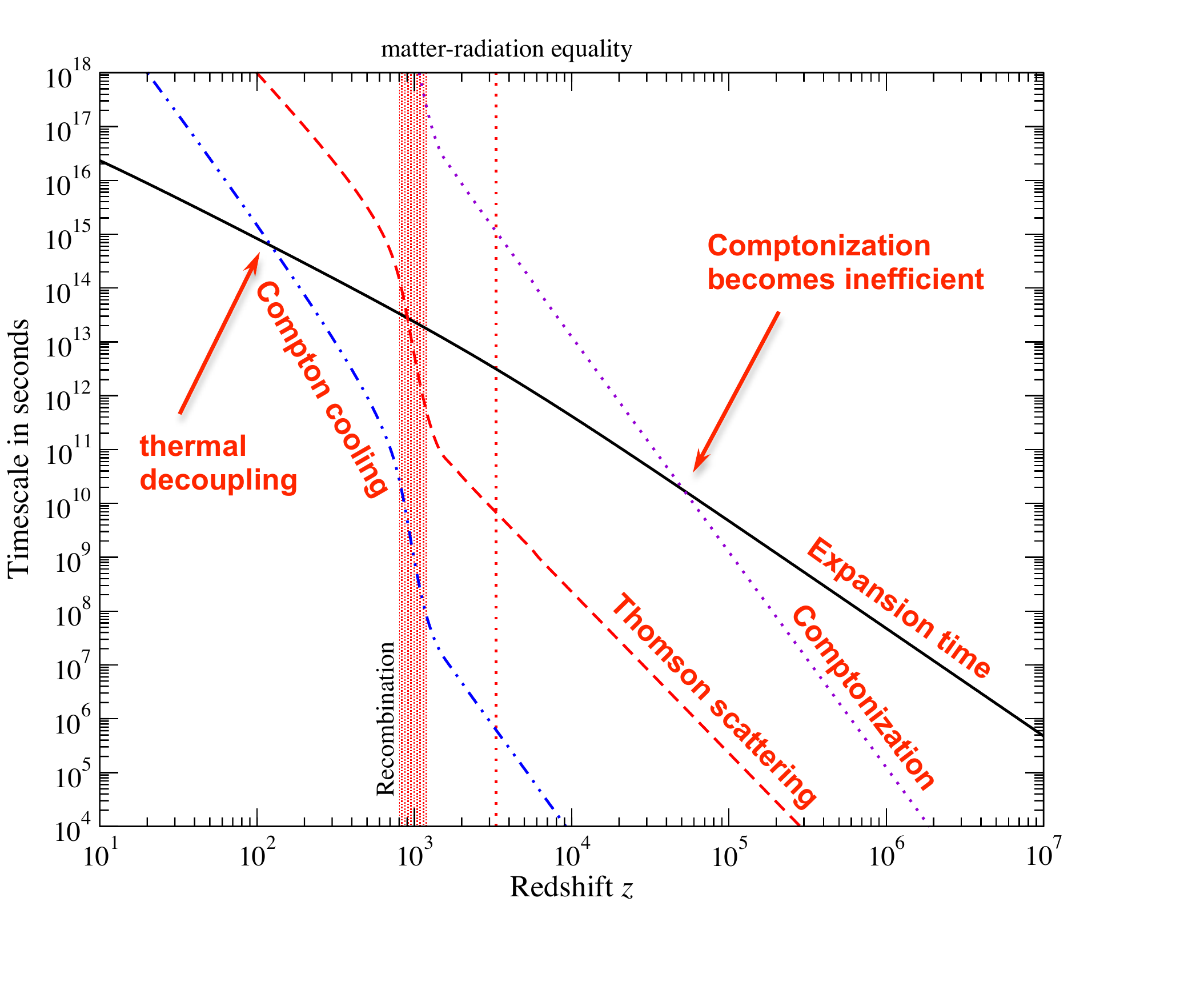}
\caption{Comparison of the Comptonization,  Compton cooling and Hubble expansion time-scale.}
\label{fig:Compton_Hubble_times}
\end{figure}

The most important aspect of Comptonization is energy exchange between electrons and photons. The time-scale on which electrons transfer energy to the photons is \citep{Zeldovich1969, Hu1995PhD}
\vspace{-1mm}
\beal{
\label{eq:t_egamma}
t_{\rm e\gamma} \approx \frac{t_{\rm T}}{4 \The} \simeq \pot{4.9}{5} \, t_{\rm T} \,\left[\frac{1+z}{1100}\right]^{-1}\simeq \pot{1.2}{29} (1+z)^{-4}\,\sec.
}
In simple words, the time-scale for scattering is $t_{\rm T}\simeq [\Ne \sigT c]^{-1}$ and per scattering the fractional energy-exchange between photons and electrons is $\Delta \nu/\nu\simeq 4 \The$.
Comparing $t_{\rm e\gamma}$ with the Hubble rate one finds that at $z_{\mu y}\simeq \pot{5}{4}$, {\it Comptonization} becomes inefficient (see Fig.~\ref{fig:Compton_Hubble_times}). At this redshift, the characteristic of spectral distortions changes, transitioning from a so-called $\mu$-distortion to a $y$-type distortion (see below). Evidently, the transition is not abrupt and the characteristic shape of the distortion changes over a range of redshift between $z\simeq 10^4 - \pot{3}{5}$ (e.g., see \citep{Chluba2013Green}).

The Comptonization time-scale is quite long compared to the time-scale over which electrons are heated by photons. The big difference is that every electron has $\simeq \pot{1.9}{9}$ photons to scatter with, making the number of interactions much larger. This fact influences many phases in the history of the Universe. For example, the cosmological recombination process is delayed until the temperature of the CMB has dropped below $k\Tg \simeq 0.26\,{\rm eV}$, which is two orders of magnitude smaller than the ionization potential, $E_{\rm ion}\simeq 13.4\,{\rm eV}$. Similarly, BBN occurs significantly later than what is expected from naively assuming $kT \simeq m c^2$.

From $\rho_{\rm th} = (3/2) \sum_i N_i k \Te=(3/2) N_{\rm H} (1+f_{\rm He}+X_{\rm e})\,k \Te$ for the thermal energy of the plasma, by comparison with the energy density of photons, we have 
\vspace{-1.5mm}
\beal{
\label{eq:t_egamma}
t_{\rm \gamma e} = \frac{\rho_{\rm th}}{\rho_\gamma} \, t_{\rm e \gamma} 
\simeq\frac{3 N_{\rm H} (1+f_{\rm He}+X_{\rm e})}{8\rho_\gamma/(\me c^2)} \, t_{\rm T} 
\simeq 0.31 \, t_{\rm T} \,(1+z)^{-1}\simeq \pot{7.3}{19} (1+z)^{-4}\,\sec.
}
where for the estimate we used $X_{\rm e}=1+2f_{\rm He}$ (fully ionized) and $f_{\rm He}\approx \Yp/[4(1-\Yp)]\approx 0.079$. Before recombination, the Compton cooling time is about $\simeq \pot{1.6}{9}$ times shorter than the Comptonization time. This means that electrons and baryons (through Coulomb scatterings) remain in full thermal contact with the photon field until very late. From Fig.~\eqref{fig:Compton_Hubble_times} one can see that thermal decoupling is expected to happen somewhere around $z\simeq 100-200$ \citep{Zeldovich68}. This is when the earliest signals from the 21cm era are produced \citep{Pritchard2008}.

\vspace{-1.5mm}
\subsection{Bremsstrahlung and double Compton emission}
So far we have only considered the redistribution of photons in energy. As discussed above, this alone is insufficient for thermalizing the radiation field. In addition, we need to adjust the photon number, which in the expanding early Universe is achieved by thermal Bremsstrahlung (BR) and double Compton (DC) emission. BR is the first and most obvious suspect for photon production and absorption in the early Universe. However, it turns out that in our Universe DC emission is much more important \citep{Danese1982}. Nevertheless, at late times BR has to be included for accurate computations \citep{Burigana1991, Hu1993, Chluba2011therm}. 

The collision term for BR and DC emission can be expressed as
\vspace{-1.5mm}
\bsub
\label{eq:dn_em_abs}
\beal{
\left.\frac{\partial n(\tau, x)}{\partial \tau}\right|_{\rm em/abs}
&= 
\frac{K_{\rm BR}\,\expf{-\xe}+K_{\rm DC}\,\expf{-2x}}{x^3}\!\left[ 1 - n(\tau, x) \, (\expf{\xe}-1)\right]
\\[0.5mm]
K_{\rm BR}
=\frac{\alpha}{2\pi}\frac{\lambda_{\rm e}^3}{\sqrt{6 \pi}\,\The^{7/2}}\,\left(\frac{\Te}{\Tg}\right)^3 \sum_i Z_i^2 &N_i \, \bar{g}_{\rm ff}(Z_i, \Te, \Tg, \xe),
\quad
K_{\rm DC}=\frac{4\alpha}{3\pi}\,\Thg^2\,I_{\rm dc} \,g_{\rm dc}(\Te, \Tg, x)
\\[0.5mm]
&\!\!\!\!\!\!\!\!\!\!\!\!\!\!\!\!\!\!\!\!\!\!\!\!\!\!\!\!\!\!\!\!\!\!\!\!\!\!\!\!\!\!\!\!\!\!\!\!\!\!\!\!\!\!\!\!\!\!\!\!\!\!\!\!\!\!
\bar{g}_{\rm ff}(\xe)\approx 
\begin{cases}
\frac{\sqrt{3}}{\pi}\ln\left(\frac{2.25}{\xe}\right)&\text{for}\quad \xe\leq 0.37
\\
1 &\text{otherwise}
\end{cases},
\qquad
g_{\rm dc}\approx 
\frac{1+\frac{3}{2}x+\frac{29}{24} x^2+\frac{11}{16} x^3+\frac{5}{12} x^4}{1+19.739\Thg-5.5797\The}.
}
\esub
where $\alpha\approx 1/137$, $I_{\rm dc}=\int x^4 n (1+n)\id x\approx 4\pi^4/15$ and $\lambda_{\rm e} =h/\me c\simeq \pot{2.43}{-10}\,\cm$. The approximation for the DC Gaunt factor, $g_{\rm dc}$, was taken from \cite{Chluba2011therm} and is based on \cite{Chluba2007a}. It should provide a very good approximation for our purpose. For the BR Gaunt factors, $\bar{g}_{\rm ff}$, we normally use fits from \citet{Itoh2000} in numerical calculations or the above approximation for estimates. 

One can already see that both BR and DC push the radiation field into equilibrium with a blackbody at the temperature of the electrons, $n_{\rm e}=1/(\expf{\xe}-1)$. Also, due to the $1/x^3$ scaling of the emissivity it is clear that BR and DC emission both are most important at low frequencies. Inserting typical numbers for $z\gtrsim 10^3$ and assuming $\Te\approx \Tg$, we have 
\vspace{-1.5mm}
\bsub
\beal{
\label{eq:BR_DC_estimates}
K_{\rm BR}&\simeq \pot{1.4}{-6}\,\left[\frac{\bar{g}_{\rm ff}}{3.0}\right] \left[\frac{\Omega_{\rm b} h^2}{0.022}\right]\,(1+z)^{-1/2}
\\[0mm]
K_{\rm DC}&\simeq \pot{1.7}{-20}\,(1+z)^{2}.
}
\esub
This implies that at $z_{\rm dc,br}\simeq \pot{3.7}{5}\left(\left[\frac{\bar{g}_{\rm ff}}{3.0}\right] \left[\frac{\Omega_{\rm b} h^2}{0.022}\right]\right)^{2/5}$ BR and DC emission are similarly important \citep{Burigana1991, Hu1993}. At $z>z_{\rm dc,br}$, DC emission is more crucial, while at lower redshifts BR dominates.

\section{Types of spectral distortions from energy release}
%
%
We are now in the position to discuss the main types of spectral distortions created by energy release. In section~\ref{sec:CS_phys}, we learned that around $z_{\mu y}\simeq \pot{5}{4}$ the Comptonization time-scale (transfer of energy from electrons to photons) becomes longer than the Hubble time. It is clear that this marks an important transition in the efficiency of Compton scattering and redistribution of photons. Let us try to quantify this a little better by looking at the photon evolution equation, for now neglecting photon emission
\beal{
\label{eq:CS_Kompa}
\frac{\partial n}{\partial \tau}
\equiv \frac{\The}{x^2}\frac{\partial}{\partial x} x^4 \left[\frac{\partial}{\partial x} n +\frac{\Tg}{\Te}\,n(1+n) \right],
}
setting $n=n(\tau, x)$. This equation has no general analytic approximation, but we can solve it for limiting cases. As we explain next, the Compton-$y$ distortion is created by scatterings with inefficient energy exchange between electrons and photons, while the chemical potential $\mu$-distortion is formed in the regime of extremely efficient energy exchange.

\subsection{Scattering of CMB photons in the limit of small $y$-parameter}
Assuming that at $\tau=0$ we start with $n=n_{\rm bb}=1/(\expf{x}-1)$, then after a very short time $\Delta \tau\ll 1$ we find
\beal{
\label{eq:def_YSZ}
\Delta n
&\approx   \frac{\Delta \tau \,\The}{x^2}\frac{\partial}{\partial x} x^4 \left[\frac{\partial}{\partial x} n_{\rm bb} +\frac{\Tg}{\Te}\,n_{\rm bb}(1+n_{\rm bb}) \right]
\approx 
\frac{\Delta \tau\,(\Thg - \The)}{x^2}\frac{\partial}{\partial x} x^4 n_{\rm bb}(1+n_{\rm bb})
\nonumber \\[1mm]
&\approx  
\Delta \tau \,(\Thg - \The) \left[ 4 x n_{\rm bb}(1+n_{\rm bb}) - x^2 n_{\rm bb}(1+n_{\rm bb}) (1+2 n_{\rm bb})\right]
\nonumber \\[1mm]
&\approx  
\Delta \tau\,(\The - \Thg) \,G(x)\,\left[ x \,\frac{\expf{x}+1}{\expf{x}-1}-4\right]\equiv  \Delta \tau\, (\The - \Thg)\,Y_{\rm SZ}(x),
}
where we used $\partial_x n_{\rm bb}=-n_{\rm bb}(1+n_{\rm bb})=-\expf{x}/(\expf{x}-1)^2=-G(x)/x$ and $(1+2 n_{\rm bb})=(\expf{x}+1)/(\expf{x}-1)=\coth(x/2)$. This is the definition of the so called {\it Compton-$y$} distortion, $Y_{\rm SZ}(x)$, which arises in the limit of scatterings with inefficient energy exchange. This distortion of the CMB was first studied by \citet{Zeldovich1969} and then applied to hot electrons residing inside the potential wells of clusters of galaxies, giving rise to the {\it thermal Sunyaev-Zeldovich (SZ) effect}. The important variable is the {\it Compton-$y$ parameter}
\beal{
\label{eq:def_y_parameter}
y=\int_0^\tau \frac{k(\Te-\Tg)}{\me c^2} \id \tau'=\int_0^t \frac{k(\Te-\Tg)}{\me c^2} \sigT \Ne c \id t',
}
which depends on the number of scattering (related to $\tau$) and the net energy exchange\footnote{To some extent it would be better to immediately write $y^*=\int_0^\tau 4\frac{k(\Te-\Tg)}{\me c^2} \id \tau'$, so that $y^*=4y=\Delta \rho_\gamma/\rho_\gamma$ evidently gives the total amount of energy transfer.}, $\Delta \nu/\nu\simeq 4(\The-\Thg)\ll 1$, per scattering. Clearly, for $\Te\equiv \Tg$ one has $y=0$ and $\Delta n=0$, no matter how many scattering  actually take place! The solution Eq.~\eqref{eq:def_YSZ} for the distortion is thus valid as long as $|y|\ll 1$. This also ensures that the electron temperature does not change much by the scattering. 
One possible way to violate this condition even if the number of scattering is tiny ($\tau\ll 1$) is by having a very large difference in the electron and photon temperature. Note, however, that $\The\ll 1$ is needed since otherwise relativistic corrections to the Compton process appear, which are not accounted for by the Kompaneets equation \cite{Chluba2014}. For the cosmological thermalization problem, we are always in the situation that the $y$-parameter is increased beyond unity by increasing the number of scatterings. In this case, Compton scattering pushes electrons and photons into kinetic equilibrium until a $\mu$-distortion is formed (Sect.~\ref{sec:mu_dist_section}).

\begin{figure}
\centering
\includegraphics[width=0.8\columnwidth]{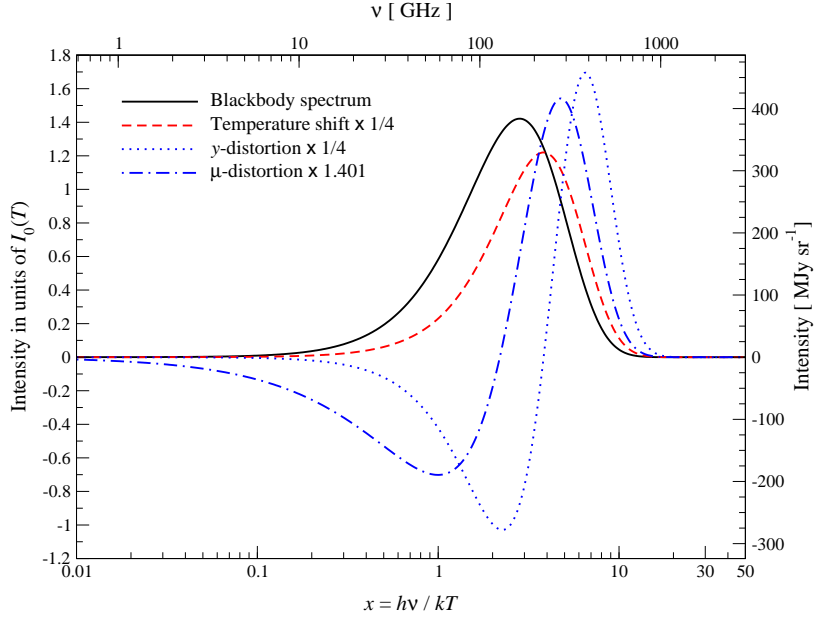}
\caption{Comparison of a Compton $y$-distortion, $Y_{\rm SZ}(x)$, and $\mu$-distortion, $M(x)$, with the blackbody spectrum and temperature shift, $G(x)$. For convenience, we plot the spectrum as a function of $x=h\nu/kT$ and normalize the left $y$-axis by $I_0(T)=(2h/c^2)(kT/h)^3\approx 270\,\MJysr (T/2.726\Kel)^3$. The $y$-distortion has its crossover frequency around $x\simeq 3.830\,(\equiv 217\GHz)$, while the $\mu$-distortion has its zero around $x\simeq 2.192\,(\equiv 124\GHz)$. The upper $x$-axis and right $y$-axis also give the corresponding frequency and spectral intensity for $T=2.726\,\Kel$.}
\label{fig:Y_SZ_and_mu_distortion}
\end{figure}

Assuming that we are in the regime $|y|\ll 1$, there are two cases of interest:
\begin{itemize}

\item $y>0$: overall energy is transferred from the electrons to the photons $\rightarrow$ {\it Comptonization}

\item $y<0$: energy flows from the photons to the electrons \hspace{23.2mm} $\rightarrow$ {\it Compton cooling}

\end{itemize}
For most conditions in our Universe, $y>0$ is relevant, since most processes tend to heat the matter in the Universe. Therefore {\it negative $y$-distortions} are usually not being considered, however, the adiabatic cooling of matter in the expanding Universe (in the absence of heating) allows $\Te<\Tg$, so that $y<0$ does occur \citep{Chluba2005, Chluba2011therm, Yacine2015DM}.

In Fig.~\ref{fig:Y_SZ_and_mu_distortion}, we illustrate frequency dependence of the $y$-distortion for $T_0=2.725\,\Kel$. It has a very characteristic shape, with a deficit of photons in the Rayleigh-Jeans part and an increment of photon in the Wien tail of the CMB spectrum. The limiting behaviors are
\beal{
\label{eq:def_YSZ_limits}
Y_{\rm SZ}(x)=G(x)\,\left[ x \,\frac{\expf{x}+1}{\expf{x}-1}-4\right]
\approx
\begin{cases}
-\frac{2}{x} &\text{for}\qquad x\ll 1
\\
x(x-4)\expf{-x} &\text{for}\qquad x\gg 1.
\end{cases}
}
This corresponds to $\Delta I/I\simeq \Delta T/T \simeq -2 y$ for $x\ll 1$ and $\Delta T/T\simeq (x-4) y$ for $x\gg 1$. The $y$-distortion vanishes close to $\nu\simeq 217\GHz\,(\equiv x\simeq 3.830)$, which in principle makes it distinguishable from the $\mu$-distortion (Sect.~\ref{sec:mu_dist_section}). 
One can easily verify that for a $y$-distortion $\Delta N_\gamma = 0 \propto \int x^2 Y_{\rm SZ}(x)$ and $\Delta \rho_\gamma =4 y\,\rho_{\gamma}^{\rm Pl} \propto \int x^3 Y_{\rm SZ}(x) \id x$. Clearly, Compton scattering should not change the number of photons, as reflected by $\Delta N_\gamma = 0$. The second relation means that $4 y \equiv \Delta \rho_\gamma/\rho_{\gamma}^{\rm Pl}$ defines the fractional energy exchange of the electrons with the initial blackbody spectrum. Thus, starting from a pure blackbody, by computing $y=(1/4)\Delta \rho_\gamma/\rho_\gamma^{\rm Pl}\ll 1$ one can directly give a simple approximation for the distortion \citep{Zeldovich1969}. In detail, it may be a little more involved to compute $\Delta \rho_\gamma/\rho_\gamma^{\rm Pl}$ for some process, but all one really needs to know is how much energy was pumped into the CMB by energy exchange with the thermal electrons.

\subsubsection{Thermal Sunyaev-Zeldovich effect}
Clusters of galaxies are the largest virialized objects in our Universe, with typical masses $M\simeq (10^{13}-10^{14}) M_\odot$ ($M_\odot\approx \pot{2}{33}\,{\rm g}$) and up to $\simeq 10^3$ galaxies. Cluster also host a hot plasma with free electrons at temperature $\Te \simeq \pot{\rm few}{7}\Kel \,(\equiv {\rm few}\times\keV)$ at typical densities $\Ne \simeq 10^{-3}\,\cm^{-3}$. We know this already for a while since clusters show a X-ray glow produced by thermal Bremsstrahlung.
The hot electrons can scatter CMB photons and create a Compton-$y$ distortion. The typical $y$-parameter of massive clusters is $y\simeq \pot{\rm few}{-5}-10^{-4}$ with $\The \simeq \pot{\rm few}{-2}$ and $\tau\simeq\pot{\rm few}{-3}$. Because for clusters $\Te \gg \Tg$, the $y$-parameter reduces to 
\beal{
\label{eq:def_y_parameter}
y=\int_0^\tau \frac{k\Te}{\me c^2} \id \tau'\approx \The \, \tau
}
and thus directly probes the {\it integrated electron pressure}, ${\bar P_{\rm e}\simeq \int \Ne \Te \id l}$, through the cluster medium. More than $10^3$ clusters have been now seen using the SZ effect \cite{Planck2014SZ}.

One of the great properties of the thermal SZ effect that is it independent of redshift (ignoring evolutionary effects) \citep{Sunyaev1980, Rephaeli1995ARAA, Carlstrom2002}. The reason is that CMB temperature increases $\propto (1+z)$ with redshift, so that the `{\it light bulb}' illuminating the hot electrons residing inside the cluster becomes brighter the higher the redshift. The cosmological redshift dimming of the signals, which for example reduces the X-ray fluxes for high redshift clusters, is therefore compensated since the CMB itself brightens, and no matter what the redshift of the cluster is it will have the same signal relative to the CMB. 
%
%
The redshift-independence of the SZ signal makes SZ clusters a powerful cosmological probe, since one can in principle track the growth of structures out to high redshifts ($z\simeq 1-2$) and thus constrain cosmological parameters and the evolution of dark energy \citep{Birkinshaw1999, Carlstrom2002, Bolliet2018}.

But the thermal SZ effect is even more rich. For a cluster with $k\Te=5\keV$, the thermal velocities of the electrons are $\varv_{\rm th}\simeq \sqrt{2 \The}c\simeq 0.14c$. That is quite fast and relativistic corrections become important. In this regime the Kompaneets equation is no longer valid and one has to include higher order corrections \citep{Challinor1998, Sazonov1998, Itoh98, Chluba2012SZpack}. In addition, if the cluster is moving with respect to the CMB, the Doppler kick adds a change in the CMB temperature towards the cluster by $\Delta I\simeq \beta_{\rm c} \tau \,T\partial_T B_\nu(T)$, also knows as {\it kinematic SZ effect} \citep{Sunyaev1980}. This can in principle be used to study large-scale bulk flows in the Universe.

\subsection{Chemical potential or $\mu$-distortion}
\label{sec:mu_dist_section}
We now understand that for inefficient energy exchange between electrons and photons (i.e., $y\ll 1$) the shape of the distortion is determined by the $y$-parameter and has a spectral dependence, $Y_{\rm SZ}(x)=G(x)[x \coth(x/2)-4]$, shown in Fig.~\ref{fig:Y_SZ_and_mu_distortion}. Let us now consider the other extreme, when many scatterings are taking place and the redistribution of photons in frequency is very efficient (i.e., $y\gg 1$). In the early Universe, this regime is found at $z\gtrsim \pot{5}{4}$ and the distortion is given by a $\mu$-distortion.

\subsubsection{Compton equilibrium solution}
When many scatterings occur, the spectrum is driven towards an equilibrium with respect to Compton scattering. Neglecting emission and absorption processes, the kinetic equation thus becomes quasi-stationary
\beal{
0 \approx  \frac{\The}{x^2}\frac{\partial}{\partial x} x^4 \left[\frac{\partial}{\partial x} n +\frac{\Tg}{\Te}\,n(1+n) \right].
}
One solution of this equation is $n_{\rm bb}=1/(\expf{x}-1)$ if $\Te\equiv \Tg$, since $\partial_x n_{\rm bb}=-n_{\rm bb}(1+n_{\rm bb})$, as it should be for full equilibrium. However, this is not the general solution of the problem. To find a more general solution we have to solve the equation $\partial_x n=-\frac{\Tg}{\Te}\,n(1+n)$. The factor $\Tg/\Te$ can be absorbed by redefining the frequency scaling $x\rightarrow \xe$ so that this becomes $\partial_{\xe} n=-n(1+n)$. This can be integrated to $\ln(1+n)-\ln(n)\equiv \xe+{\rm const}$, or 
\beal{
n_{\rm BE}=\frac{1}{\expf{\xe+\mu_0}-1},
}
where we introduced the integration constant $\mu_0$. This is a {\it Bose-Einstein spectrum} with constant chemical potential\footnote{Notice that the sign is different from the normal convention used in thermodynamics.} $\mu_0$. Let's pause for a moment. Photons have no rest mass, so the chemical potential should vanish, shouldn't it? This statement is only true if we are in full equilibrium, i.e., we have a blackbody at the temperature of the medium. More generally, for fixed photon number and energy densities the chemical potential can be non-zero. 

The chemical potential can in principle be both positive or negative:
\begin{itemize}

\item $\mu_0>0$: {\it fewer} photons than a in blackbody at $\Te$ $\rightarrow$ {\it energy release / photon destruction}

\item $\mu_0\equiv 0$: blackbody at temperature $\Te$ 
\qquad\;\;\,\,\quad\hspace{1.2mm} $\rightarrow$ {\it full equilibrium}

\item $\mu_0<0$: {\it more} photons than a in blackbody at $\Te$ 
\,$\rightarrow$ {\it energy extraction / photon injection}

\end{itemize}
In practice, the solution $\mu_0<0$ is unphysical unless $\mu_0$ is actually a function of frequency. The reason is that $\xe + \mu_0$ can vanish at $\xe=-\mu_0>0$, but this state is never reached or even passed through during the evolution, since instead excess photons would form a Bose-condensate at $x=0$ with $\mu_0=0$ elsewhere \citep{Levich1969, Zeldovich1972shock}. In a real plasma, BR and DC emission will prevent this from happening though \citep{Illarionov1975, Khatri2011BE}.

\subsubsection{Definition of the $\mu$-distortion}
In the previous section, we found that $n=1/(\expf{\xe+\mu_0}-1)$ is approached for many scatterings in the plasma. But how do we fix the constant $\mu_0$ and what is the definition of the distortion really? Let us assume we start with a blackbody and electrons at temperature $\Tg=\Te=T_i$. Let us change the number and energy densities of the photon field by some $\epsilon_N=\Delta N_\gamma/N_\gamma^{\rm Pl}(T_i)$ and $\epsilon_\rho=\Delta \rho_\gamma/\rho_\gamma^{\rm Pl}(T_i)$, respectively, and then wait until everything has equilibrated by Compton scattering. This means
\vspace{-1.5mm}
\bsub
\label{eq:BE_conditions}
\beal{
N^{\rm BE}_\gamma=N_\gamma^{\rm Pl}(T_i)(1+\epsilon_N)\equiv 
\frac{N_\gamma^{\rm Pl}(T_f)}{G_2^{\rm Pl}}
\int \frac{x_{\rm f}^2\id x_{\rm f}}{\expf{x_{\rm f}+\mu_0}-1}
\\
\rho^{\rm BE}_\gamma=\rho_\gamma^{\rm Pl}(T_i)(1+\epsilon_\rho) \equiv 
\frac{\rho_\gamma^{\rm Pl}(T_f)}{G_3^{\rm Pl}}
\int \frac{x_{\rm f}^3\id x_{\rm f}}{\expf{x_{\rm f}+\mu_0}-1},
}
\esub
where $T_f$ is the final electron temperature in the distorted (Bose-Einstein spectrum) radiation field, $x_{\rm f}=h\nu/kT_f$ and $G_2^{\rm Pl}\approx 2.404$ and $G_3^{\rm Pl}\approx 6.494$. These two equations allow us to fix $T_f$ and $\mu_0$ as a function of the parameters $\epsilon_N$ and $\epsilon_\rho$. Assuming that all changes are small we have 
\vspace{-1.5mm}
\bsub
\beal{
N^{\rm BE}_\gamma &\approx N_\gamma^{\rm Pl}(T_f)\left[1 - \mu_0 \mathcal{M}^{\rm c}_2\right]
\approx N_\gamma^{\rm Pl}(T_i)\left[1 + 3\frac{\Delta T}{T_i}- \mu_0 \mathcal{M}^{\rm c}_2\right]
\\
\rho^{\rm BE}_\gamma &\approx  \rho_\gamma^{\rm Pl}(T_f)\left[1 - \mu_0 \mathcal{M}^{\rm c}_3\right]
\,\approx\;  \rho_\gamma^{\rm Pl}(T_i)\left[1 + 4\frac{\Delta T}{T_i} - \mu_0 \mathcal{M}^{\rm c}_3\right],
}
\esub
where $\mathcal{M}^{\rm c}_2 \approx 1.3684$ and $\mathcal{M}^{\rm c}_3 \approx 1.1106$. With the conditions Eq.~\eqref{eq:BE_conditions}, we then find
\citep{Sunyaev1970mu, Hu1995PhD}
\vspace{-1.5mm}
\bsub
\label{eq:sols_mu_DT_T}
\beal{
\label{eq:sols_mu_DT_T_a}
\mu_0&\approx 
\frac{3}{\kappa^{\rm c}}\left[\frac{\Delta \rho_\gamma}{\rho_\gamma}-\frac{4}{3}\frac{\Delta N_\gamma}{N_\gamma}\right]
\approx 1.401\left[\frac{\Delta \rho_\gamma}{\rho_\gamma}-\frac{4}{3}\frac{\Delta N_\gamma}{N_\gamma}\right]
\\[1mm]
\label{eq:sols_mu_DT_T_b}
\frac{\Delta T}{T_i}
&\approx 
\frac{\mathcal{M}^{\rm c}_2}{\kappa^{\rm c}}\frac{\Delta \rho_\gamma}{\rho_\gamma}
-
\frac{\mathcal{M}^{\rm c}_3}{\kappa^{\rm c}}\frac{\Delta N_\gamma}{N_\gamma}
\approx 0.6389 \frac{\Delta \rho_\gamma}{\rho_\gamma} - 0.5185 \frac{\Delta N_\gamma}{N_\gamma}
\approx 0.4561 \mu_0 + \frac{1}{3} \frac{\Delta N_\gamma}{N_\gamma}
}
\esub
with $\kappa^{\rm c}=4\mathcal{M}^{\rm c}_2-3\mathcal{M}^{\rm c}_3\approx 2.1419$. From Eq.~\eqref{eq:sols_mu_DT_T_a} we see that for $\Delta \rho_\gamma/\rho_\gamma\equiv (4/3)\Delta N_\gamma/N_\gamma$ we have no distortion ($\mu_0=0$), as we already understood from the adiabatic condition, Eq.~\eqref{eq:blackbody_cond}. In this case, only the temperature of the blackbody is increased\footnote{we neglect the small heat capacity of the electrons and baryons.} after Compton scattering redistributed all photons, $\Delta T/T_i\approx \frac{1}{3} \Delta N_\gamma/N_\gamma$.

\begin{figure}
\centering
\includegraphics[width=0.8\columnwidth]{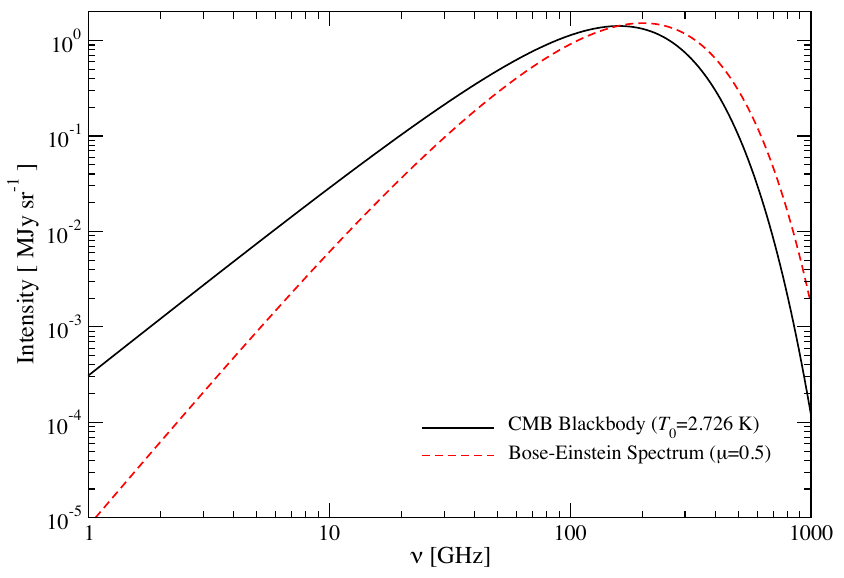}
\caption{Bose-Einstein spectrum for large chemical potential $\mu=0.5$ and $T_i=T_0=2.726\,\Kel$. Only energy was added to the photon field, but the number of photons was not changed with respect to the initial CMB spectrum. For large chemical potential, the cross over frequency shifts towards higher frequencies according to $\nu_{\mu}\approx 124\,\GHz\,(1-0.304\,\mu\ln\mu)\approx 158\,\GHz$. The figure was taken from \citet{Chluba2014}.}
\label{fig:BE_spectrum}
\end{figure}
In Fig.~\ref{fig:BE_spectrum} we illustrate a Bose-Einstein spectrum with $\mu_0=0.5$ and $T_i=T_0=2.726\,\Kel$. Only energy was added to the photons but the number of photons was not changed with respect to the initial CMB spectrum. One can see that in the Rayleigh-Jeans tail of the CMB the Bose-Einstein spectrum shows a deficit of photons, while in the Wien tail more photons than in the CMB blackbody spectrum are present. We have $n_{\rm BE}\approx n_{\rm bb}$ at $\nu_{\mu}\approx 124\,\GHz$ although for large chemical potential $\nu_{\mu}\approx 124\,\GHz\,(1-0.304\,\mu\ln\mu)$ is more accurate \citep{Chluba2014}.

\subsubsection{But how do we define the distortion} To derive the expressions from above, we used 
\beal{
n_{\rm BE}=\frac{1}{\expf{\xe+\mu_0}-1}\approx \frac{1}{\expf{\xe}-1}- \frac{G(\xe)}{\xe}\mu_0+\mathcal{O}(\mu_0^2)
}
for $\mu_0\ll 1$. This suggest that $\Delta n=-G(\xe) \, \mu_0/\xe$ could be called the distortion with respect to the blackbody part at temperature $\Te$ and in fact this definition has been used frequently. However, since also the final electron temperature, $\Te=T_f$, depends on $\mu_0$, this definition does not separate the distortion cleanly. Motivated by the fact that Compton scattering conserves photon number, one natural definition is to fix the $\mu$-distortion such that $\int x^2 M(x)\id x=0$. Integrating $\Delta n$ gives $\int x^2 \Delta n \id x=-2\mu_0\int x \id x /(\expf{x}-1)=-\mu_0\,\pi^2/3 \approx -3.2899\,\mu_0$, so that $M(x)=G(x)[\alpha_\mu - 1/x]$ with $\alpha_\mu=\pi^2/18\zeta(3)\approx 0.4561$ fulfills $\int x^2 M(x)\id x=0$. If in addition we now normalize the relative change of the photon energy density to unity $(\Delta \rho_M/\rho^{\rm Pl}=1)$, we obtain the spectral shape of the {$\mu$-distortion}
\beal{
\label{eq:def_mu_distortion_limits}
M^\ast(x)=\frac{3}{\kappa^{\rm c}} M(x)
\approx 1.401 G(x)\left[0.4561 -\frac{1}{x}\right]
\approx
\begin{cases}
-\frac{1.401}{x^2} &\text{for}\quad x\ll 1
\\
0.6390 \,x \,\expf{-x} &\text{for}\quad x\gg 1,
\end{cases}}
where $3/\kappa^{\rm c}\approx 1.401$. This implies $\Delta I/I\simeq \Delta T/T \simeq - \mu_0/x$ for $x\ll 1$ and $\Delta T/T\simeq 0.4561\,\mu_0$ at $x\gg 1$. The frequency dependence of $M(x)$ is illustrated in Fig.~\eqref{fig:Y_SZ_and_mu_distortion} in comparison with the $y$-distortion and spectrum of a temperature shift. The important feature of a $\mu$-distortion is that it is shifted towards lower frequencies with respect to the $y$-distortion. This makes it in principle distinguishable and observing a $\mu$-distortion is a clear indication for a signal created in the pre-recombination era, deep into the thermal history of our Universe.

\begin{figure}
\centering
\includegraphics[width=0.74\columnwidth]{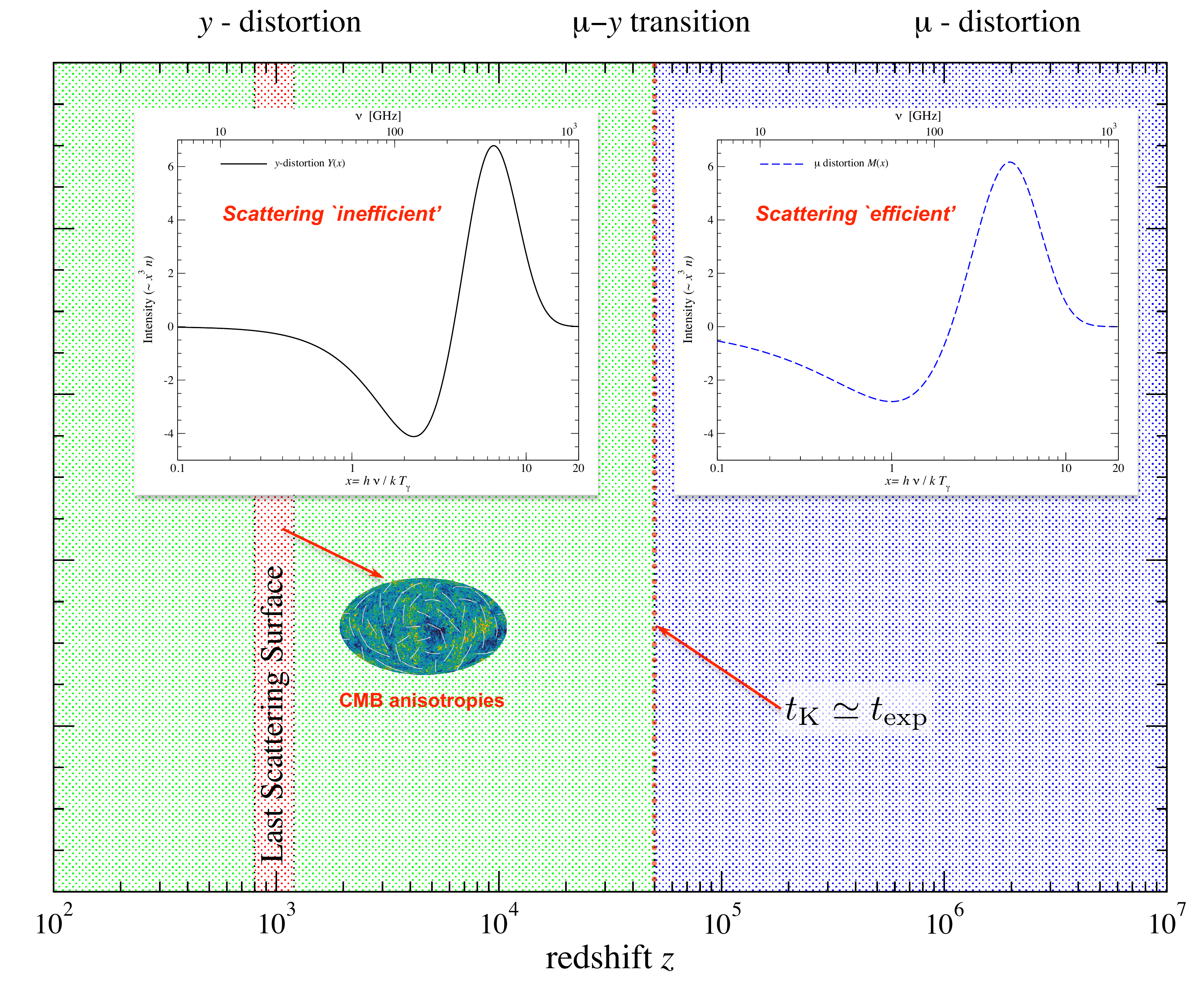}
\caption{Simplest zeroth order picture for the formation of primordial distortions. At low redshifts ($z\lesssim \pot{5}{4}$), a $y$-distortion is formed, while at high redshifts we expect a $\mu$-distortion. At this point we have not included any photon production and we will see that this strongly attenuates the amplitude of the $\mu$-distortion at $z\gtrsim \pot{2}{6}$.}
\label{fig:zeroth_order_picture}
\end{figure}

\vspace{-2mm}
\subsection{Simple description of primordial distortions}
\label{sec:zeroth_order_picture}
We now have all the pieces for a simplest, zeroth order description of primordial distortions. At late times, ($z\lesssim z_{\mu y}\simeq \pot{5}{4}$), the redistribution of photons by Compton scattering becomes inefficient and a $y$-type distortion is formed, in the other extreme we have a $\mu$-distortion (see Fig.~\ref{fig:zeroth_order_picture}) with the approximations \cite{Zeldovich1969, Sunyaev1970mu}
\vspace{-1.5mm}
\bsub
\beal{
y&\approx \frac{1}{4} \left.\frac{\Delta \rho_\gamma}{\rho_\gamma}\right|_y
\\
\label{eq:approx_y_mu_standard_b}
\mu_0&\approx 1.401 \left[\left.\frac{\Delta \rho_\gamma}{\rho_\gamma}\right|_\mu-\frac{4}{3}\left.\frac{\Delta N_\gamma}{N_\gamma}\right|_\mu\right],
}
\esub
such that the total distortion is given by $\Delta n\approx Y_{\rm SZ} \, y + M(x)\, \mu_0$. Here, we indicate that to estimate the distortion one needs to consider the {\it partial energy release} and {\it photon production} relative to the CMB blackbody in the respective $y$- and $\mu$-era. If extra photons are injected in the $y$-era (e.g., by particle decay), the distortion generally is not just a $y$-distortion, since these extra photons are not redistributed very efficiently, but in the $\mu$-era they can be ingested and modify the effective chemical potential. Photon injection during the $y$-era was considered in detail in \cite{Chluba2015GreenII}.

Two important aspects are still missing. Firstly, we have not included any thermal photon production by BR or DC but assumed that only Compton scattering changes the photon field. Photon production will be mostly relevant for the evolution of $\mu$-distortions, implying that not all energy release or photon production is eventually visible as a distortion. That is, at very early phases the {\it distortion visibility} (see explanation below) is smaller than unity because thermalization reduces the effective amount of energy release that survives as a distortion. This is implicitly hidden in the definition of $\Delta \rho_\gamma/\rho_\gamma|_\mu$ and $\Delta N_\gamma/N_\gamma|_\mu$. We will consider this problem in Sect.~\ref{sec:distortion_vis}. The second point is that the transition between $\mu$ and $y$ distortions is not abrupt at $z\simeq \pot{5}{4}$ but occurs over a range of redshifts, where in the intermediate regime the distortion is not only given by the superposition of $\mu$ and $y$-distortion. This makes the distortion signal much richer, as pointed out only recently \citep{Chluba2011therm, Khatri2012mix, Chluba2013Green}. We will consider this problem in Sect.~\ref{sec:transition_mu_y}.

\subsubsection{Inclusion of photon production in the $\mu$-era}
\label{sec:distortion_vis}
It is straightforward to approximately include the effect of BR and DC in the $\mu$-distortion era. For a detailed discussion of the approximations and its limitations we refer the interested reader to \citet{Chluba2014}.
Since scattering is efficient, we can assume that the spectrum evolves along a sequence of quasi-stationary stages. However, now we also have to account for emission and absorption, such that 
\beal{
0 \approx  \frac{\The}{x^2}\frac{\partial}{\partial x} x^4 \left[\frac{\partial}{\partial x} n +\frac{\Tg}{\Te}\,n(1+n) \right] + \frac{K}{x^3}\Big[1-n(\expf{\xe}-1)\Big]
}
determines the CMB spectrum. Inserting $n\approx 1/(\expf{\xe}-1)-\mu(z, \xe)\,G(\xe)/\xe$ and assuming $\xe\ll 1$ yields a simple differential equation for $\mu(z, \xe)\ll 1$, which has the approximate solution 
\vspace{-1mm}
\beal{
\label{eq:mu_x}
\mu(z, \xe)\approx \mu_0(z) \,\expf{-\xc(z)/\xe}.
}
This solution was first derived by \citet{Sunyaev1970mu}. Including both DC and BR, the critical frequency, $\xc$, which is determined by the competition between photon emission and absorption and Compton up-scattering of photons, is usually $\xc(z)\simeq 10^{-3}-10^{-2}$ during the thermalization period \citep{Burigana1991, Hu1993}.

Equation~\eqref{eq:mu_x} shows that at $x\gg \xc$, the chemical potential becomes constant, $\mu(z, \xe)\approx \mu_0(z)$, while at low frequencies it vanishes exponentially, returning to a blackbody at the temperature of the electrons, with a smooth transition between these regimes around $x\simeq \xc$. The solution has the expected limiting behavior, even if strictly speaking it is only valid at low frequencies. Indeed, the correct high-frequency behavior is $\mu(z, x)\simeq \mu^\ast_0(z)+{\rm C}(z) \ln x$, where the coefficient, ${\rm C}(z)$, is related to the time derivative of the electron temperature \citep{Chluba2014}.

With Eq.~\eqref{eq:mu_x}, one can now compute the total photon production rate at any redshift. From that one can estimate how the high-frequency photon chemical potential is affected by photon production. This essentially boils down to a differential equation for $\mu_0(z)$, which for single energy release $\Delta \rho_\gamma/\rho_\gamma|_i$ has the solution
\vspace{-1mm}
\beal{
\label{eq:mu_sol_sing}
\mu_0(z)
&\approx 
1.401\left.\frac{\Delta\rho_\gamma}{\rho_\gamma} \right|_i
\expf{-(z_i/z_\mu)^{5/2}+(z/z_\mu)^{5/2}}=\mu_{i}\,\Jbb(z_i, z)
}
with $z_{\rm dc}\approx \pot{1.98}{6}$ \citep{Burigana1991, Hu1993}. Here, we defined $\mu_i=1.401\Delta \rho_\gamma/\rho_\gamma|_i$. The factor $\Jbb(z_i, z)$ defines the {\it spectral distortion visibility} between the injection redshift $z_i$ and $z$ (with $z<z_i$). It determines the fraction of energy injected at $z_i$ that is still visible as a distortion at $z$. For $\Jbb(z_i, z)\simeq 1$, most of the energy is still stored in the distortion, while for $\Jbb(z_i, z)\ll 1$, most of the energy was thermalized and converted into a temperature shift. 
This implies that after a single energy release event, today's remaining chemical potential is heavily suppressed if the energy injection happens at $z\gtrsim z_{\rm dc}\approx \pot{1.98}{6}$ or some $\simeq 3$ months after the big bang. 
For continuous energy release in the $\mu$-era, we can then estimate the final distortion measured today using \citep{Sunyaev1970mu, Burigana1991, Hu1993}
\vspace{-1mm}
\beal{
\label{eq:mu_improved}
\mu_0 &\approx 1.401 \left.\frac{\Delta \rho_\gamma}{\rho_\gamma}\right|_\mu 
\approx 1.401 \int_{z_{\mu y}}^\infty \frac{\id ({Q}/\rho_\gamma)}{\id z}\Jbb(z', 0) \id z'.
}
Here, $\id ({Q}/\rho_\gamma)/\id z$ describes the energy release relative to the CMB blackbody and depends on the specific energy release mechanism (see Sect.~\ref{sec:mechanisms}). We neglected any extra photon production, but refer to \cite{Chluba2015GreenII} for additional discussion.

\subsubsection{The importance of double Compton emission} In the above, the DC process dominated in 	the definition of the thermalization redshift \citep{Danese1982, Burigana1991, Hu1993}
\vspace{-1mm}
\beal{
\zmudc\approx \pot{1.98}{6} \left[\frac{\Omega_{\rm b} h^2}{0.022}\right]^{-2/5} \left[\frac{T_0}{2.725 \Kel}\right]^{1/5}
\left[\frac{(1-\Yp/2)}{0.88}\right]^{-2/5},
}
assuming $N_{\rm eff}=3.046$. At $z\gg\zmudc $, thermalization is very efficient and the distortion visibility drops exponentially. If alternatively we only include BR emission, we find \citep{Sunyaev1970mu, Danese1982, Hu1993}
\vspace{-1mm}
\beal{
\mathcal{J}_{\rm BR}(\zh)=\exp\left(-[\zh/z_{\rm br}]^{1.328}\right)
}
with $z_{\rm br}\approx \pot{5.27}{6}$. In the classical result, given first by \citet{Sunyaev1970mu}, the power-law coefficient is $5/4=1.25$ because a different approximation for the BR Gaunt factor was utilized.
This shows that the thermalization redshift is significantly higher when only BR is included. In addition, the distortion visibility function drops less steeply at $z\gtrsim \pot{5.27}{6}$.

\begin{figure}
\centering
\includegraphics[width=0.7\columnwidth]{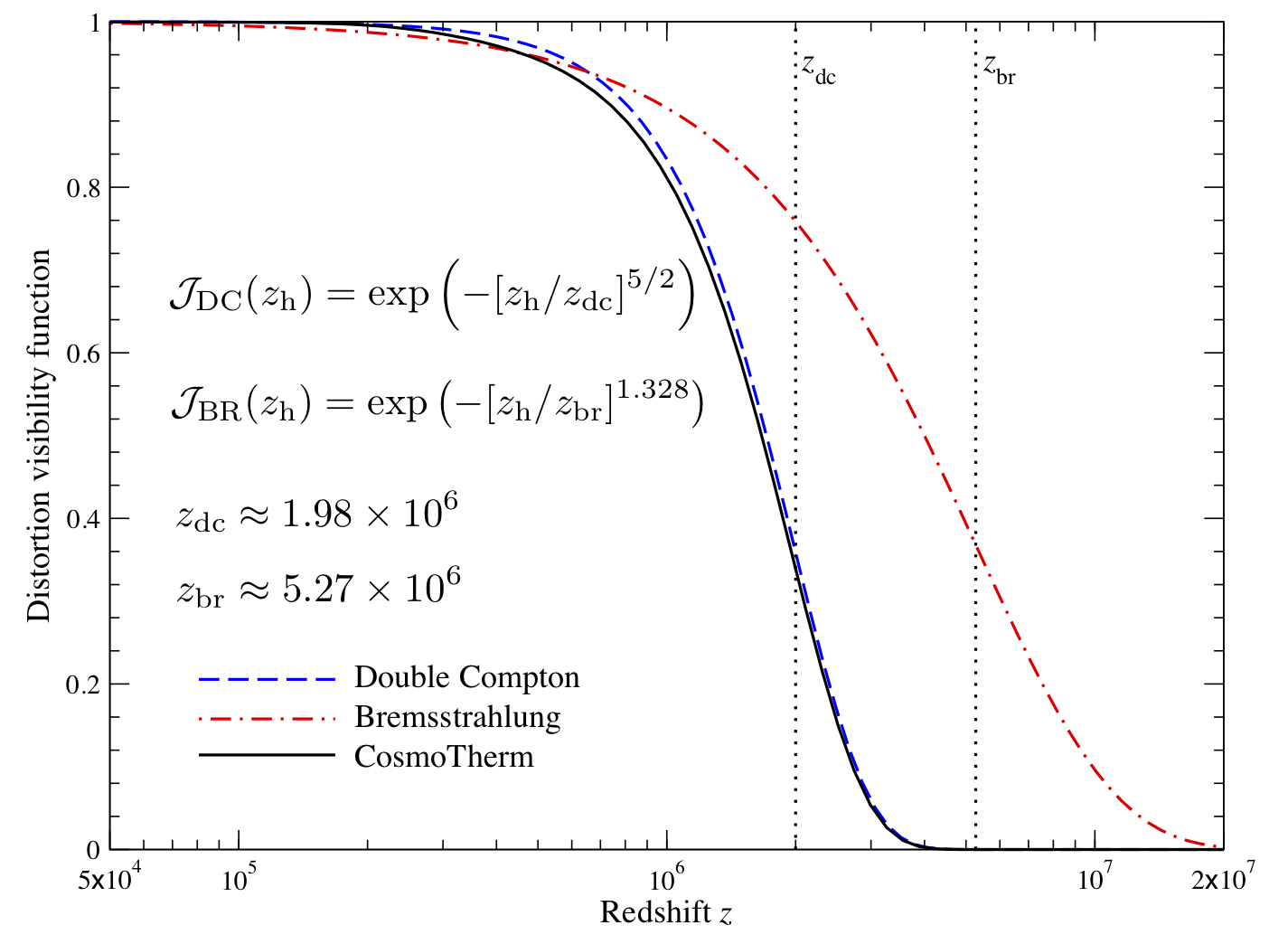}
\caption{Distortion visibility function (adapted from \cite{Chluba2014}). We compare $\mathcal{J}_{\rm DC}(\zh)$, $\mathcal{J}_{\rm BR}(\zh)$ and the numerical result obtained with {\sc CosmoTherm}. DC emission significantly change the thermalization efficiency.}
\label{fig:Vis_DC_BR}
\end{figure}

In Fig.~\ref{fig:Vis_DC_BR} we compare the distortion visibility functions for DC and BR only with the full numerical result for the distortion visibility obtained from {\sc CosmoTherm} \citep{Chluba2011therm, Chluba2014}. Clearly, DC emission increases the thermalization efficiency significantly. If only BR were taken into account, we would still expect to see some small distortion even from the tail of the electron-positron annihilation era around $z\simeq \pot{2}{7}$! In full detail, this would be quite complicated to compute, but luckily the distortion visibility is exceedingly small, even if only DC is included, providing a rough but tiny upper limit. Comparing with the full numerical result, $\mathcal{J}_{\rm DC}(\zh)=\exp\left(-[\zh/z_{\rm dc}]^{5/2}\right)$, provides a very good approximation, which, for simple estimates, is more than sufficient. Improvements to $\mathcal{J}$ can be added analytically \cite{Chluba2014, Khatri2012b}, but for refined computations it is easier to simply use the Green's function method (e.g., \citep{Chluba2013Green, Chluba2015GreenII}), described in the next Section.

\begin{figure} 
   \centering
   \includegraphics[width=0.75\columnwidth]{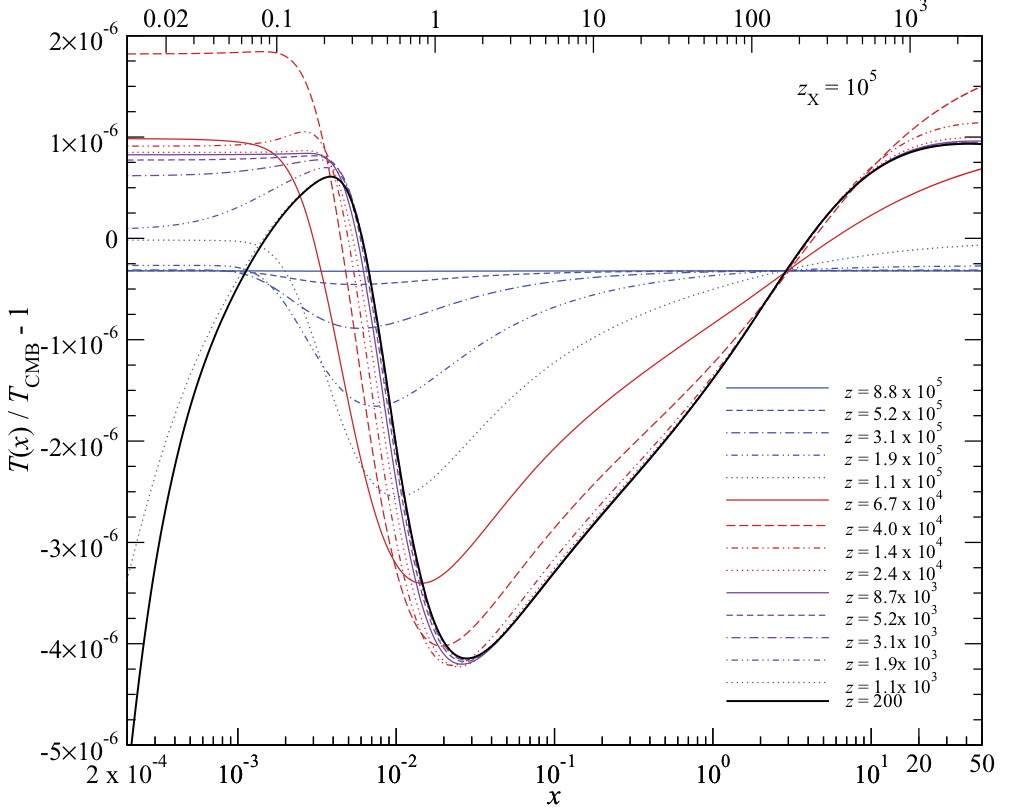}
   \caption{Spectral distortion signal (in terms of brightness temperature) caused by a decaying particle scenario at different stages of the evolution (Figure taken from \cite{Chluba2011therm}). The total energy release was $\Delta \rho_\gamma/\rho_\gamma \simeq \pot{1.3}{-6}$ assuming a particle lifetime $t_X\simeq \pot{2.4}{9}\,{\rm s}$ or $z_X= 10^5$. The final distortion is not described by a simple superposition of $\mu$ and $y$ and thus contains valuable time-dependent information teaching us about the lifetime of the partcle.}
   \label{fig:Decay_example}
\end{figure}

\begin{figure} 
   \centering
   \includegraphics[width=0.9\columnwidth]{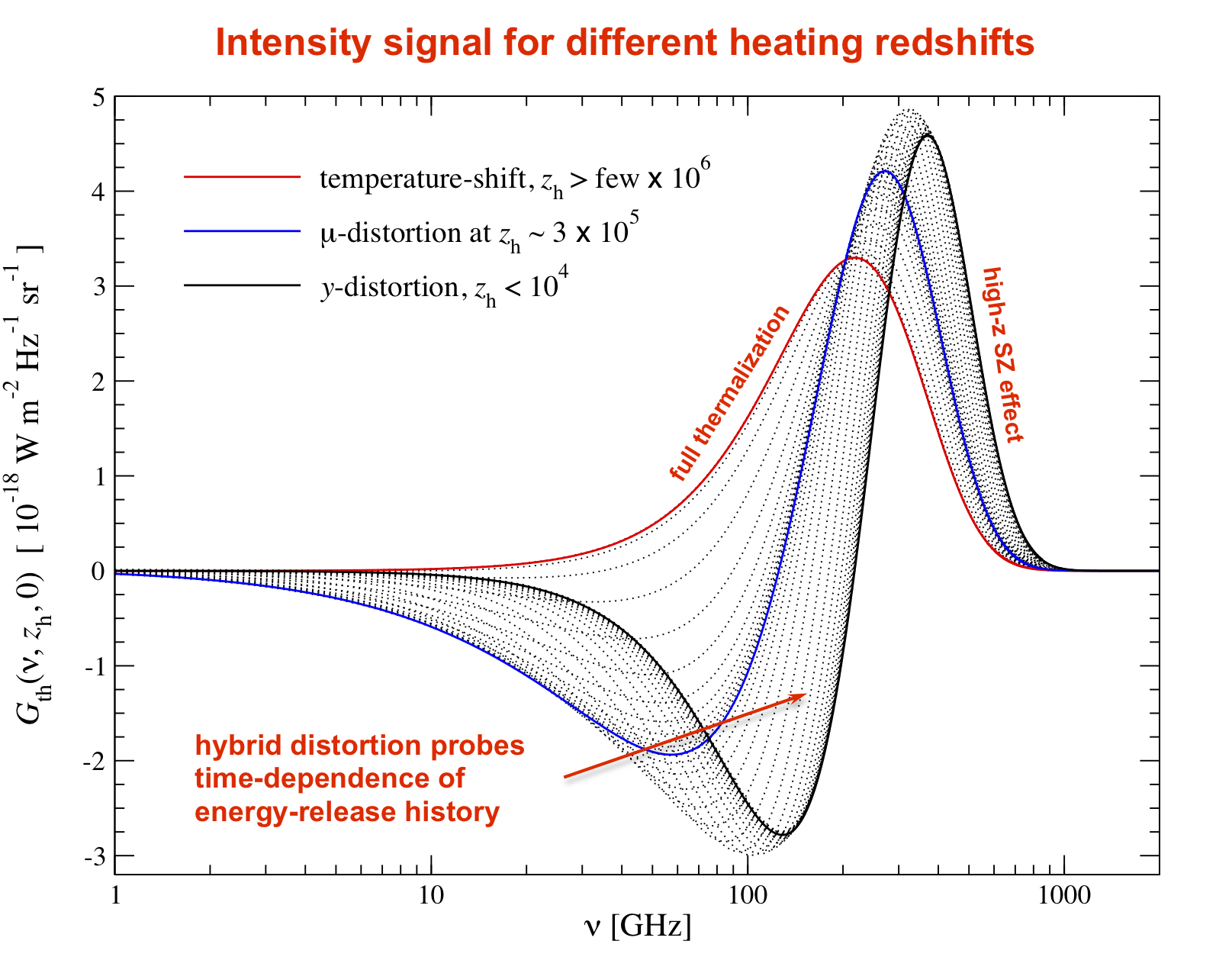}
   \\[2mm]
   \includegraphics[width=0.9\columnwidth]{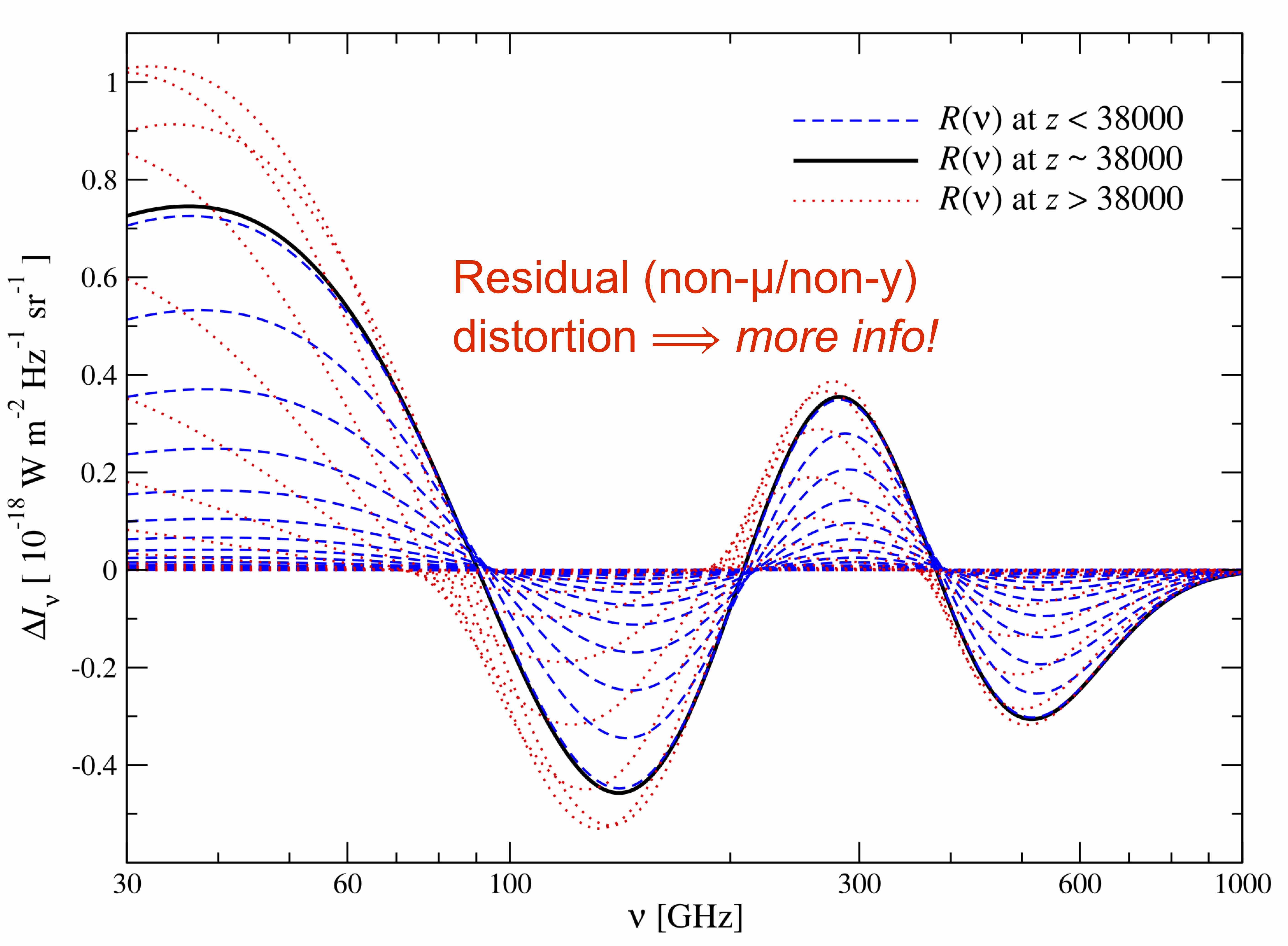}
   \caption{Change in the CMB spectrum after a single energy release at different heating redshifts, $z_{\rm h}$. Top panel: thermalization Green's function, $G_{\rm th}(\nu, z_{\rm h})$ -- lower panel: residual distortion. At $z\gtrsim \pot{\rm few}{6}$, a temperature shift is created. Around $z \simeq \pot{3}{5}$ a pure $\mu$-distortion appears, while at $z\lesssim 10^4$ a pure $y$-distortion is formed. At all intermediate stages, the signal is given by a superposition of these extreme cases with a small residual (non-$\mu$/non-$y$) distortion that contains information about the time-dependence of the energy-release process (Figures adapted from \cite{Chluba2013Green} and \cite{Chluba2013PCA}).}
   \label{fig:Greens}
\end{figure}
\subsection{Modeling the transition between $\mu$ and $y$}
\label{sec:transition_mu_y}
In what was presented so far we modeled the transition between $\mu$ and $y$-era as a simple step-function around $z\simeq z_{\mu y}$. Even in the early studies of the evolution of distortions it was realized that this is not quite correct and that the transition is much more gradual \citep{Illarionov1975, Burigana1991, Hu1995PhD}. However, only more recently it was explicitly highlighted that the distortion in the intermediate-regime ($z\simeq 10^4 - \pot{3}{5}$), contains valuable additional information, allowing us in principle to distinguish different types of distortions \cite{Chluba2011therm, Khatri2012mix, Chluba2013fore, Chluba2013PCA}.

In Fig.~\ref{fig:Decay_example}, we illustrate the shape of the distortion caused by a decaying particle scenario at different stages of the evolution. The final distortion is not simply given by the sum of $\mu$ and $y$ distortions and thus could allow determining the lifetime of the particle \citep{Chluba2011therm}. The description of the distortion in the intermediate regime was later refined by \citep{Khatri2012mix} and \citep{Chluba2013Green}. For single energy release, the distortion response ($\leftrightarrow$ Green's function) is illustrated in Fig.~\ref{fig:Greens}. Eliminating the leading order $\mu$ and $y$-distortion contributions, one is left with a smaller signal, the so called {\it residual distortion} or $r$-type distortion (see Fig.~\ref{fig:Greens}), which can be conveniently parametrized using distortion eigenmodes \citep{Chluba2013PCA}. The $r$-type distortion is what contains the extra time-dependent information and detection limits for different energy release scenarios are presented in \citep{Chluba2013PCA}.
However, the thermalization problem is even richer when including the effect of pre-recombination ($z\gtrsim 10^3$) atomic transitions \cite{Liubarskii83, Chluba2008c}. This might allow us to reach even deeper into the $\mu$- and $y$-eras by using spectral features of the cosmological recombination radiation \cite{Sunyaev2009}.

\begin{figure}
\centering
\includegraphics[width=0.85\columnwidth]{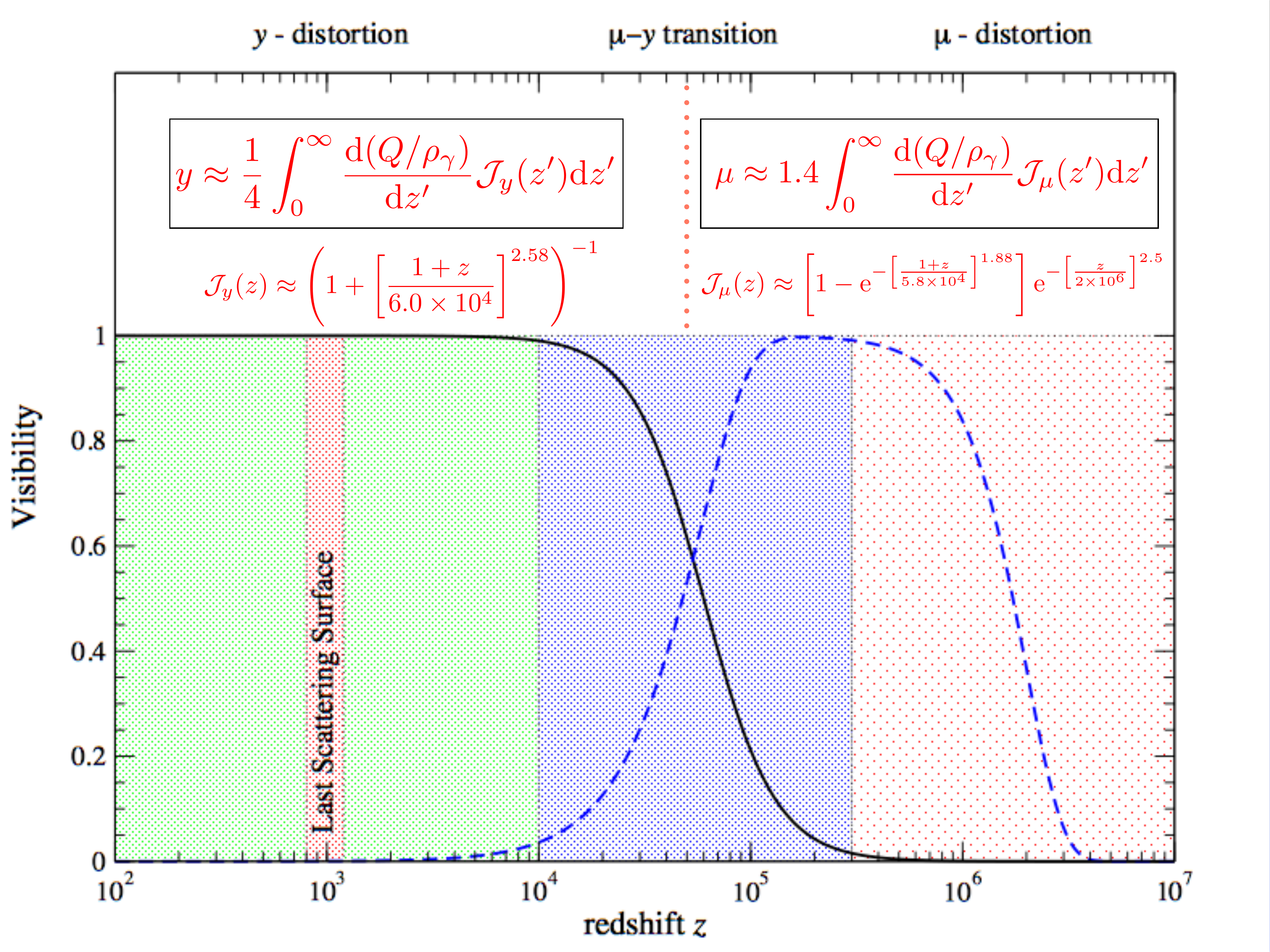}
\caption{Improved picture for the formation of primordial distortions. At low redshifts ($z\lesssim z_{\mu y}\simeq \pot{5}{4}$), a $y$-distortion is formed with distortion visibility close to unity, while at high redshifts a $\mu$-distortion appears. The energy release has to be weighted with distortion visibility function which drops exponentially at $\zmudc\gtrsim \pot{2}{6}$, leading to a pure temperature shift in that regime from inside the {\it cosmic photosphere}.}
\label{fig:improved_picture}
\end{figure}

For computational purposes, efficient modeling of the $r$-type distortion is best handled using the Green's function method \citep{Chluba2013Green}. For even more accurate results, the flexible thermalization code {\tt CosmoTherm} \citep{Chluba2011therm} can be used, which now runs in $\simeq 30\,{\rm s}$ for a given model on a standard laptop.
However, we can still improve the analytical description of the $\mu$ and $y$-distortion contributions using simple representations of the total distortion. From the full Green's function response, one can determine the best-fitting $\mu$ and $y$ distortion representation. The obtained approximation can then be used to improve the distortion visibility functions in the different regimes. This approach was used in \cite{Chluba2013Green} and can be summarized using
\vspace{-1mm}
\bsub
\label{eq:Greens_approx_improved}
\begin{align}
y&=\frac{1}{4}\left.\frac{\Delta \rho_\gamma}{\rho_\gamma}\right|_y = \frac{1}{4}\int_0^\infty \mathcal{J}_y(z')\,\frac{\id (Q/\rho_\gamma)}{\id z'} \id z'
\\
\mu&=1.401\left.\frac{\Delta \rho_\gamma}{\rho_\gamma}\right|_\mu =1.401 \int_0^\infty \mathcal{J}_\mu(z') \frac{\id (Q/\rho_\gamma)}{\id z'} \id z'
\end{align}
\esub
with the distortion visibilities
\vspace{-1mm}
\bsub
\begin{align}
\label{eq:branching_approx_improved}
\mathcal{J}_y(z)
&\approx
\begin{cases}
\left(1+\left[\frac{1+z}{\pot{6}{4}}\right]^{2.58}\right)^{-1}
 & \text{for}\; z_{\rm rec}\simeq 10^3 \leq z 
\\
0 & \text{otherwise}
\end{cases}
\\
\mathcal{J}_\mu(z) 
&\approx\mathcal{J}_{\rm bb}(z)\,\left[1-\exp\left(-\left[\frac{1+z}{\pot{5.8}{4}}\right]^{1.88}\right)\right].
\end{align}
\esub
These expression should represent the exact fractions of $\mu$ and $y$ to $\simeq 10\%-20\%$ precision. To ensure full energy conservation (no leakage of energy to the $r$-distortion), instead one can use $\mathcal{J}_\mu(z)\approx [1- \mathcal{J}_y(z)]\,\mathcal{J}_{\rm bb}(z)$. These approximation were presented in \cite{Chluba2015IJMPD} and \cite{Chluba2016}. The distortion visibilities are illustrated in Fig.~\ref{fig:improved_picture}.

\begin{figure}
\centering
\includegraphics[width=0.95\columnwidth]{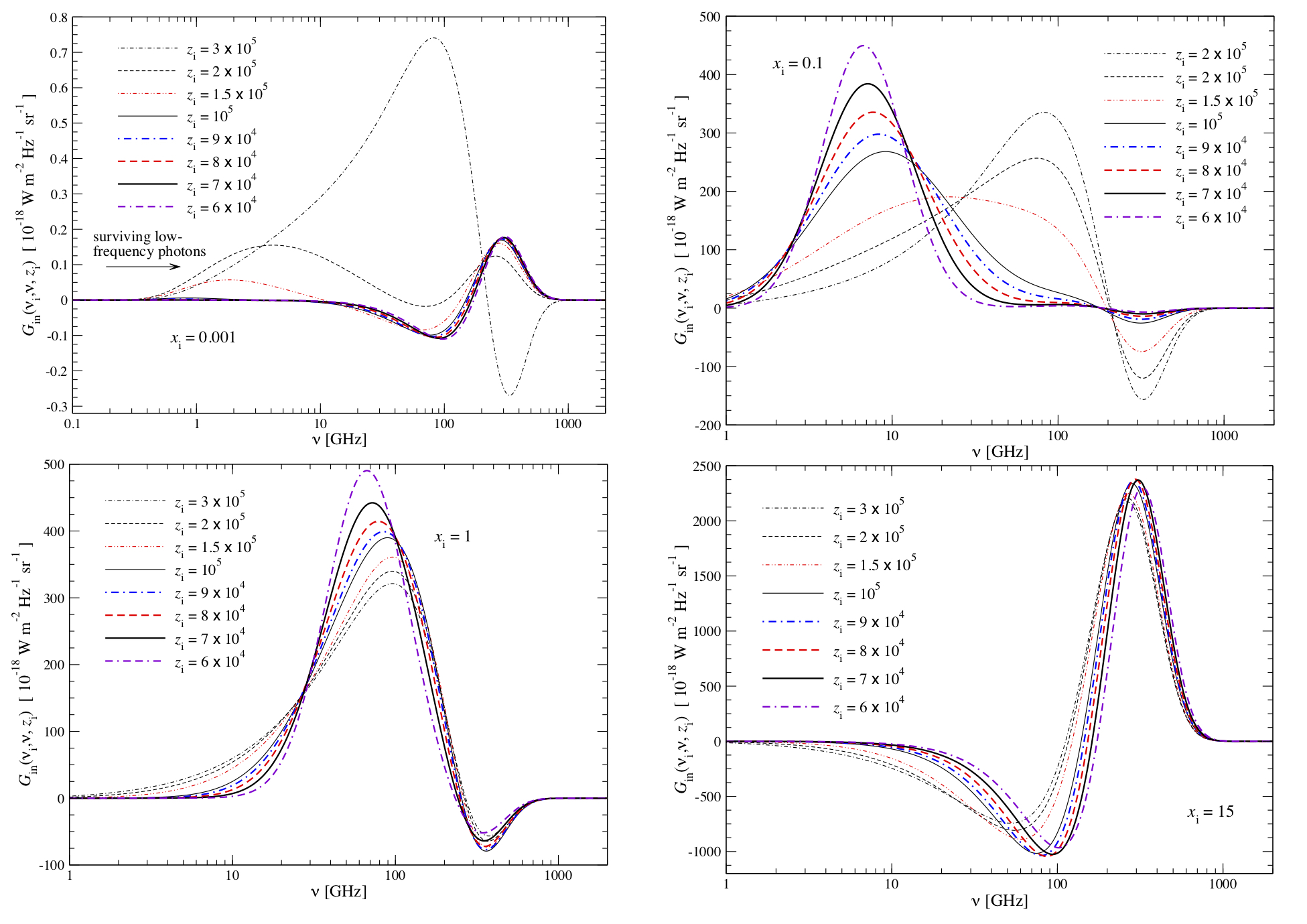}
\caption{Spectral distortions created by photon injection at different frequencies and initial redshifts. The Figure is taken from \cite{Chluba2015GreenII}.}
\label{fig:photon_inj_dist}
\end{figure}

\vspace{-1mm}
\subsection{Distortions from photon injection}
\label{sec:photon_injection}
To finish our discussion of spectral distortion physics, we briefly mention distortions created by photon injection. As shown by \cite{Chluba2015GreenII}, these can have a much more rich phenomenology than just the simple and broad $\mu$ and $y$-distortion created by energy release. This is illustrated in Fig.~\ref{fig:photon_inj_dist} for several cases, showing that the final distortion depends on both the injection time and frequency.

In terms of physics, distortions created by photon injection do not directly heat the electrons or baryons. Only once Comptonization becomes relevant do the electrons start heating or cooling. The net effect depends on the injection frequency of the photons. For frequencies $x_i \gtrsim 3.6-3.8$, photons on average loose energy heating the matter. This causes a broad $\mu$- and $y$-type contribution to the total distortion signal, which for extremely high frequency injection, $x_i \gtrsim 10$, can dominate. At lower frequencies, cooling of the medium occurs since photons are on average up-scattered. This can create negative $\mu$ and $y$-type contributions \citep{Chluba2015GreenII}.

Photon injection distortions are by no means exotic. For example, the cosmological recombination radiation \citep{Sunyaev2009}, one of the standard $\Lambda$CDM distortions, is created by photon injection. Injection of photons can also occur in decaying or annihilating particle scenarios or evaporation of primordial black holes. In light of recent measurements of EDGES \citep{Bowman2018} and the ARCADE low-frequency excess \citep{exp7, arcade2, Fixsen2011}, photon injection distortions of the CMB could become very interesting. This is because these observations potentially point towards a connection with photon injection (or absorption) from decaying or annihilating particles and their low energy by-products in form of non-thermal Bremsstrahlung or synchrotron emission \citep{Chluba2015GreenII, Feng2018, Hektor2018, Moroi2018}.

\begin{figure}
\centering 
\includegraphics[width=0.95\columnwidth]{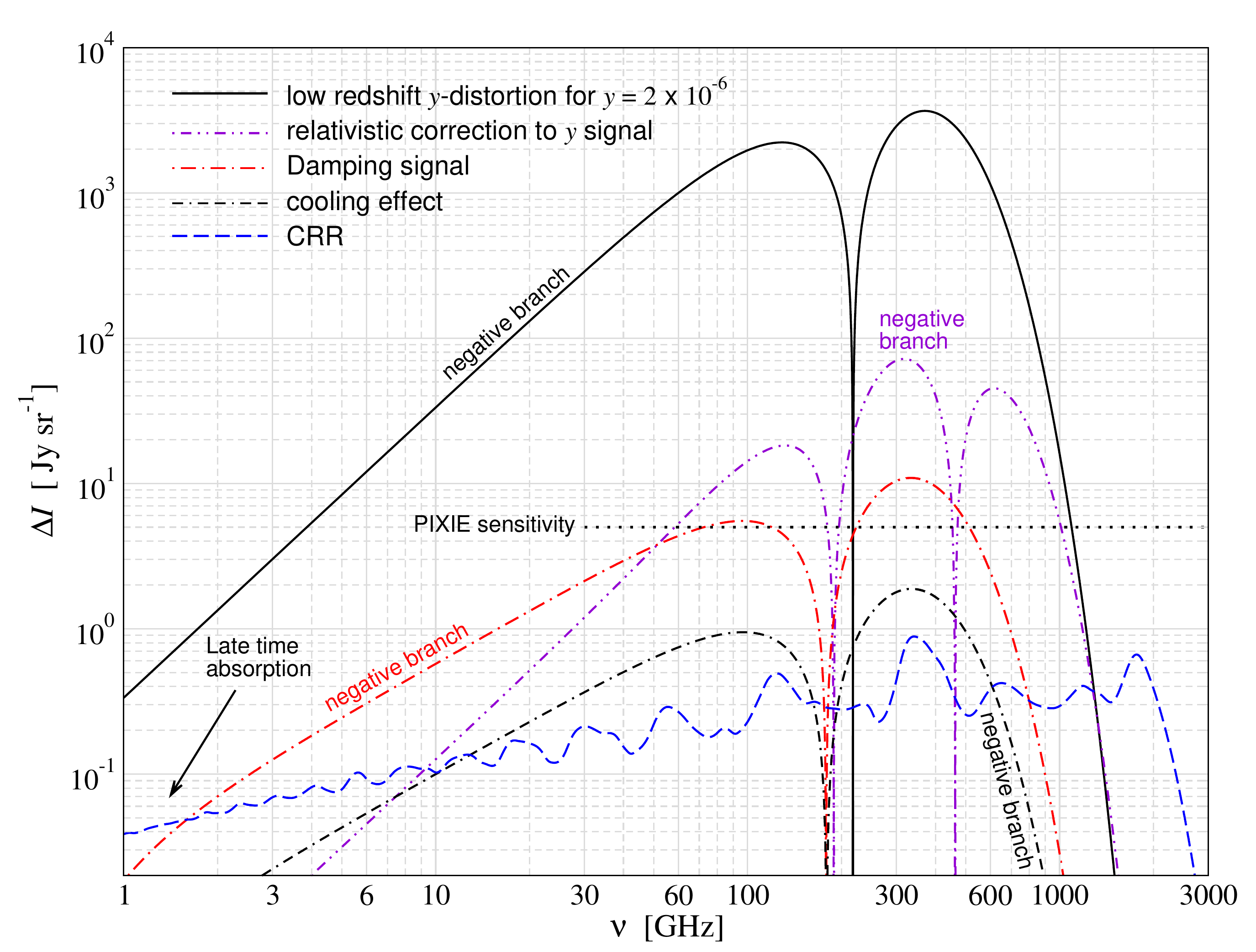}
\caption{Comparison of several CMB monopole distortion signals produced in the standard $\Lambda$CDM cosmology. 
The low-redshift distortion created by reionization and structure formation is close to a pure Compton-$y$ distortion with $y\simeq \pot{2}{-6}$. Contributions from the hot gas in low mass haloes give rise to a noticeable relativistic temperature correction, which is taken from \citet{Hill2015}. The damping and adiabatic cooling signals were explicitly computed using {\tt CosmoTherm} \citep{Chluba2011therm}. The cosmological recombination radiation (CRR) was obtained with {\tt CosmoSpec} \citep{Chluba2016Rec}. The estimated sensitivity ($\Delta I_\nu \approx 5\, {\rm Jy/sr}$) of PIXIE is shown for comparison (dotted line). The figure was taken from \cite{Chluba2016}.}
\label{fig:signals}
\end{figure}

\section{CMB spectral distortion signals from various scenarios}
\label{sec:mechanisms}
Several exhaustive reviews on various spectral distortion scenarios exist \citep{Chluba2011therm, Sunyaev2013, Chluba2013fore, Tashiro2014, deZotti2015, Chluba2016}, covering both standard and non-standard processes. Here we highlight some of the main distortion signals expected within $\Lambda$CDM and only briefly mention more exotic sources of distortions. A summary of the relevant $\Lambda$CDM distortions is shown in Fig.~\ref{fig:signals}. The distortion templates are available at {\tt www.Chluba.de/CosmoTherm}.

\subsection{Reionization and structure formation}
\label{sect:reion}
The first sources of radiation during reionization \citep{Hu1994pert, Barkana2001}, supernova feedback \citep{Oh2003} and structure formation shocks \citep{Sunyaev1972b, Cen1999, Refregier2000, Miniati2000} heat the intergalactic medium at low redshifts ($z\lesssim 10$), producing hot electrons (in a wide range of temperatures $\Te\simeq 10^4\,{\rm K}-10^6\,{\rm K}$) that partially up-scatter CMB photons, causing a Compton $y$-distortion \citep{Zeldovich1969}. 
Although this is the {\it largest} expected average distortion of the CMB caused within $\Lambda$CDM, its amplitude is quite uncertain and depends on the detailed structure and temperature of the medium, as well as scaling relations (e.g., between halo mass and temperature).
Several estimates for this contribution were obtained, yielding values for the total $y$-parameter at the level $y\simeq \pot{\rm few}{-6}$ \citep{Refregier2000, Zhang2004, Hill2015, Dolag2015, deZotti2015}. 

Following  \citet{Hill2015}, we use a fiducial value of $y=\pot{2}{-6}$ (see Fig.~\ref{fig:signals}). This is dominated by the low-mass end of the halo function ($M\simeq 10^{13}\,M_\odot$) and the signal should be detectable with a {PIXIE}-type experiment at more than $10^3\,\sigma$. The detection significance reduces to a few hundred $\sigma$ when including estimates for the CMB foregrounds \cite{Abitbol2017}, but still this provide a sensitive probe of reionization and structure formation physics. Future CMB imagers (e.g., CORE and PICO) furthermore have the potential to separate the spatially varying signature caused by the warm hot intergalactic medium (often referred to as WHIM) and proto-clusters \cite{Refregier2000, Zhang2004}, if the challenge of accurate channel intercalibration can be overcome.

\begin{figure} 
   \centering
   \includegraphics[width=0.9\columnwidth]{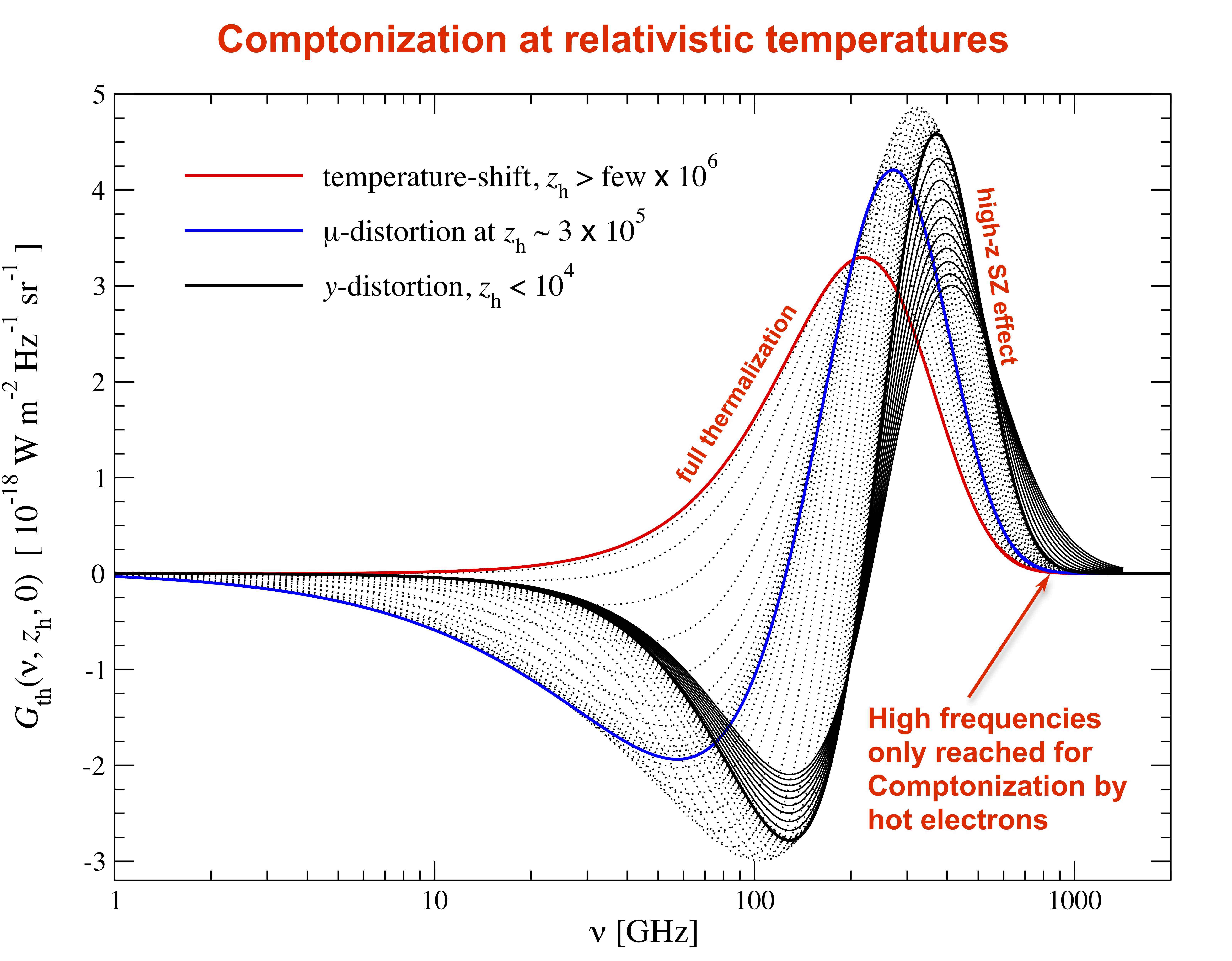}
   \caption{Illustration for the effect of relativistic temperature corrections on the distortion signal. In the primordial Universe, electrons hardly reach temperatures $\simeq 1\,{\rm keV}$ during the thermalization era ($z\lesssim 10^6$). Therefore even repeated Compton scattering can never push the distortion signals beyond the standard non-relativistic $y$-distortion signal. Inside clusters of galaxies, electrons can have temperatures $k\Te\gtrsim 1\,{\rm keV}$. In this case, the distortion signals can extend to much higher frequencies.}
   \label{fig:Greens_rel}
\end{figure}

Because the signal is so easily detectable, small corrections due to the high gas temperature ($k\Te \simeq 1\,\keV$) become noticeable \citep{Hill2015}. The relativistic correction can be computed using the temperature moment method of {\tt SZpack} \citep{Chluba2012SZpack, Chluba2012moments} and differs from the distortions produced in the early Universe (see Fig.~\ref{fig:Greens_rel}). This correction should be detectable with {PIXIE} at $\simeq 10-20\,\sigma$ \citep{Hill2015, Abitbol2017} and could teach us about the average temperature of the intergalactic medium, promising a way to solve the missing baryon problem \cite{Cen1999}. Both distortion signals are illustrated in Fig.~\ref{fig:signals}.

\begin{figure} 
   \centering
   \includegraphics[width=0.95\columnwidth]{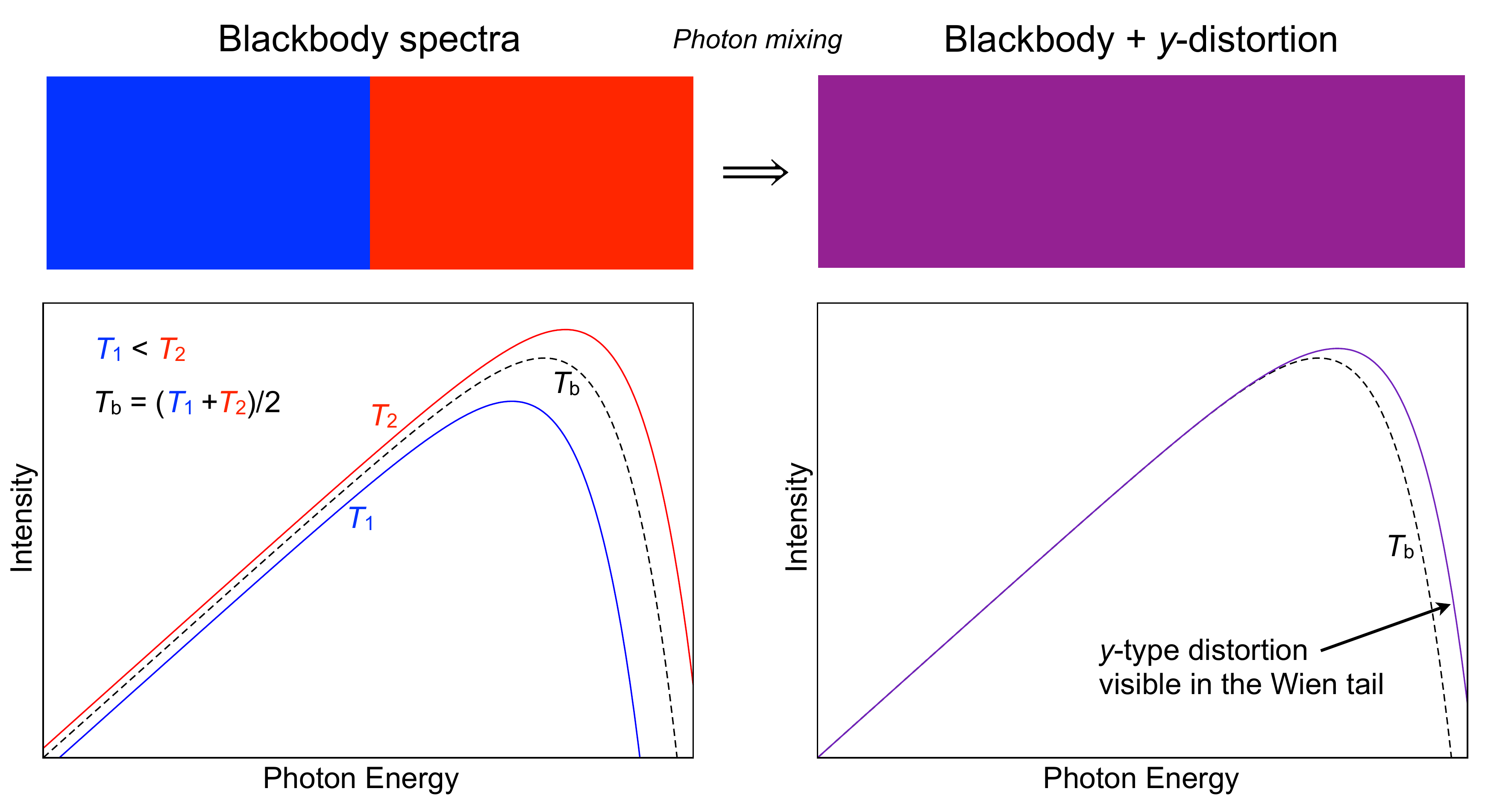}
   \caption{llustration for the superposition of blackbodies. We envision blackbody photons inside a box at two temperatures $T_1$ and $T_2$, and mean $T_{\rm b}=\frac{1}{2}(T_1+T_2)$ initially (left panel). Thomson scattering mixes the two photon distributions without changing the photon number or energy. The averaged distribution is not a pure blackbody but at second order in the temperature difference exhibits a $y$-type distortion in the Wien tail (right panel). This then starts the thermalization process and repeated Compton scattering slowly converts the distortion to a $\mu$-distortion.}
   \label{fig:Superposition}
\end{figure}

\begin{figure} 
   \centering
   \includegraphics[width=\columnwidth]{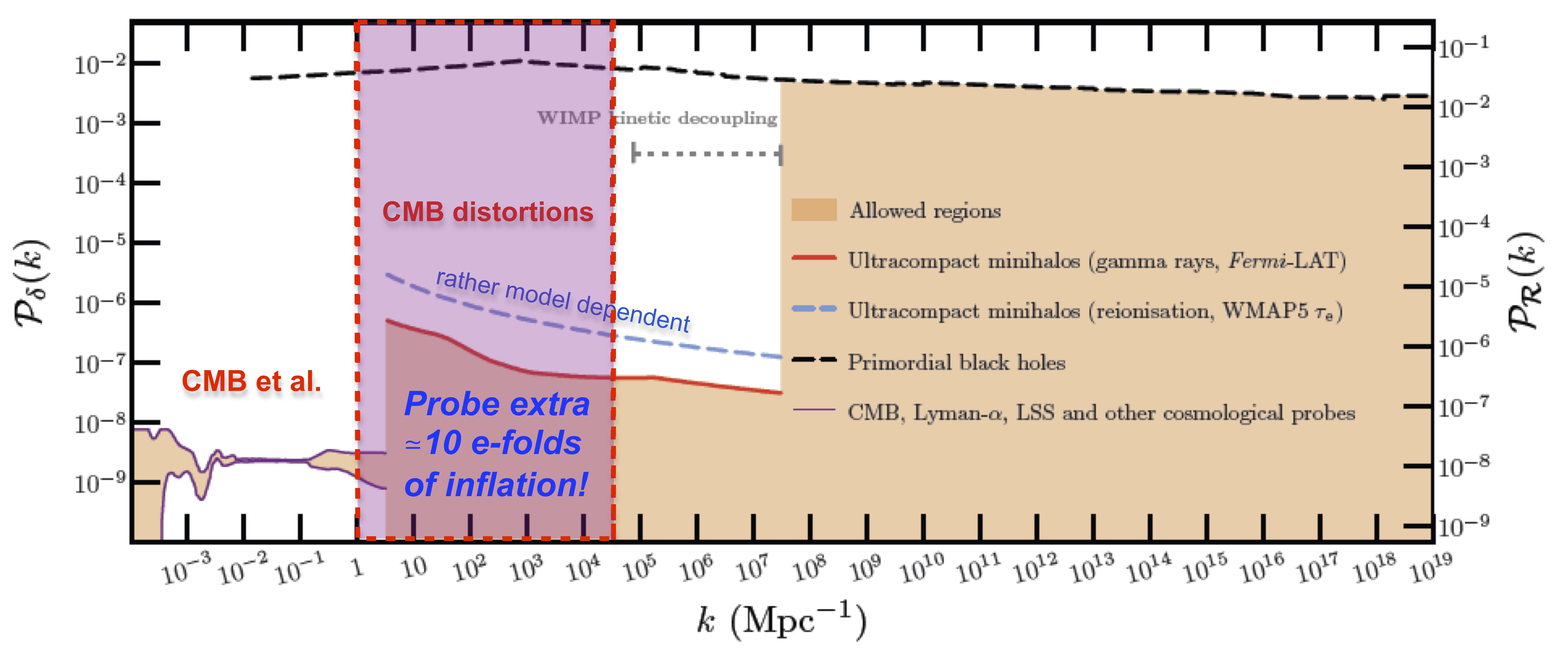}
   \caption{Current constraints on the small-scale power spectrum. At large scales ($k\lesssim 3\,{\rm Mpc}^{-1}$), CMB anisotropies and large scale structure measurements provide very stringent limits on the amplitude and shape of the primordial power spectrum. At smaller scales, the situation is much more uncertain and at $3\,{\rm Mpc}^{-1}\lesssim k\lesssim 10^4\,{\rm Mpc}^{-1}$ which can be targeted with CMB spectral distortion measurements wiggle room of at least two orders of magnitude is present. CMB distortion measurements could improve these limits to a level similar to the large-scale constraints. The figure is adapted from \cite{BSA11}.} 
   \label{fig:constraints}
\end{figure}

\vspace{-0mm}
\subsection{Damping of primordial small-scale perturbations}
\label{sec:damp}
The damping of small-scale fluctuations of the CMB temperature set up by inflation at wavelength $\lambda<1\,\Mpc$  causes another inevitable distortion of the CMB spectrum \citep{Sunyaev1970diss, Daly1991, Barrow1991, Hu1994, Hu1994isocurv}. The idea behind this mechanism is extremely simple and just based on the mixing of blackbodies with varying temperatures through Thomson scattering (see Fig.~\ref{fig:Superposition}). However, the process was only recently described rigorously \citep{Chluba2012, Pajer2012b}, allowing us to perform detailed computations of the associated distortion signal for different early-universe models \citep{Chluba2012, Chluba2012inflaton, Dent2012, Khatri2013forecast, Chluba2013fore, Clesse2014, Cabass2016}. The distortion is sensitive to the amplitude and shape of the power spectrum at very small scales (wavenumbers $1\,\Mpc^{-1}\lesssim k \lesssim \pot{2}{4}\,\Mpc^{-1}$ corresponding to multipoles $10^5\lesssim \ell \lesssim 10^8$) and thus provides a promising new way for constraining inflation while modes are still evolving in the linear regime. 

In the early days of CMB cosmology, this effect was already used to derive first upper limits on the spectral index of scalar perturbations, yielding $n_{\rm S}\lesssim 1.6$ from COBE/FIRAS \citep{Hu1994}. Perturbation modes with $1\,\Mpc^{-1}\lesssim k \lesssim 50\,\Mpc^{-1}$ create $y$-distortions, while modes with $50\,\Mpc^{-1}\lesssim k \lesssim \pot{2}{4}\,\Mpc^{-1}$ yield $\mu$-distortions. These scales are hard to access by any other means but spectral distortions provide a new sensitive probe in this regime (Fig.~\ref{fig:constraints}).

For a given initial power spectrum of perturbations, the effective heating rate in general has to be computed numerically \cite{Chluba2012}. However, at high redshifts the tight coupling approximation can be used to simplify the calculation. An excellent approximation for the effective heating rate can be obtained using\footnote{Here, we define the heating rate such that $\int_z^\infty \frac{\id (Q/\rho_\gamma)}{\id z}\id z>0$.} \citep{Chluba2012, Chluba2013iso}
\begin{align}
\label{eq:adiabatic_damping}
\frac{\id (Q/\rho_\gamma)}{\id z}&\approx 4 A^2 \partial_z \kD^{-2} \int^\infty_{k_{\rm min}} \frac{k^4\id k}{2\pi^2} P_\zeta(k)\,\expf{-2k^2/\kD^2},
\end{align}
where $P_\zeta(k)=2\pi^2\,A_{\rm s}\,k^{-3}\,(k/k_0)^{\nS-1 + \frac{1}{2}\,\nrun \ln(k/k_0)}$ defines the usual curvature power spectrum of scalar perturbations and $\kD$ is the photon damping scale \citep{Weinberg1971, Kaiser1983}, which early on scales as $\kD\approx \pot{4.048}{-6}\,(1 + z)^{3/2} \Mpc^{-1}$. For adiabatic modes, we obtain a heating efficiency $A^2\approx (1+4R_\nu/15)^{-2}\approx 0.813$, where $R_\nu\approx 0.409$ for $N_{\rm eff}=3.046$. The $k$-space integral is truncated at $k_{\rm min}\approx 0.12\,\Mpc^{-1}$, which reproduces the full heating rate across the recombination era quite well \citep{Chluba2013fore}. With this we can directly compute the associated distortion using {\tt CosmoTherm} \citep{Chluba2011therm}. The various isocurvature perturbations can be treated in a similar manner \citep{Chluba2013iso}; however, in the standard inflation model these should be small. Tensor perturbations also contribute to the dissipation process, but the associated heating rate is orders of magnitudes lower than for adiabatic modes even for very blue tensor power spectra and thus can be neglected \citep{Ota2014, Chluba2015}.

For $A_{\rm s}=\pot{2.207}{-9}$, $\nS=0.9645$ and $\nrun=0$ \citep{Planck2015params}, we present the result in Fig.~\ref{fig:signals}. The adiabatic cooling distortion (see Sect.~\ref{sec:ad_cool}) was simultaneously included. The signal is uncertain to within $\simeq 10\%$ in $\Lambda$CDM, simply because of the remaining uncertainties in the measurement of $A_{\rm s}$ and $\nS$. It is described by a sum of $\mu$- and $y$-distortion with $\mu\approx \pot{2.0}{-8}$ and $y\approx \pot{3.6}{-9}$ and a non-vanishing overall residual at the level of $\simeq 20\%-30\%$ \citep{Chluba2016}. 
In terms of raw sensitivity, this signal is close to the detection limit of a PIXIE-like experiment; however, foregrounds in particular at low frequencies make a detection more challenging \citep{Abitbol2017}. Still, a PIXIE-like experiment could place interesting upper limits on the amplitude of scalar fluctuations around $k\simeq 10^3\,\Mpc^{-1}$ \citep{Chluba2012inflaton, Chluba2013PCA}, potentially helping to shed light on the small-scale crisis \citep{Nakama2017} and rule out models of inflation with increased small-scale power \citep{Clesse2014, Clesse2015}.

The damping signal is also sensitive to primordial non-Gaussianity in the squeezed-limit, leading to a spatially varying spectral signal 
that correlates with CMB temperature anisotropies as large angular scales \cite{Pajer2012, Ganc2012}.
This effect therefore provides a unique way for studing the scale-dependence of $f_{\rm NL}$ \cite{Biagetti2013, Razi2015, Chluba2017, Ravenni2017, Remazeilles2018}.
CMB spectral distortions hence deliver a complementary and independent probe of early-Universe physics, which allows capitalizing on the synergies with large-scale $B$-mode polarization measurements.

\subsection{Adiabatic cooling for baryons}
\label{sec:ad_cool}
The adiabatic cooling of ordinary matter continuously extracts energy from the CMB photon bath by Compton scattering, leading to another small but guaranteed distortion that directly depends on the baryon density and helium abundance. The distortion is characterized by {\it negative} $\mu$- and $y$-parameters at the level of $\simeq \pot{\rm few}{-9}$ \citep{Chluba2005, Chluba2011therm, Khatri2011BE}. The effective energy extraction history is given by
\begin{align}
\label{eq:adiabatic_cooling}
\frac{\id (Q/\rho_\gamma)}{\id z}\!=\!-\frac{3}{2}\,\frac{N_{\rm tot} k\Tg}{\rho_\gamma (1+z)} 
\!\approx\! -\frac{\pot{5.7}{-10}}{(1+z)}\,
\!\left[\frac{(1-\Yp)}{0.75}\right]
\!\left[\frac{\Omega_{\rm b}h^2}{0.022}\right]
\left[\frac{(1+f_{\rm He}+X_{\rm e})}{2.25}\right]\left[\frac{T_0}{2.726\,\Kel}\right]^{-3}
\end{align}
where $N_{\rm tot}=N_{\rm H}(1+f_{\rm He}+X_{\rm e})$ is the number density of all thermally coupled baryons and electrons; $N_{\rm H}\approx \pot{1.881}{-6}\,(1+z)^3\,\cm^{-3}$ is the number density of hydrogen nuclei; $f_{\rm He}\approx \Yp/4(1-\Yp)\approx 0.0819$ and $X_{\rm e}=\Ne/N_{\rm H}$ is the free electron fraction, which can be computed accurately with {\tt CosmoRec} \citep{Chluba2010b}. For {\it Planck} 2015 parameters, the signal is shown in Fig.~\ref{fig:signals}. It is uncertain at the $\simeq 1\%$ level in $\Lambda$CDM and cancels part of the damping signal; however, it is roughly one order of magnitude weaker and cannot be separated at the currently expected level of sensitivity of next generation CMB spectrometers.

Additional interactions of dark matter with photons, electrons or protons could further increase the cooling distortion \citep{Yacine2015DM}. This allows placing interesting constraints on the nature of dark matter and its interactions with the standard sectors. The recent EDGES measurements \citep{Bowman2018} have spurred increased interest in this possiblity \citep{Barkana2018, Julian2018, Berlin2018}.

\begin{figure} 
   \centering
   \includegraphics[width=0.95\columnwidth]{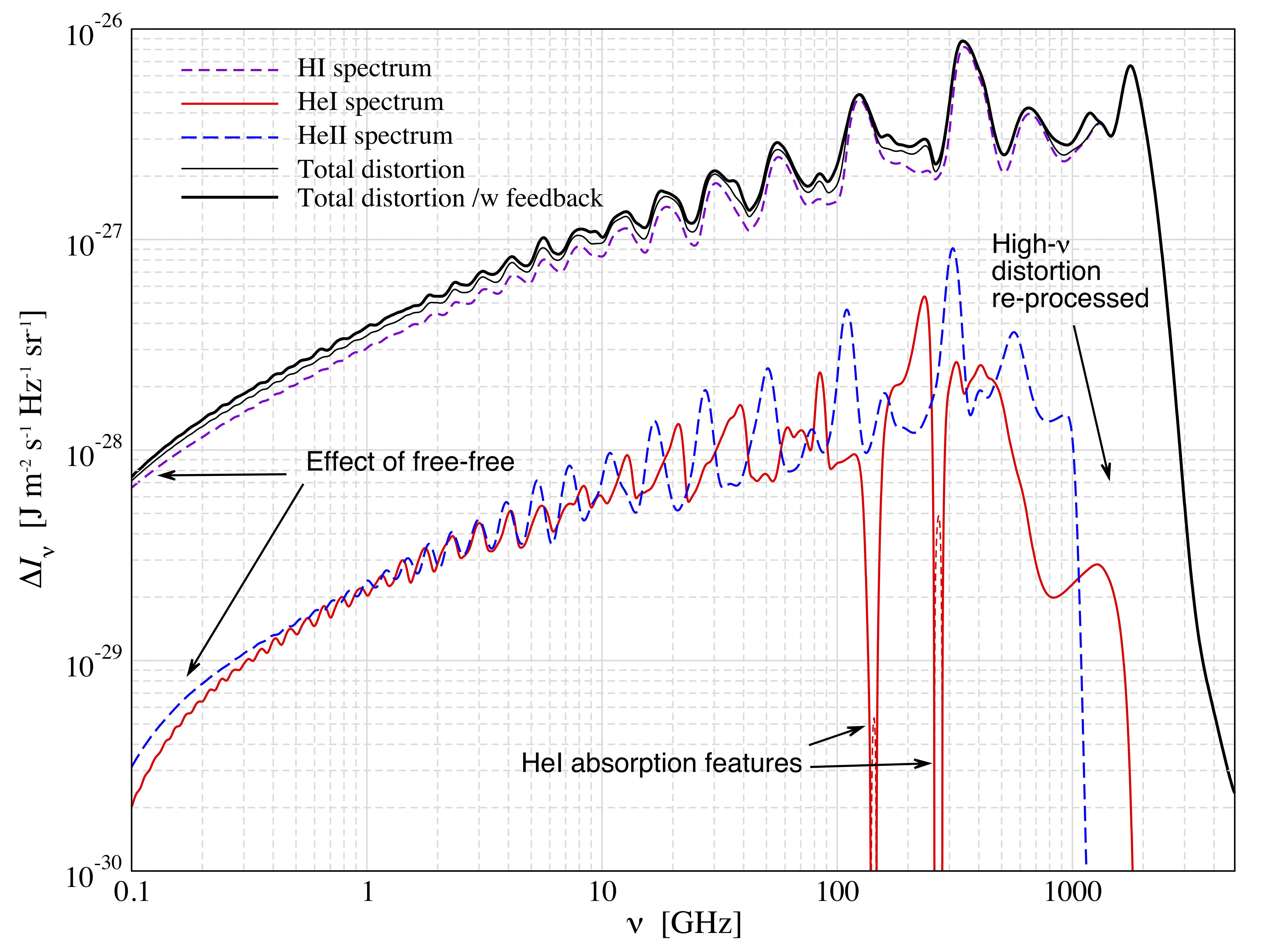}
   \caption{CRR from hydrogen and helium for 500-shell calculations. The different curves show individual contributions (without feedback) as well as the total distortion with and without feedback processes. At low frequencies, free-free absorption becomes noticeable. The effect is stronger for the contributions from helium due to the larger free-free optical depth before recombination ends at $z\simeq 10^3$. In total, some $6.1\gamma$ are emitted per hydrogen atom when all emission and feedback are included. Hydrogen alone contributes about $5.4\gamma /N_{\rm H}$ and helium $\simeq 0.7\gamma /N_{\rm H}$ ($\simeq 8.9 \gamma /N_{\rm He}$). The Figure was taken from \citep{Chluba2016Rec}.} 
   \label{fig:recombination}
\end{figure}
\subsection{The cosmological recombination radiation}
The cosmological recombination process is associated with the emission of photons in free-bound and bound-bound transitions of hydrogen and helium \citep{Zeldovich68, Peebles68, Dubrovich1975}. This causes a small distortion of the CMB and the redshifted recombination photons should still be visible as the cosmological recombination radiation (CRR), a tiny spectral distortion ($\simeq$ nK-$\mu$K level) present at mm to dm wavelength (for overview see \cite{Sunyaev2009}). The amplitude of the CRR depends directly on the number density of baryons in the Universe. The helium abundance furthermore affects the detailed shape of the recombination lines, while the number of neutrinos has a minor effect \citep{Chluba2016Rec}. Finally, the line positions and widths depend on when and how fast the Universe recombined. The CRR thus provides an independent way to constrain cosmological parameters and  map the recombination history \citep{Chluba2008T0}. 

Several computations of this CRR have been carried out in the past \citep{Rybicki93, DubroVlad95, Dubrovich1997, Kholu2005, Jose2006, Chluba2006b, Chluba2007, Jose2008, Chluba2009c, Chluba2010}. These calculations were very time-consuming, taking a few days of supercomputer time for one cosmology \cite{Chluba2007, Chluba2010}. This big computational challenge was recently overcome \citep{Yacine2013RecSpec, Chluba2016Rec}, today allowing us to compute the CRR in about 15 seconds on a standard laptop using {\tt CosmoSpec}\footnote{{\tt www.Chluba.de/CosmoSpec}} \citep{Chluba2016Rec}. 
The {\it fingerprint} from the recombination era shows several distinct spectral features that encode valuable information about the recombination process (Fig.~\ref{fig:signals}). Many subtle radiative transfer and atomic physics processes \citep{Chluba2007, Chluba2009c, Chluba2010b, Yacine2010c} can now be included by {\tt CosmoSpec}, yielding the most detailed and accurate predictions of the CRR in the standard $\Lambda$CDM model to date (see Fig.~\ref{fig:recombination}). In $\Lambda$CDM, the CRR is uncertain at the level of a few percent, with the error being dominated by atomic physics rather than cosmological parameter values \cite{Chluba2016Rec}.

The CRR is currently roughly $\simeq 6$ times below the estimated detection limit of PIXIE (cf. Fig.~\ref{fig:signals}) and a detection from space will require several times higher sensitivity \citep{Vince2015}. In the future, this could be achieved by experimental concepts similar to {PRISM} \citep{PRISM2013WPII} or {Millimetron} \citep{Smirnov2012}. At low frequencies ($1\,\GHz\lesssim \nu\lesssim 10\,\GHz$), the significant spectral variability of the CRR may also allow us to detect it from the ground with APSERa \citep{Mayuri2015}.
This could open a new way for directly studying the conditions of the Universe at $z\simeq 10^3$ ({\sc HI}-recombination), $z\simeq 2000$ ({\sc HeI}-recombination) and $z\simeq  6000$ ({\sc HeII}-recombination). Furthermore, if something unexpected happened during different stages of the recombination epoch, atomic species will react to this \cite{Chluba2008c} and produce additional distortion features that can exceed those of the normal recombination process. This will provide a unique way to distinguish pre- from post-recombination energy release \cite{Chluba2008c, Chluba2010a}.

To appreciate the importance of the cosmological recombination process at $z\simeq 10^3$ a little more, consider that today measurements of the CMB anisotropies are sensitive to uncertainties of the ionization history at a level of $\simeq 0.1\%-1\%$ \cite{Jose2010, Shaw2011}. For a precise interpretation of CMB data, uncertainties present in the original recombination calculations had to be reduced by including several previously omitted atomic physics and radiative transfer effects \cite{Fendt2009, Jose2010}. This led to the development of the new recombination modules {\sc CosmoRec} \cite{Chluba2010b} and {\sc HyRec} \cite{Yacine2010c} which are used in the analysis of Planck data \cite{Planck2013params}. Without these improve treatments of the recombination calculation the value for $\nS$ would be biased by $\Delta \nS\simeq -0.01$ to $\nS\simeq 0.95$ instead of $\simeq 0.96$ \cite{Shaw2011}. We would be discussing different inflation models \cite{Planck2013iso} without these corrections taken into account! Conversely, this emphasizes how important it is to experimentally confirm the recombination process and CMB spectral distortions provide a way to do so.

\begin{figure} 
   \centering
   \includegraphics[width=0.8\columnwidth]{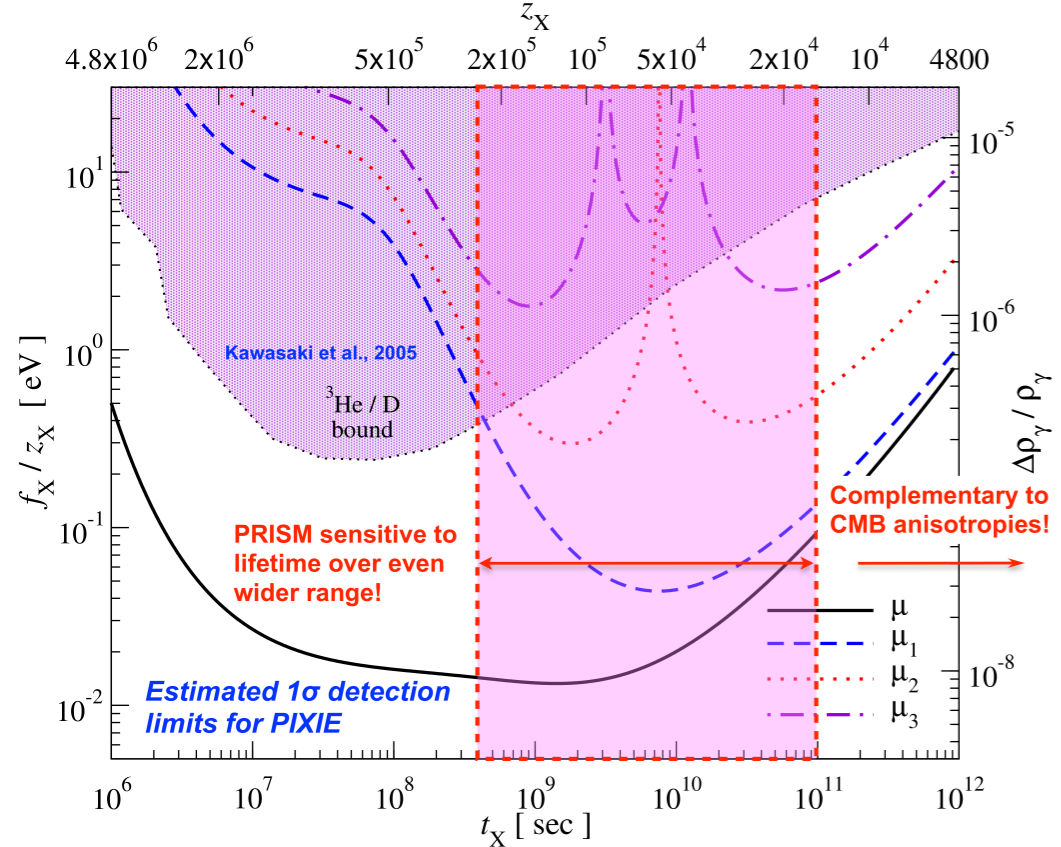}
   \caption{Decaying particle detection limits ($1\sigma$) for a PIXIE-like experiment. The eigenamplitudes $\mu_i$ characterize the non-$\mu$/non-$y$ distortion signal \cite{Chluba2013PCA}, which provides time-dependent information of the energy release history. CMB distortion limits could be $\simeq 50$ times tighter than those derived from light element abundances \cite{Kawasaki2005}. A separate determination of lifetime and particle abundance could be possible for lifetimes $t_{\rm X} \simeq 10^{8}\,{\rm sec}-10^{11}\,{\rm sec}$, being complementary to constraints derived using the CMB anisotropies \cite{Chen2004, Zhang2007}. The figure is adapted from \cite{Chluba2013PCA}.} 
   \label{fig:decay}
\end{figure}

\subsection{Dark matter annihilation}
Today, cold dark matter is a well-established constituent of our Universe \citep{WMAP_params, Planck2013params, Planck2015params}. However, the nature of dark matter is still unclear and many groups are trying to gather any new clue to help unravel this big puzzle \citep{Adriani2009, Galli2009, CDMS2010, Zavala2011, Huetsi2011, BSA11, Aslanyan2015}. Similarly, it is unclear how dark matter was produced, however, within $\Lambda$CDM, the WIMP scenario provides one viable solution \citep{Jungman1996, Bertone2005}. In this case, dark matter should annihilate at a low level throughout the history of the Universe and even today.

For specific dark matter models, the level of annihilation around the recombination epoch is tightly constrained with the CMB anisotropies \citep{Galli2009, Cirelli2009, Huetsi2009, Slatyer2009, Huetsi2011, Giesen2012, Diamanti2014, Planck2015params}. The annihilation of dark matter can cause changes in the ionization history around last scattering ($z\simeq 10^3$), which in turn can lead to changes of the CMB temperature and polarization anisotropies \citep{Chen2004, Padmanabhan2005, Zhang2006}. Albeit significant dependence on the interaction of the annihilation products with the primordial plasma \citep{Shull1985, Slatyer2009, Valdes2010, Galli2013, Slatyer2015}, the same process should lead to distortions of the CMB \citep{McDonald2001, Chluba2010a, Chluba2011therm}. 
Sadly, it turns out that for the standard WIMP scenario with s-wave annihilation cross section, the expected signal is even smaller than the adiabatic cooling distortion \citep{Chluba2013fore}. We will thus not go into more details here.

\subsection{Decaying particle scenarios}
The CMB spectrum also allows us to place stringent limits on decaying particles in the pre-recombination epoch \cite{Hu1993b, McDonald2001, Chluba2010a, Chluba2011therm}. This is especially interesting for decaying particles with lifetimes $t_{\rm X} \simeq 10^{8}\,{\rm sec}-10^{11}\,{\rm sec}$ \cite{Chluba2013fore, Chluba2013PCA}, as the exact shape of the distortion encodes when the decay occurred.
Decays associated with significant low-energy photon production could furthermore create a unique spectral signature that can be distinguished from simple energy release \cite{Chluba2015GreenII}.
This would provide an unprecedented probe of early-universe particle physics (e.g., dark matter in excited states \cite{Pospelov2007, Finkbeiner2007}), with many natural particle candidates found in supersymmetric models \cite{Feng2003, Feng2010}. This could also shed light on gravitinos physics \citep{Dimastrogiovanni2015}, axions \citep{Tashiro2013} and primordial black holes \citep{Carr2010, Poulin2017}.

The expected $1\sigma$ detection limits for a PIXIE-like experiment are illustrated in Fig.~\ref{fig:decay}. The bounds obtained from measurements of light-elements \citep{Kawasaki2005} could be superseded by more than one order of magnitude. Similar improvements from light-elements are not expected any time soon, and most recent updated only improved the limits by $\simeq 10\%$ \citep{Kawasaki2018}. Spectral distortions thus provide a powerful new probe of particle physics.

\subsection{Anisotropic CMB distortions}
To close the discussion of different distortion signals, we briefly mention anisotropic ($\leftrightarrow$ {\it spectral-spatial}) CMB distortions.
Even in the standard $\Lambda$CDM cosmology, anisotropies in the spectrum of the CMB are expected. The largest source of anisotropies is due to the Sunyaev-Zeldovich effect caused by the hot plasma inside clusters of galaxies \citep{Zeldovich1969, Sunyaev1980, Birkinshaw1999, Carlstrom2002}, as mentioned above. The $y$-distortion power spectrum has already been measured directly by {Planck} \citep{Planck2013ymap, Planck2015ymap} and encodes valuable information about the atmospheres of clusters \citep{Refregier2000, Komatsu2002, Diego2004, Battaglia2010, Shaw2010, Munshi2013, Dolag2015}. Similarly, the warm hot intergalactic medium contributes and should become visible \citep{Zhang2004, Dolag2015}.

In the primordial Universe, anisotropies in the $\mu$ and $y$ distortions are expected to be tiny (relative perturbations $\lesssim 10^{-4}$, e.g., see \cite{Pitrou2010}) unless strong spatial variations in the primordial heating mechanism are expected \citep{Chluba2012}. As mentioned above, this could in principle be caused by non-Gaussianity of perturbations in the squeezed limit \citep{Pajer2012, Ganc2012, Biagetti2013, Razi2015, Chluba2017, Ravenni2017}; however, a present detectable levels of non-Gaussianity are beyond $\Lambda$CDM cosmology (see \cite{Remazeilles2018} for discussion of some of the foreground issues) and will not be considered further.

Another guaranteed anisotropic signal is due to Rayleigh scattering of CMB photons in the Lyman-series resonances of hydrogen around the recombination era \citep{Yu2001, Lewis2013}. The signal is strongly frequency dependent, can be  modeled precisely and may be detectable with future CMB imagers (e.g., {COrE+}) or possibly {PIXIE} at large angular scales \citep{Lewis2013}. In a very similar manner, the resonant scattering of CMB photons by metals appearing in the dark ages \citep{Loeb2001, Zaldarriaga2002, Kaustuv2004, Carlos2007Pol} or scattering in the excited levels of hydrogen during recombination \citep{Jose2005, Carlos2007Pol} can lead to anisotropic distortions. To measure these signals, precise channel intercalibration and foreground rejection is required. 

Due to our motion relative to the CMB rest frame, the spectrum of the CMB dipole should also be distorted simply because the CMB monopole has a distortion \citep{Danese1981, Balashev2015}. The signal associated with the large late-time $y$-distortion could be detectable with {PIXIE} at the level of a few $\sigma$ \citep{Balashev2015}. Since for these measurements no absolute calibration is required, this effect will allow us to check for systematics. In addition, the dipole spectrum can be used to constrain monopole foregrounds \citep{Balashev2015, deZotti2015, Burigana2018}.

Finally, due to the superposition of blackbodies of different temperatures (caused by the spherical harmonic expansion of the intensity map), the CMB quadrupole spectrum is also distorted, exhibiting a $y$-distortion related to our motion \citep{Kamionkowski2003, Chluba2004}. The associated effective $y$-parameter is $y_{\rm Q}=\beta^2/6 \approx \pot{(2.525 \pm 0.012)}{-7}$ and should be noticeable with {PIXIE} and future CMB imagers \citep{Burigana2018}.

\section{Conclusions}
CMB spectral distortion measurements provide a unique way for studying physical processes leading to energy release or photon injection in the pre- and post-recombination eras. In the future, this could open a new unexplored window to early-universe and particle physics, delivering independent and complementary pieces of information about the Universe we live in. 
We highlighted several processes that should lead to distortions at a level within reach of present-day technology. Different distortion signals can be computed precisely and efficiently for various scenarios using both analytical and numerical schemes.  Time-dependent information, beyond the standard $\mu$- and $y$-type parametrization, may allow us to independently constrain lifetime and abundance of decaying relic particles, learn about the shape and amplitude of the small-scale power spectrum of primordial perturbations and shed light on dark matter. The cosmological recombination radiation will allow us to check our understanding of the recombination processes at redshifts of $z\simeq 10^3$. It furthermore should allow us to distinguish pre- from post-recombination $y$-distortions. 
All this emphasizes the immense potential of CMB spectroscopy, both in terms of {\it discovery} and {\it characterization} science, and we should make use of this invaluable source of information with the next CMB space mission and worldwide ground-based efforts.

\acknowledgments
JC cordially thanks the organizers of the Varenna International School of Physics "Enrico Fermi", Nicola Vittorio and Joseph Silk, for their great hospitality and continued encouragement to finish these notes. JC would also like to thank Boris Bolliet and Andrea Ravenni for their comments on the manuscript.
JC is supported by the Royal Society as a Royal Society University Research Fellow and an European Research Council Consolidator Grant ({\it CMBSPEC}, No.~725456) at the University of Manchester, UK.

\bibliographystyle{varenna}
\bibliography{Lit}

\begin{thebibliography}{100}
\expandafter\ifx\csname url\endcsname\relax\def\url#1{\texttt{#1}}\fi
\expandafter\ifx\csname urlprefix\endcsname\relax\def\urlprefix{URL }\fi

\bibitem{Smoot1992}
\NAME{{Smoot} G.~F., {Bennett} C.~L., {Kogut} A., {Wright} E.~L., {Aymon} J.,
  {Boggess} N.~W., {Cheng} E.~S., {de Amici} G., {Gulkis} S., {Hauser} M.~G.,
  {Hinshaw} G., {Jackson} P.~D., {Janssen} M., {Kaita} E., {Kelsall} T. \atque
  {Keegstra} P.}, \IN{\apjl}{396}{1992}{L1}.

\bibitem{WMAP_params}
\NAME{{Bennett} C.~L., {Halpern} M., {Hinshaw} G., {Jarosik} N., {Kogut} A.,
  {Limon} M., {Meyer} S.~S., {Page} L., {Spergel} D.~N., {Tucker} G.~S.,
  {Wollack} E., {Wright} E.~L., {Barnes} C., {Greason} M.~R., {Hill} R.~S.,
  {Komatsu} E., {Nolta} M.~R., {Odegard} N., {Peiris} H.~V. \atque {Verde} L.},
  \IN{\apjs}{148}{2003}{1}.

\bibitem{Planck2013params}
\NAME{{Planck Collaboration}, {Ade} P.~A.~R., {Aghanim} N., {Armitage-Caplan}
  C., {Arnaud} M., {Ashdown} M., {Atrio-Barandela} F., {Aumont} J.,
  {Baccigalupi} C., {Banday} A.~J. \atque et~al.}, \IN{\aap}{571}{2014}{A16}.

\bibitem{PRISM2013WPII}
\NAME{{Andr{\'e}} P., {Baccigalupi} C., {Banday} A., {Barbosa} D., {Barreiro}
  B., {Bartlett} J., {Bartolo} N., {Battistelli} E., {Battye} R., {Bendo} G.,
  {Beno{\#523}t} A., {Bernard} J.-P., {Bersanelli} M., {B{\'e}thermin} M.,
  {Bielewicz} P., {Bonaldi} A., {Bouchet} F., {Boulanger} F., {Brand} J.,
  {Bucher} M., {Burigana} C., {Cai} Z.-Y., {Camus} P., {Casas} F., {Casasola}
  V., {Castex} G., {Challinor} A., {Chluba} J., {Chon} G., {Colafrancesco} S.,
  {Comis} B., {Cuttaia} F., {D'Alessandro} G., {Da Silva} A., {Davis} R., {de
  Avillez} M., {de Bernardis} P., {de Petris} M., {de Rosa} A., {de Zotti} G.,
  {Delabrouille} J., {D{\'e}sert} F.-X., {Dickinson} C., {Diego} J.~M.,
  {Dunkley} J., {En{\ss}lin} T., {Errard} J., {Falgarone} E., {Ferreira} P.,
  {Ferri{\`e}re} K., {Finelli} F., {Fletcher} A., {Fosalba} P., {Fuller} G.,
  {Galli} S., {Ganga} K., {Garc{\'{\i}}a-Bellido} J., {Ghribi} A., {Giard} M.,
  {Giraud-H{\'e}raud} Y., {Gonzalez-Nuevo} J., {Grainge} K., {Gruppuso} A.,
  {Hall} A., {Hamilton} J.-C., {Haverkorn} M., {Hernandez-Monteagudo} C.,
  {Herranz} D., {Jackson} M., {Jaffe} A., {Khatri} R., {Kunz} M., {Lamagna} L.,
  {Lattanzi} M., {Leahy} P., {Lesgourgues} J., {Liguori} M., {Liuzzo} E.,
  {Lopez-Caniego} M., {Macias-Perez} J., {Maffei} B., {Maino} D., {Mangilli}
  A., {Martinez-Gonzalez} E., {Martins} C., {Masi} S., {Massardi} M.,
  {Matarrese} S., {Melchiorri} A., {Melin} J.-B., {Mennella} A., {Mignano} A.,
  {Miville-Desch{\^e}nes} M.-A., {Monfardini} A., {Murphy} A., {Naselsky} P.,
  {Nati} F., {Natoli} P., {Negrello} M., {Noviello} F., {O'Sullivan} C., {Paci}
  F., {Pagano} L., {Paladino} R., {Palanque-Delabrouille} N., {Paoletti} D.,
  {Peiris} H., {Perrotta} F., {Piacentini} F., {Piat} M., {Piccirillo} L.,
  {Pisano} G., {Polenta} G., {Pollo} A., {Ponthieu} N., {Remazeilles} M.,
  {Ricciardi} S., {Roman} M., {Rosset} C., {Rubino-Martin} J.-A., {Salatino}
  M., {Schillaci} A., {Shellard} P., {Silk} J., {Starobinsky} A., {Stompor} R.,
  {Sunyaev} R., {Tartari} A., {Terenzi} L., {Toffolatti} L., {Tomasi} M.,
  {Trappe} N., {Tristram} M., {Trombetti} T., {Tucci} M., {Van de Weijgaert}
  R., {Van Tent} B., {Verde} L., {Vielva} P., {Wandelt} B., {Watson} R. \atque
  {Withington} S.}, \IN{\jcap}{2}{2014}{6}.

\bibitem{Abazajian2015Inf}
\NAME{{Abazajian} K.~N., {Arnold} K., {Austermann} J., {Benson} B.~A.,
  {Bischoff} C., {Bock} J., {Bond} J.~R., {Borrill} J., {Buder} I., {Burke}
  D.~L., {Calabrese} E., {Carlstrom} J.~E., {Carvalho} C.~S., {Chang} C.~L.,
  {Chiang} H.~C., {Church} S., {Cooray} A., {Crawford} T.~M., {Crill} B.~P.,
  {Dawson} K.~S., {Das} S., {Devlin} M.~J., {Dobbs} M., {Dodelson} S.,
  {Dor{\'e}} O., {Dunkley} J., {Feng} J.~L., {Fraisse} A., {Gallicchio} J.,
  {Giddings} S.~B., {Green} D., {Halverson} N.~W., {Hanany} S., {Hanson} D.,
  {Hildebrandt} S.~R., {Hincks} A., {Hlozek} R., {Holder} G., {Holzapfel}
  W.~L., {Honscheid} K., {Horowitz} G., {Hu} W., {Hubmayr} J., {Irwin} K.,
  {Jackson} M., {Jones} W.~C., {Kallosh} R., {Kamionkowski} M., {Keating} B.,
  {Keisler} R., {Kinney} W., {Knox} L., {Komatsu} E., {Kovac} J., {Kuo} C.-L.,
  {Kusaka} A., {Lawrence} C., {Lee} A.~T., {Leitch} E., {Linde} A., {Linder}
  E., {Lubin} P., {Maldacena} J., {Martinec} E., {McMahon} J., {Miller} A.,
  {Mukhanov} V., {Newburgh} L., {Niemack} M.~D., {Nguyen} H., {Nguyen} H.~T.,
  {Page} L., {Pryke} C., {Reichardt} C.~L., {Ruhl} J.~E., {Sehgal} N., {Seljak}
  U., {Senatore} L., {Sievers} J., {Silverstein} E., {Slosar} A., {Smith}
  K.~M., {Spergel} D., {Staggs} S.~T., {Stark} A., {Stompor} R., {Vieregg}
  A.~G., {Wang} G., {Watson} S., {Wollack} E.~J., {Wu} W.~L.~K., {Yoon} K.~W.,
  {Zahn} O. \atque {Zaldarriaga} M.}, \IN{Astroparticle Physics}{63}{2015}{55}.

\bibitem{Abazajian2015}
\NAME{{Abazajian} K.~N., {Arnold} K., {Austermann} J., {Benson} B.~A.,
  {Bischoff} C., {Bock} J., {Bond} J.~R., {Borrill} J., {Calabrese} E.,
  {Carlstrom} J.~E., {Carvalho} C.~S., {Chang} C.~L., {Chiang} H.~C., {Church}
  S., {Cooray} A., {Crawford} T.~M., {Dawson} K.~S., {Das} S., {Devlin} M.~J.,
  {Dobbs} M., {Dodelson} S., {Dor{\'e}} O., {Dunkley} J., {Errard} J.,
  {Fraisse} A., {Gallicchio} J., {Halverson} N.~W., {Hanany} S., {Hildebrandt}
  S.~R., {Hincks} A., {Hlozek} R., {Holder} G., {Holzapfel} W.~L., {Honscheid}
  K., {Hu} W., {Hubmayr} J., {Irwin} K., {Jones} W.~C., {Kamionkowski} M.,
  {Keating} B., {Keisler} R., {Knox} L., {Komatsu} E., {Kovac} J., {Kuo} C.-L.,
  {Lawrence} C., {Lee} A.~T., {Leitch} E., {Linder} E., {Lubin} P., {McMahon}
  J., {Miller} A., {Newburgh} L., {Niemack} M.~D., {Nguyen} H., {Nguyen} H.~T.,
  {Page} L., {Pryke} C., {Reichardt} C.~L., {Ruhl} J.~E., {Sehgal} N., {Seljak}
  U., {Sievers} J., {Silverstein} E., {Slosar} A., {Smith} K.~M., {Spergel} D.,
  {Staggs} S.~T., {Stark} A., {Stompor} R., {Vieregg} A.~G., {Wang} G.,
  {Watson} S., {Wollack} E.~J., {Wu} W.~L.~K., {Yoon} K.~W. \atque {Zahn} O.},
  \IN{Astroparticle Physics}{63}{2015}{66}.

\bibitem{Abazajian2016S4SB}
\NAME{{Abazajian} K.~N., {Adshead} P., {Ahmed} Z., {Allen} S.~W., {Alonso} D.,
  {Arnold} K.~S., {Baccigalupi} C., {Bartlett} J.~G., {Battaglia} N., {Benson}
  B.~A., {Bischoff} C.~A., {Borrill} J., {Buza} V., {Calabrese} E., {Caldwell}
  R., {Carlstrom} J.~E., {Chang} C.~L., {Crawford} T.~M., {Cyr-Racine} F.-Y.,
  {De Bernardis} F., {de Haan} T., {di Serego Alighieri} S., {Dunkley} J.,
  {Dvorkin} C., {Errard} J., {Fabbian} G., {Feeney} S., {Ferraro} S.,
  {Filippini} J.~P., {Flauger} R., {Fuller} G.~M., {Gluscevic} V., {Green} D.,
  {Grin} D., {Grohs} E., {Henning} J.~W., {Hill} J.~C., {Hlozek} R., {Holder}
  G., {Holzapfel} W., {Hu} W., {Huffenberger} K.~M., {Keskitalo} R., {Knox} L.,
  {Kosowsky} A., {Kovac} J., {Kovetz} E.~D., {Kuo} C.-L., {Kusaka} A., {Le
  Jeune} M., {Lee} A.~T., {Lilley} M., {Loverde} M., {Madhavacheril} M.~S.,
  {Mantz} A., {Marsh} D.~J.~E., {McMahon} J., {Meerburg} P.~D., {Meyers} J.,
  {Miller} A.~D., {Munoz} J.~B., {Nguyen} H.~N., {Niemack} M.~D., {Peloso} M.,
  {Peloton} J., {Pogosian} L., {Pryke} C., {Raveri} M., {Reichardt} C.~L.,
  {Rocha} G., {Rotti} A., {Schaan} E., {Schmittfull} M.~M., {Scott} D.,
  {Sehgal} N., {Shandera} S., {Sherwin} B.~D., {Smith} T.~L., {Sorbo} L.,
  {Starkman} G.~D., {Story} K.~T., {van Engelen} A., {Vieira} J.~D., {Watson}
  S., {Whitehorn} N. \atque {Kimmy Wu} W.~L.}, \IN{ArXiv:1610.02743}{}{2016}{}.

\bibitem{Calabrese2013}
\NAME{{Calabrese} E., {Hlozek} R.~A., {Battaglia} N., {Battistelli} E.~S.,
  {Bond} J.~R., {Chluba} J., {Crichton} D., {Das} S., {Devlin} M.~J., {Dunkley}
  J., {D{\"u}nner} R., {Farhang} M., {Gralla} M.~B., {Hajian} A., {Halpern} M.,
  {Hasselfield} M., {Hincks} A.~D., {Irwin} K.~D., {Kosowsky} A., {Louis} T.,
  {Marriage} T.~A., {Moodley} K., {Newburgh} L., {Niemack} M.~D., {Nolta}
  M.~R., {Page} L.~A., {Sehgal} N., {Sherwin} B.~D., {Sievers} J.~L.,
  {Sif{\'o}n} C., {Spergel} D.~N., {Staggs} S.~T., {Switzer} E.~R. \atque
  {Wollack} E.~J.}, \IN{\prd}{87}{2013}{103012}.

\bibitem{Planck2015params}
\NAME{{Planck Collaboration}, {Ade} P.~A.~R., {Aghanim} N., {Arnaud} M.,
  {Ashdown} M., {Aumont} J., {Baccigalupi} C., {Banday} A.~J., {Barreiro}
  R.~B., {Bartlett} J.~G. \atque et~al.}, \IN{\aap}{594}{2016}{A13}.

\bibitem{Fixsen1996}
\NAME{{Fixsen} D.~J., {Cheng} E.~S., {Gales} J.~M., {Mather} J.~C., {Shafer}
  R.~A. \atque {Wright} E.~L.}, \IN{\apj}{473}{1996}{576}.

\bibitem{Fixsen2009}
\NAME{{Fixsen} D.~J.}, \IN{\apj}{707}{2009}{916}.

\bibitem{Chluba2011therm}
\NAME{{Chluba} J. \atque {Sunyaev} R.~A.}, \IN{\mnras}{419}{2012}{1294}.

\bibitem{Sunyaev2013}
\NAME{{Sunyaev} R.~A. \atque {Khatri} R.}, \IN{International Journal of Modern
  Physics D}{22}{2013}{30014}.

\bibitem{Chluba2013fore}
\NAME{{Chluba} J.}, \IN{\mnras}{436}{2013}{2232}.

\bibitem{Tashiro2014}
\NAME{{Tashiro} H.}, \IN{Prog. of Theo. and Exp. Physics}{2014}{2014}{060000}.

\bibitem{deZotti2015}
\NAME{{De Zotti} G., {Negrello} M., {Castex} G., {Lapi} A. \atque {Bonato} M.},
  \IN{\jcap}{3}{2016}{047}.

\bibitem{Chluba2016}
\NAME{{Chluba} J.}, \IN{\mnras}{460}{2016}{227}.

\bibitem{Fixsen2002}
\NAME{{Fixsen} D.~J. \atque {Mather} J.~C.}, \IN{\apj}{581}{2002}{817}.

\bibitem{Kogut2011PIXIE}
\NAME{{Kogut} A., {Fixsen} D.~J., {Chuss} D.~T., {Dotson} J., {Dwek} E.,
  {Halpern} M., {Hinshaw} G.~F., {Meyer} S.~M., {Moseley} S.~H., {Seiffert}
  M.~D., {Spergel} D.~N. \atque {Wollack} E.~J.}, \IN{\jcap}{7}{2011}{25}.

\bibitem{Kogut2016SPIE}
\NAME{{Kogut} A., {Chluba} J., {Fixsen} D.~J., {Meyer} S. \atque {Spergel} D.},
  \TITLE{{The Primordial Inflation Explorer (PIXIE)}}, in proc. of \TITLE{SPIE
  Conference Series}, Vol. 9904 of \emph{Proc.SPIE} 2016, p. 99040W.

\bibitem{Chluba2014Science}
\NAME{{Silk} J. \atque {Chluba} J.}, \IN{Science}{344}{2014}{586}.

\bibitem{Mayuri2015}
\NAME{{Sathyanarayana Rao} M., {Subrahmanyan} R., {Udaya Shankar} N. \atque
  {Chluba} J.}, \IN{\apj}{810}{2015}{3}.

\bibitem{Roadmap2014}
\NAME{{Kouveliotou} C., {Agol} E., {Batalha} N., {Bean} J., {Bentz} M.,
  {Cornish} N., {Dressler} A., {Figueroa-Feliciano} E., {Gaudi} S., {Guyon} O.,
  {Hartmann} D., {Kalirai} J., {Niemack} M., {Ozel} F., {Reynolds} C.,
  {Roberge} A., {Straughn} K.~S.~A., {Weinberg} D. \atque {Zmuidzinas} J.},
  \IN{ArXiv:1401.3741}{}{2014}{}.

\bibitem{Zeldovich1969}
\NAME{{Zeldovich} Y.~B. \atque {Sunyaev} R.~A.}, \IN{\apss}{4}{1969}{301}.

\bibitem{Sunyaev1970mu}
\NAME{{Sunyaev} R.~A. \atque {Zeldovich} Y.~B.}, \IN{\apss}{7}{1970}{20}.

\bibitem{Illarionov1974}
\NAME{{Illarionov} A.~F. \atque {Sunyaev} R.~A.}, \IN{Astronomicheskii
  Zhurnal}{51}{1974}{1162}.

\bibitem{Illarionov1975}
\NAME{{Illarionov} A.~F. \atque {Sunyaev} R.~A.}, \IN{Soviet
  Astronomy}{18}{1975}{413}.

\bibitem{Danese1977}
\NAME{{Danese} L. \atque {de Zotti} G.}, \IN{Nuovo Cimento Rivista
  Serie}{7}{1977}{277}.

\bibitem{Danese1982}
\NAME{{Danese} L. \atque {de Zotti} G.}, \IN{\aap}{107}{1982}{39}.

\bibitem{Burigana1991}
\NAME{{Burigana} C., {Danese} L. \atque {de Zotti} G.},
  \IN{\aap}{246}{1991}{49}.

\bibitem{Hu1993}
\NAME{{Hu} W. \atque {Silk} J.}, \IN{\prd}{48}{1993}{485}.

\bibitem{Jose2006}
\NAME{{Rubi{\~n}o-Mart{\'{\i}}n} J.~A., {Chluba} J. \atque {Sunyaev} R.~A.},
  \IN{\mnras}{371}{2006}{1939}.

\bibitem{Sunyaev2009}
\NAME{{Sunyaev} R.~A. \atque {Chluba} J.}, \IN{Astronomische
  Nachrichten}{330}{2009}{657}.

\bibitem{Chluba2016Rec}
\NAME{{Chluba} J. \atque {Ali-Ha{\"i}moud} Y.}, \IN{\mnras}{456}{2016}{3494}.

\bibitem{Carlstrom2002}
\NAME{{Carlstrom} J.~E., {Holder} G.~P. \atque {Reese} E.~D.},
  \IN{\araa}{40}{2002}{643}.

\bibitem{Pajer2012}
\NAME{{Pajer} E. \atque {Zaldarriaga} M.}, \IN{Physical Review
  Letters}{109}{2012}{021302}.

\bibitem{Ganc2012}
\NAME{{Ganc} J. \atque {Komatsu} E.}, \IN{\prd}{86}{2012}{023518}.

\bibitem{Chluba2013Green}
\NAME{{Chluba} J.}, \IN{\mnras}{434}{2013}{352}.

\bibitem{Chluba2015GreenII}
\NAME{{Chluba} J.}, \IN{\mnras}{454}{2015}{4182}.

\bibitem{Chluba2010b}
\NAME{{Chluba} J. \atque {Thomas} R.~M.}, \IN{\mnras}{412}{2011}{748}.

\bibitem{Pozdniakov1979}
\NAME{{Pozdniakov} L.~A., {Sobol} I.~M. \atque {Sunyaev} R.~A.},
  \IN{\aap}{75}{1979}{214}.

\bibitem{Sunyaev1980Comptonization}
\NAME{{Sunyaev} R.~A. \atque {Titarchuk} L.~G.}, \IN{\aap}{86}{1980}{121}.

\bibitem{Chluba2012}
\NAME{{Chluba} J., {Khatri} R. \atque {Sunyaev} R.~A.},
  \IN{\mnras}{425}{2012}{1129}.

\bibitem{Chluba2014mSZII}
\NAME{{Chluba} J. \atque {Dai} L.}, \IN{\mnras}{438}{2014}{1324}.

\bibitem{Kompa56}
\NAME{Kompaneets A.}, \IN{Sov.Phys. JETP}{31}{1956}{876}.

\bibitem{Rybicki1979}
\NAME{{Rybicki} G.~B. \atque {Lightman} A.~P.}, \TITLE{{Radiative processes in
  astrophysics}} (New York, Wiley-Interscience, 1979.~393 p.) 1979.

\bibitem{Sazonov2000}
\NAME{{Sazonov} S.~Y. \atque {Sunyaev} R.~A.}, \IN{\apj}{543}{2000}{28}.

\bibitem{Itoh98}
\NAME{{Itoh} N., {Kohyama} Y. \atque {Nozawa} S.}, \IN{\apj}{502}{1998}{7}.

\bibitem{Sazonov1998}
\NAME{{Sazonov} S.~Y. \atque {Sunyaev} R.~A.}, \IN{\apj}{508}{1998}{1}.

\bibitem{Challinor1998}
\NAME{{Challinor} A. \atque {Lasenby} A.}, \IN{\apj}{499}{1998}{1}.

\bibitem{ChlubaSZpack}
\NAME{{Chluba} J., {Nagai} D., {Sazonov} S. \atque {Nelson} K.},
  \IN{\mnras}{426}{2012}{510}.

\bibitem{Chluba2012moments}
\NAME{{Chluba} J., {Switzer} E., {Nelson} K. \atque {Nagai} D.},
  \IN{\mnras}{430}{2013}{3054}.

\bibitem{Hu1995PhD}
\NAME{{Hu} W.}, \IN{arXiv:astro-ph/9508126}{}{1995}{}.

\bibitem{Zeldovich68}
\NAME{{Zeldovich} Y.~B., {Kurt} V.~G. \atque {Syunyaev} R.~A.},
  \IN{ZETF}{55}{1968}{278}.

\bibitem{Pritchard2008}
\NAME{{Pritchard} J.~R. \atque {Loeb} A.}, \IN{\prd}{78}{2008}{103511}.

\bibitem{Chluba2007a}
\NAME{{Chluba} J., {Sazonov} S.~Y. \atque {Sunyaev} R.~A.},
  \IN{\aap}{468}{2007}{785}.

\bibitem{Itoh2000}
\NAME{{Itoh} N., {Sakamoto} T., {Kusano} S., {Nozawa} S. \atque {Kohyama} Y.},
  \IN{\apjs}{128}{2000}{125}.

\bibitem{Chluba2014}
\NAME{{Chluba} J.}, \IN{\mnras}{440}{2014}{2544}.

\bibitem{Chluba2005}
\NAME{{Chluba} J.}, \TITLE{{Spectral Distortions of the Cosmic Microwave
  Background}}, Ph.D. thesis, LMU M{\"u}nchen (Mar. 2005).

\bibitem{Yacine2015DM}
\NAME{{Ali-Ha{\"i}moud} Y., {Chluba} J. \atque {Kamionkowski} M.}, \IN{Physical
  Review Letters}{115}{2015}{071304}.

\bibitem{Planck2014SZ}
\NAME{{Planck Collaboration}, {Ade} P.~A.~R., {Aghanim} N., {Armitage-Caplan}
  C., {Arnaud} M., {Ashdown} M., {Atrio-Barandela} F., {Aumont} J., {Aussel}
  H., {Baccigalupi} C. \atque et~al.}, \IN{\aap}{571}{2014}{A29}.

\bibitem{Sunyaev1980}
\NAME{{Sunyaev} R.~A. \atque {Zeldovich} I.~B.}, \IN{\mnras}{190}{1980}{413}.

\bibitem{Rephaeli1995ARAA}
\NAME{{Rephaeli} Y.}, \IN{\araa}{33}{1995}{541}.

\bibitem{Birkinshaw1999}
\NAME{{Birkinshaw} M.}, \IN{Phys. Reports}{310}{1999}{97}.

\bibitem{Bolliet2018}
\NAME{{Bolliet} B., {Comis} B., {Komatsu} E. \atque {Mac{\'{\i}}as-P{\'e}rez}
  J.~F.}, \IN{\mnras}{477}{2018}{4957}.

\bibitem{Chluba2012SZpack}
\NAME{{Chluba} J., {Nagai} D., {Sazonov} S. \atque {Nelson} K.},
  \IN{\mnras}{426}{2012}{510}.

\bibitem{Levich1969}
\NAME{{Zeldovich} Y.~B. \atque {Levich} E.~V.}, \IN{SJETP}{28}{1969}{1287}.

\bibitem{Zeldovich1972shock}
\NAME{{Zeldovich} Y.~B. \atque {Sunyaev} R.~A.}, \IN{Zhurnal Eksperimentalnoi i
  Teoreticheskoi Fiziki}{62}{1972}{153}.

\bibitem{Khatri2011BE}
\NAME{{Khatri} R., {Sunyaev} R.~A. \atque {Chluba} J.},
  \IN{\aap}{540}{2012}{A124}.

\bibitem{Khatri2012mix}
\NAME{{Khatri} R. \atque {Sunyaev} R.~A.}, \IN{\jcap}{9}{2012}{16}.

\bibitem{Khatri2012b}
\NAME{{Khatri} R. \atque {Sunyaev} R.~A.}, \IN{\jcap}{6}{2012}{38}.

\bibitem{Chluba2013PCA}
\NAME{{Chluba} J. \atque {Jeong} D.}, \IN{\mnras}{438}{2014}{2065}.

\bibitem{Liubarskii83}
\NAME{{Lyubarsky} Y.~E. \atque {Sunyaev} R.~A.}, \IN{\aap}{123}{1983}{171}.

\bibitem{Chluba2008c}
\NAME{{Chluba} J. \atque {Sunyaev} R.~A.}, \IN{\aap}{501}{2009}{29}.

\bibitem{Chluba2015IJMPD}
\NAME{{Chluba} J., {Hamann} J. \atque {Patil} S.~P.}, \IN{International Journal
  of Modern Physics D}{24}{2015}{1530023}.

\bibitem{Bowman2018}
\NAME{{Bowman} J.~D., {Rogers} A.~E.~E., {Monsalve} R.~A., {Mozdzen} T.~J.
  \atque {Mahesh} N.}, \IN{\nat}{555}{2018}{67}.

\bibitem{exp7}
\NAME{{Fixsen} D.~J., {Kogut} A., {Levin} S., {Limon} M., {Lubin} P., {Mirel}
  P., {Seiffert} M. \atque {Wollack} E.}, \IN{\apj}{612}{2004}{86}.

\bibitem{arcade2}
\NAME{{Seiffert} M., {Fixsen} D.~J., {Kogut} A., {Levin} S.~M., {Limon} M.,
  {Lubin} P.~M., {Mirel} P., {Singal} J., {Villela} T., {Wollack} E. \atque
  {Wuensche} C.~A.}, \IN{\apj}{734}{2011}{6}.

\bibitem{Fixsen2011}
\NAME{{Fixsen} D.~J., {Kogut} A., {Levin} S., {Limon} M., {Lubin} P., {Mirel}
  P., {Seiffert} M., {Singal} J., {Wollack} E., {Villela} T. \atque {Wuensche}
  C.~A.}, \IN{\apj}{734}{2011}{5}.

\bibitem{Feng2018}
\NAME{{Feng} C. \atque {Holder} G.}, \IN{\apjl}{858}{2018}{L17}.

\bibitem{Hektor2018}
\NAME{{Hektor} A., {H{\"u}tsi} G., {Marzola} L., {Raidal} M., {Vaskonen} V.
  \atque {Veerm{\"a}e} H.}, \IN{ArXiv:1803.09697}{}{2018}{}.

\bibitem{Moroi2018}
\NAME{{Moroi} T., {Nakayama} K. \atque {Tang} Y.},
  \IN{ArXiv:1804.10378}{}{2018}{}.

\bibitem{Hill2015}
\NAME{{Hill} J.~C., {Battaglia} N., {Chluba} J., {Ferraro} S., {Schaan} E.
  \atque {Spergel} D.~N.}, \IN{Physical Review Letters}{115}{2015}{261301}.

\bibitem{Hu1994pert}
\NAME{{Hu} W., {Scott} D. \atque {Silk} J.}, \IN{\prd}{49}{1994}{648}.

\bibitem{Barkana2001}
\NAME{{Barkana} R. \atque {Loeb} A.}, \IN{Physics Reports}{349}{2001}{125}.

\bibitem{Oh2003}
\NAME{{Oh} S.~P., {Cooray} A. \atque {Kamionkowski} M.},
  \IN{\mnras}{342}{2003}{L20}.

\bibitem{Sunyaev1972b}
\NAME{{Sunyaev} R.~A. \atque {Zeldovich} Y.~B.}, \IN{\aap}{20}{1972}{189}.

\bibitem{Cen1999}
\NAME{{Cen} R. \atque {Ostriker} J.~P.}, \IN{\apj}{514}{1999}{1}.

\bibitem{Refregier2000}
\NAME{{Refregier} A. \atque {et al.}}, \IN{\prd}{61}{2000}{123001}.

\bibitem{Miniati2000}
\NAME{{Miniati} F., {Ryu} D., {Kang} H., {Jones} T.~W., {Cen} R. \atque
  {Ostriker} J.~P.}, \IN{\apj}{542}{2000}{608}.

\bibitem{Zhang2004}
\NAME{{Zhang} P., {Pen} U.-L. \atque {Trac} H.}, \IN{\mnras}{355}{2004}{451}.

\bibitem{Dolag2015}
\NAME{{Dolag} K., {Komatsu} E. \atque {Sunyaev} R.},
  \IN{\mnras}{463}{2016}{1797}.

\bibitem{Abitbol2017}
\NAME{{Abitbol} M.~H., {Chluba} J., {Hill} J.~C. \atque {Johnson} B.~R.},
  \IN{\mnras}{471}{2017}{1126}.

\bibitem{BSA11}
\NAME{{Bringmann} T., {Scott} P. \atque {Akrami} Y.},
  \IN{\prd}{85}{2012}{125027}.

\bibitem{Sunyaev1970diss}
\NAME{{Sunyaev} R.~A. \atque {Zeldovich} Y.~B.}, \IN{\apss}{9}{1970}{368}.

\bibitem{Daly1991}
\NAME{{Daly} R.~A.}, \IN{\apj}{371}{1991}{14}.

\bibitem{Barrow1991}
\NAME{{Barrow} J.~D. \atque {Coles} P.}, \IN{\mnras}{248}{1991}{52}.

\bibitem{Hu1994}
\NAME{{Hu} W., {Scott} D. \atque {Silk} J.}, \IN{\apjl}{430}{1994}{L5}.

\bibitem{Hu1994isocurv}
\NAME{{Hu} W. \atque {Sugiyama} N.}, \IN{\apj}{436}{1994}{456}.

\bibitem{Pajer2012b}
\NAME{{Pajer} E. \atque {Zaldarriaga} M.}, \IN{\jcap}{2}{2013}{36}.

\bibitem{Chluba2012inflaton}
\NAME{{Chluba} J., {Erickcek} A.~L. \atque {Ben-Dayan} I.},
  \IN{\apj}{758}{2012}{76}.

\bibitem{Dent2012}
\NAME{{Dent} J.~B., {Easson} D.~A. \atque {Tashiro} H.},
  \IN{\prd}{86}{2012}{023514}.

\bibitem{Khatri2013forecast}
\NAME{{Khatri} R. \atque {Sunyaev} R.~A.}, \IN{\jcap}{6}{2013}{26}.

\bibitem{Clesse2014}
\NAME{{Clesse} S., {Garbrecht} B. \atque {Zhu} Y.}, \IN{\jcap}{10}{2014}{046}.

\bibitem{Cabass2016}
\NAME{{Cabass} G., {Melchiorri} A. \atque {Pajer} E.},
  \IN{\prd}{93}{2016}{083515}.

\bibitem{Chluba2013iso}
\NAME{{Chluba} J. \atque {Grin} D.}, \IN{\mnras}{434}{2013}{1619}.

\bibitem{Weinberg1971}
\NAME{{Weinberg} S.}, \IN{\apj}{168}{1971}{175}.

\bibitem{Kaiser1983}
\NAME{{Kaiser} N.}, \IN{\mnras}{202}{1983}{1169}.

\bibitem{Ota2014}
\NAME{{Ota} A., {Takahashi} T., {Tashiro} H. \atque {Yamaguchi} M.},
  \IN{\jcap}{10}{2014}{029}.

\bibitem{Chluba2015}
\NAME{{Chluba} J., {Dai} L., {Grin} D., {Amin} M.~A. \atque {Kamionkowski} M.},
  \IN{\mnras}{446}{2015}{2871}.

\bibitem{Nakama2017}
\NAME{{Nakama} T., {Chluba} J. \atque {Kamionkowski} M.},
  \IN{\prd}{95}{2017}{121302}.

\bibitem{Clesse2015}
\NAME{{Clesse} S. \atque {Garc{\'{\i}}a-Bellido} J.},
  \IN{\prd}{92}{2015}{023524}.

\bibitem{Biagetti2013}
\NAME{{Biagetti} M., {Perrier} H., {Riotto} A. \atque {Desjacques} V.},
  \IN{\prd}{87}{2013}{063521}.

\bibitem{Razi2015}
\NAME{{Emami} R., {Dimastrogiovanni} E., {Chluba} J. \atque {Kamionkowski} M.},
  \IN{\prd}{91}{2015}{123531}.

\bibitem{Chluba2017}
\NAME{{Chluba} J., {Dimastrogiovanni} E., {Amin} M.~A. \atque {Kamionkowski}
  M.}, \IN{\mnras}{466}{2017}{2390}.

\bibitem{Ravenni2017}
\NAME{{Ravenni} A., {Liguori} M., {Bartolo} N. \atque {Shiraishi} M.},
  \IN{\jcap}{9}{2017}{042}.

\bibitem{Remazeilles2018}
\NAME{{Remazeilles} M. \atque {Chluba} J.}, \IN{\mnras}{}{2018}{}.

\bibitem{Barkana2018}
\NAME{{Barkana} R.}, \IN{\nat}{555}{2018}{71}.

\bibitem{Julian2018}
\NAME{{Mu{\~n}oz} J.~B. \atque {Loeb} A.}, \IN{ArXiv:1802.10094}{}{2018}{}.

\bibitem{Berlin2018}
\NAME{{Berlin} A., {Hooper} D., {Krnjaic} G. \atque {McDermott} S.~D.},
  \IN{ArXiv:1803.02804}{}{2018}{}.

\bibitem{Peebles68}
\NAME{{Peebles} P.~J.~E.}, \IN{\apj}{153}{1968}{1}.

\bibitem{Dubrovich1975}
\NAME{{Dubrovich} V.~K.}, \IN{Soviet Astronomy Letters}{1}{1975}{196}.

\bibitem{Chluba2008T0}
\NAME{{Chluba} J. \atque {Sunyaev} R.~A.}, \IN{\aap}{478}{2008}{L27}.

\bibitem{Rybicki93}
\NAME{{Rybicki} G.~B. \atque {dell'Antonio} I.~P.}, \TITLE{{Spectral
  Distortions in the CMB from Recombination.}}, in proc. of \TITLE{ASP Conf.
  Ser. 51: Observational Cosmology}, edited by \NAME{{Chincarini} G.~L.,
  {Iovino} A., {Maccacaro} T. \atque {Maccagni} D.} 1993, pp. 548--+.

\bibitem{DubroVlad95}
\NAME{{Dubrovich} V.~K. \atque {Stolyarov} V.~A.}, \IN{\aap}{302}{1995}{635}.

\bibitem{Dubrovich1997}
\NAME{{Dubrovich} V.~K. \atque {Stolyarov} V.~A.}, \IN{Astronomy
  Letters}{23}{1997}{565}.

\bibitem{Kholu2005}
\NAME{Kholupenko E.~E., Ivanchik A.~V. \atque Varshalovich D.~A.},
  \IN{Gravitation and Cosmology}{11}{2005}{161}.

\bibitem{Chluba2006b}
\NAME{{Chluba} J. \atque {Sunyaev} R.~A.}, \IN{\aap}{458}{2006}{L29}.

\bibitem{Chluba2007}
\NAME{{Chluba} J., {Rubi{\~n}o-Mart{\'{\i}}n} J.~A. \atque {Sunyaev} R.~A.},
  \IN{\mnras}{374}{2007}{1310}.

\bibitem{Jose2008}
\NAME{{Rubi{\~n}o-Mart{\'{\i}}n} J.~A., {Chluba} J. \atque {Sunyaev} R.~A.},
  \IN{\aap}{485}{2008}{377}.

\bibitem{Chluba2009c}
\NAME{{Chluba} J. \atque {Sunyaev} R.~A.}, \IN{\mnras}{402}{2010}{1221}.

\bibitem{Chluba2010}
\NAME{{Chluba} J., {Vasil} G.~M. \atque {Dursi} L.~J.},
  \IN{\mnras}{407}{2010}{599}.

\bibitem{Yacine2013RecSpec}
\NAME{{Ali-Ha{\"i}moud} Y.}, \IN{\prd}{87}{2013}{023526}.

\bibitem{Yacine2010c}
\NAME{{Ali-Ha{\"i}moud} Y. \atque {Hirata} C.~M.}, \IN{\prd}{83}{2011}{043513}.

\bibitem{Vince2015}
\NAME{{Desjacques} V., {Chluba} J., {Silk} J., {de Bernardis} F. \atque
  {Dor{\'e}} O.}, \IN{\mnras}{451}{2015}{4460}.

\bibitem{Smirnov2012}
\NAME{{Smirnov} A.~V., {Baryshev} A.~M., {Pilipenko} S.~V., {Myshonkova} N.~V.,
  {Bulanov} V.~B., {Arkhipov} M.~Y., {Vinogradov} I.~S., {Likhachev} S.~F.
  \atque {Kardashev} N.~S.}, \TITLE{{Space mission Millimetron for terahertz
  astronomy}}, presented at \TITLE{SPIE Conference Series}, Vol. 8442 2012.

\bibitem{Chluba2010a}
\NAME{{Chluba} J.}, \IN{\mnras}{402}{2010}{1195}.

\bibitem{Jose2010}
\NAME{{Rubi{\~n}o-Mart{\'{\i}}n} J.~A., {Chluba} J., {Fendt} W.~A. \atque
  {Wandelt} B.~D.}, \IN{\mnras}{403}{2010}{439}.

\bibitem{Shaw2011}
\NAME{{Shaw} J.~R. \atque {Chluba} J.}, \IN{\mnras}{415}{2011}{1343}.

\bibitem{Fendt2009}
\NAME{{Fendt} W.~A., {Chluba} J., {Rubi{\~n}o-Mart{\'{\i}}n} J.~A. \atque
  {Wandelt} B.~D.}, \IN{\apjs}{181}{2009}{627}.

\bibitem{Planck2013iso}
\NAME{{Planck Collaboration}, {Ade} P.~A.~R., {Aghanim} N., {Armitage-Caplan}
  C., {Arnaud} M., {Ashdown} M., {Atrio-Barandela} F., {Aumont} J.,
  {Baccigalupi} C., {Banday} A.~J. \atque et~al.}, \IN{\aap}{571}{2014}{A22}.

\bibitem{Kawasaki2005}
\NAME{{Kawasaki} M., {Kohri} K. \atque {Moroi} T.},
  \IN{\prd}{71}{2005}{083502}.

\bibitem{Chen2004}
\NAME{{Chen} X. \atque {Kamionkowski} M.}, \IN{\prd}{70}{2004}{043502}.

\bibitem{Zhang2007}
\NAME{{Zhang} L., {Chen} X., {Kamionkowski} M., {Si} Z. \atque {Zheng} Z.},
  \IN{\prd}{76}{2007}{061301}.

\bibitem{Adriani2009}
\NAME{{Adriani} O., {Barbarino} G.~C., {Bazilevskaya} G.~A., {Bellotti} R.,
  {Boezio} M., {Bogomolov} E.~A., {Bonechi} L., {Bongi} M., {Bonvicini} V.,
  {Bottai} S., {Bruno} A., {Cafagna} F., {Campana} D., {Carlson} P., {Casolino}
  M., {Castellini} G., {de Pascale} M.~P., {de Rosa} G., {de Simone} N., {di
  Felice} V., {Galper} A.~M., {Grishantseva} L., {Hofverberg} P., {Koldashov}
  S.~V., {Krutkov} S.~Y., {Kvashnin} A.~N., {Leonov} A., {Malvezzi} V.,
  {Marcelli} L., {Menn} W., {Mikhailov} V.~V., {Mocchiutti} E., {Orsi} S.,
  {Osteria} G., {Papini} P., {Pearce} M., {Picozza} P., {Ricci} M.,
  {Ricciarini} S.~B., {Simon} M., {Sparvoli} R., {Spillantini} P., {Stozhkov}
  Y.~I., {Vacchi} A., {Vannuccini} E., {Vasilyev} G., {Voronov} S.~A., {Yurkin}
  Y.~T., {Zampa} G., {Zampa} N. \atque {Zverev} V.~G.},
  \IN{\nat}{458}{2009}{607}.

\bibitem{Galli2009}
\NAME{{Galli} S., {Iocco} F., {Bertone} G. \atque {Melchiorri} A.},
  \IN{\prd}{80}{2009}{023505}.

\bibitem{CDMS2010}
\NAME{{CDMS II Collaboration}, {Ahmed} Z., {Akerib} D.~S., {Arrenberg} S.,
  {Bailey} C.~N., {Balakishiyeva} D., {Baudis} L., {Bauer} D.~A., {Brink}
  P.~L., {Bruch} T., {Bunker} R., {Cabrera} B., {Caldwell} D.~O., {Cooley} J.,
  {Cushman} P., {Daal} M., {DeJongh} F., {Dragowsky} M.~R., {Duong} L.,
  {Fallows} S., {Figueroa-Feliciano} E., {Filippini} J., {Fritts} M., {Golwala}
  S.~R., {Grant} D.~R., {Hall} J., {Hennings-Yeomans} R., {Hertel} S.~A.,
  {Holmgren} D., {Hsu} L., {Huber} M.~E., {Kamaev} O., {Kiveni} M., {Kos} M.,
  {Leman} S.~W., {Mahapatra} R., {Mandic} V., {McCarthy} K.~A., {Mirabolfathi}
  N., {Moore} D., {Nelson} H., {Ogburn} R.~W., {Phipps} A., {Pyle} M., {Qiu}
  X., {Ramberg} E., {Rau} W., {Reisetter} A., {Saab} T., {Sadoulet} B.,
  {Sander} J., {Schnee} R.~W., {Seitz} D.~N., {Serfass} B., {Sundqvist} K.~M.,
  {Tarka} M., {Wikus} P., {Yellin} S., {Yoo} J., {Young} B.~A. \atque {Zhang}
  J.}, \IN{Science}{327}{2010}{1619}.

\bibitem{Zavala2011}
\NAME{{Zavala} J., {Vogelsberger} M., {Slatyer} T.~R., {Loeb} A. \atque
  {Springel} V.}, \IN{\prd}{83}{2011}{123513}.

\bibitem{Huetsi2011}
\NAME{{H{\"u}tsi} G., {Chluba} J., {Hektor} A. \atque {Raidal} M.},
  \IN{\aap}{535}{2011}{A26}.

\bibitem{Aslanyan2015}
\NAME{{Aslanyan} G., {Price} L.~C., {Adams} J., {Bringmann} T., {Clark} H.~A.,
  {Easther} R., {Lewis} G.~F. \atque {Scott} P.},
  \IN{ArXiv:1512.04597}{}{2015}{}.

\bibitem{Jungman1996}
\NAME{{Jungman} G., {Kamionkowski} M. \atque {Griest} K.}, \IN{Physics
  Reports}{267}{1996}{195}.

\bibitem{Bertone2005}
\NAME{{Bertone} G., {Hooper} D. \atque {Silk} J.}, \IN{Physics
  Reports}{405}{2005}{279}.

\bibitem{Cirelli2009}
\NAME{{Cirelli} M., {Iocco} F. \atque {Panci} P.}, \IN{\jcap}{10}{2009}{9}.

\bibitem{Huetsi2009}
\NAME{{H{\"u}tsi} G., {Hektor} A. \atque {Raidal} M.},
  \IN{\aap}{505}{2009}{999}.

\bibitem{Slatyer2009}
\NAME{Slatyer T.~R., Padmanabhan N. \atque Finkbeiner D.~P.}, \IN{Physical
  Review D (Particles, Fields, Gravitation, and Cosmology)}{80}{2009}{043526}.
\newline\urlprefix\url{http://link.aps.org/abstract/PRD/v80/e043526}

\bibitem{Giesen2012}
\NAME{{Giesen} G., {Lesgourgues} J., {Audren} B. \atque {Ali-Ha{\"i}moud} Y.},
  \IN{\jcap}{12}{2012}{8}.

\bibitem{Diamanti2014}
\NAME{{Diamanti} R., {Lopez-Honorez} L., {Mena} O., {Palomares-Ruiz} S. \atque
  {Vincent} A.~C.}, \IN{\jcap}{2}{2014}{017}.

\bibitem{Padmanabhan2005}
\NAME{{Padmanabhan} N. \atque {Finkbeiner} D.~P.}, \IN{\prd}{72}{2005}{023508}.

\bibitem{Zhang2006}
\NAME{{Zhang} L., {Chen} X., {Lei} Y. \atque {Si} Z.},
  \IN{\prd}{74}{2006}{103519}.

\bibitem{Shull1985}
\NAME{{Shull} J.~M. \atque {van Steenberg} M.~E.}, \IN{\apj}{298}{1985}{268}.

\bibitem{Valdes2010}
\NAME{{Vald{\'e}s} M., {Evoli} C. \atque {Ferrara} A.},
  \IN{\mnras}{404}{2010}{1569}.

\bibitem{Galli2013}
\NAME{{Galli} S., {Slatyer} T.~R., {Valdes} M. \atque {Iocco} F.},
  \IN{\prd}{88}{2013}{063502}.

\bibitem{Slatyer2015}
\NAME{{Slatyer} T.~R.}, \IN{\prd}{93}{2016}{023521}.

\bibitem{McDonald2001}
\NAME{{McDonald} P., {Scherrer} R.~J. \atque {Walker} T.~P.},
  \IN{\prd}{63}{2001}{023001}.

\bibitem{Hu1993b}
\NAME{{Hu} W. \atque {Silk} J.}, \IN{Physical Review Letters}{70}{1993}{2661}.

\bibitem{Pospelov2007}
\NAME{{Pospelov} M. \atque {Ritz} A.}, \IN{Physics Letters B}{651}{2007}{208}.

\bibitem{Finkbeiner2007}
\NAME{{Finkbeiner} D.~P. \atque {Weiner} N.}, \IN{\prd}{76}{2007}{083519}.

\bibitem{Feng2003}
\NAME{{Feng} J.~L., {Rajaraman} A. \atque {Takayama} F.},
  \IN{\prd}{68}{2003}{063504}.

\bibitem{Feng2010}
\NAME{{Feng} J.~L.}, \IN{\araa}{48}{2010}{495}.

\bibitem{Dimastrogiovanni2015}
\NAME{{Dimastrogiovanni} E., {Krauss} L.~M. \atque {Chluba} J.},
  \IN{\prd}{94}{2016}{023518}.

\bibitem{Tashiro2013}
\NAME{{Tashiro} H., {Silk} J. \atque {Marsh} D.~J.~E.},
  \IN{\prd}{88}{2013}{125024}.

\bibitem{Carr2010}
\NAME{{Carr} B.~J., {Kohri} K., {Sendouda} Y. \atque {Yokoyama} J.},
  \IN{\prd}{81}{2010}{104019}.

\bibitem{Poulin2017}
\NAME{{Poulin} V., {Serpico} P.~D., {Calore} F., {Clesse} S. \atque {Kohri}
  K.}, \IN{\prd}{96}{2017}{083524}.

\bibitem{Kawasaki2018}
\NAME{{Kawasaki} M., {Kohri} K., {Moroi} T. \atque {Takaesu} Y.},
  \IN{\prd}{97}{2018}{023502}.

\bibitem{Planck2013ymap}
\NAME{{Planck Collaboration}, {Ade} P.~A.~R., {Aghanim} N., {Armitage-Caplan}
  C., {Arnaud} M., {Ashdown} M., {Atrio-Barandela} F., {Aumont} J.,
  {Baccigalupi} C., {Banday} A.~J. \atque et~al.}, \IN{ArXiv
  e-prints}{}{2013}{}.

\bibitem{Planck2015ymap}
\NAME{{Planck Collaboration}, {Aghanim} N., {Arnaud} M., {Ashdown} M., {Aumont}
  J., {Baccigalupi} C., {Banday} A.~J., {Barreiro} R.~B., {Bartlett} J.~G.,
  {Bartolo} N. \atque et~al.}, \IN{ArXiv:1502.01596}{}{2015}{}.

\bibitem{Komatsu2002}
\NAME{{Komatsu} E. \atque {Seljak} U.}, \IN{\mnras}{336}{2002}{1256}.

\bibitem{Diego2004}
\NAME{{Diego} J.~M. \atque {Majumdar} S.}, \IN{\mnras}{352}{2004}{993}.

\bibitem{Battaglia2010}
\NAME{{Battaglia} N., {Bond} J.~R., {Pfrommer} C., {Sievers} J.~L. \atque
  {Sijacki} D.}, \IN{\apj}{725}{2010}{91}.

\bibitem{Shaw2010}
\NAME{{Shaw} L.~D., {Nagai} D., {Bhattacharya} S. \atque {Lau} E.~T.},
  \IN{\apj}{725}{2010}{1452}.

\bibitem{Munshi2013}
\NAME{{Munshi} D., {Joudaki} S., {Smidt} J., {Coles} P. \atque {Kay} S.~T.},
  \IN{\mnras}{429}{2013}{1564}.

\bibitem{Pitrou2010}
\NAME{{Pitrou} C., {Bernardeau} F. \atque {Uzan} J.-P.},
  \IN{\jcap}{7}{2010}{19}.

\bibitem{Yu2001}
\NAME{{Yu} Q., {Spergel} D.~N. \atque {Ostriker} J.~P.},
  \IN{\apj}{558}{2001}{23}.

\bibitem{Lewis2013}
\NAME{{Lewis} A.}, \IN{\jcap}{8}{2013}{053}.

\bibitem{Loeb2001}
\NAME{{Loeb} A.}, \IN{\apjl}{555}{2001}{L1}.

\bibitem{Zaldarriaga2002}
\NAME{{Zaldarriaga} M. \atque {Loeb} A.}, \IN{\apj}{564}{2002}{52}.

\bibitem{Kaustuv2004}
\NAME{{Basu} K., {Hern{\'a}ndez-Monteagudo} C. \atque {Sunyaev} R.~A.},
  \IN{\aap}{416}{2004}{447}.

\bibitem{Carlos2007Pol}
\NAME{{Hern{\'a}ndez-Monteagudo} C., {Rubi{\~n}o-Mart{\'{\i}}n} J.~A. \atque
  {Sunyaev} R.~A.}, \IN{\mnras}{380}{2007}{1656}.

\bibitem{Jose2005}
\NAME{{Rubi{\~n}o-Mart{\'{\i}}n} J.~A., {Hern{\'a}ndez-Monteagudo} C. \atque
  {Sunyaev} R.~A.}, \IN{\aap}{438}{2005}{461}.

\bibitem{Danese1981}
\NAME{{Danese} L. \atque {de Zotti} G.}, \IN{\aap}{94}{1981}{L33}.

\bibitem{Balashev2015}
\NAME{{Balashev} S.~A., {Kholupenko} E.~E., {Chluba} J., {Ivanchik} A.~V.
  \atque {Varshalovich} D.~A.}, \IN{\apj}{810}{2015}{131}.

\bibitem{Burigana2018}
\NAME{{Burigana} C., {Carvalho} C.~S., {Trombetti} T., {Notari} A., {Quartin}
  M., {Gasperis} G.~D., {Buzzelli} A., {Vittorio} N., {De Zotti} G., {de
  Bernardis} P., {Chluba} J., {Bilicki} M., {Danese} L., {Delabrouille} J.,
  {Toffolatti} L., {Lapi} A., {Negrello} M., {Mazzotta} P., {Scott} D.,
  {Contreras} D., {Ach{\'u}carro} A., {Ade} P., {Allison} R., {Ashdown} M.,
  {Ballardini} M., {Banday} A.~J., {Banerji} R., {Bartlett} J., {Bartolo} N.,
  {Basak} S., {Bersanelli} M., {Bonaldi} A., {Bonato} M., {Borrill} J.,
  {Bouchet} F., {Boulanger} F., {Brinckmann} T., {Bucher} M., {Cabella} P.,
  {Cai} Z.-Y., {Calvo} M., {Castellano} M.~G., {Challinor} A., {Clesse} S.,
  {Colantoni} I., {Coppolecchia} A., {Crook} M., {D'Alessandro} G., {Diego}
  J.-M., {Di Marco} A., {Di Valentino} E., {Errard} J., {Feeney} S.,
  {Fern{\'a}ndez-Cobos} R., {Ferraro} S., {Finelli} F., {Forastieri} F.,
  {Galli} S., {G{\'e}nova-Santos} R., {Gerbino} M., {Gonz{\'a}lez-Nuevo} J.,
  {Grandis} S., {Greenslade} J., {Hagstotz} S., {Hanany} S., {Handley} W.,
  {Hern{\'a}ndez-Monteagudo} C., {Hervias-Caimapo} C., {Hills} M., {Hivon} E.,
  {Kiiveri} K., {Kisner} T., {Kitching} T., {Kunz} M., {Kurki-Suonio} H.,
  {Lamagna} L., {Lasenby} A., {Lattanzi} M., {Lesgourgues} J., {Liguori} M.,
  {Lindholm} V., {Lopez-Caniego} M., {Luzzi} G., {Maffei} B., {Mandolesi} N.,
  {Martinez-Gonzalez} E., {Martins} C.~J.~A.~P., {Masi} S., {Matarrese} S.,
  {McCarthy} D., {Melchiorri} A., {Melin} J.-B., {Molinari} D., {Monfardini}
  A., {Natoli} P., {Paiella} A., {Paoletti} D., {Patanchon} G., {Piat} M.,
  {Pisano} G., {Polastri} L., {Polenta} G., {Pollo} A., {Poulin} V.,
  {Remazeilles} M., {Roman} M., {Rubi{\~n}o-Mart{\'{\i}}n} J.-A., {Salvati} L.,
  {Tartari} A., {Tomasi} M., {Tramonte} D., {Trappe} N., {Tucker} C.,
  {V{\"a}liviita} J., {Van de Weijgaert} R., {van Tent} B., {Vennin} V.,
  {Vielva} P., {Young} K. \atque {Zannoni} M.}, \IN{\jcap}{4}{2018}{021}.

\bibitem{Kamionkowski2003}
\NAME{{Kamionkowski} M. \atque {Knox} L.}, \IN{\prd}{67}{2003}{063001}.

\bibitem{Chluba2004}
\NAME{{Chluba} J. \atque {Sunyaev} R.~A.}, \IN{\aap}{424}{2004}{389}.

\end{thebibliography}

\end{document}